\documentclass[12pt]{article}
\pdfoutput=1
\usepackage{putex}
\usepackage{amsmath}
\usepackage{amssymb}
\usepackage{graphicx}
\usepackage{epstopdf}
\usepackage{enumerate}
\usepackage{cite}
\usepackage{tensor}
\usepackage{slashed}
\usepackage{feynmf}
\usepackage{hyperref}

\usepackage{caption}
\usepackage{subcaption}

\usepackage[utf8x]{inputenc}
\usepackage{braket}
\usepackage{titlesec}
\usepackage{fixltx2e}
\usepackage{cleveref}
\hypersetup{breaklinks}
\numberwithin{equation}{section}

\def\vwvw{{\la V W(t)VW(t)\ra_{\b}\over \la VV\ra_\b\la W(t)W(t)\ra_\b}}
\def\W{{\cal W}}
\def\Gc{{\cal G}}
\def\Reg{{\rm Regge}}

\def\H{{\cal H}}
\def\ab{\bar\alpha}
\def\2{{\bf 20'}}
\def\winf{W_{\infty}[\lambda]}
\def\finf{{\cal F}_{{\rm vac},\infty}^{(1)}}
\def\qvs{q_v^{(s)}}
\def\qws{q_w^{(s)}}

\def\ssec{\subsection}
\def\sssec{\subsubsection}
\def\sec{\section}
\newcommand{\Farg}[3]{\left( \begin{array}{c} #1 \\ #2\end{array} \Big| #3 \right)}
\def\i{\infty}
\newcommand{\G}[1] {\Gamma\left({#1}\right)}

\def\F{{\cal F}}
\def\O{{\cal O}}
\def\ra{\rangle}
\def\la{\langle}
\def\pb{\bar \partial}
\def\Db{\bar{D}}
\def\t{\tau}
\def\vac{{\rm vac}}

\def\hsl{hs[$\lambda$]}

\def\Oc{{\cal O}}
\def\vs{\vskip .1 in}
\def\z{Zamolodchikov}
\def\D{\Delta}
\def\hyp{{}_2F_1}

\def\g{\gamma}
\def\a{\alpha}

\newcommand {\be} {\begin {equation}}
\newcommand {\ee} {\end {equation}}

\newcommand {\bes} {\begin {equation*}}
\newcommand {\ees} {\end {equation*}}

\newcommand{\es}[2] {\begin{equation} \label{#1} \begin{split} #2 \end{split} \end{equation}}
\newcommand{\e}[2] {\begin{equation} \label{#1} #2 \end{equation}}

\newcommand{\Z}{\mathbb{Z}}
\newcommand{\N}{\mathbb{N}}
\newcommand{\R}{\mathbb{R}}

\newcommand{\beq}{\begin{equation}}
\newcommand{\eeq}{\end{equation}}

\newcommand{\cF}{\mathcal{F}}

\newcommand{\Li}{\mathrm{Li}}

\newcommand{\p}{\partial}

\def\be{ \begin{equation} }
\def\ee{ \end{equation} }

\def\half{{1\over  2}}

\def\subsubsec{\subsubsection}
\def\subsec{\subsection}
\def\N{{\cal N}}
\def\A{{\cal A}}

\def\cF{{\cal F}}

\def\eqr{\eqref}

\def\b{\beta}
\def\l{\lambda}

\renewcommand{\Im}{\textrm{Im}\,}
\renewcommand{\Re}{\textrm{Re}\,}

\newcommand\zb{\bar{z}}

\def\rar{\rightarrow}

\def\hb{\overline{h}}
\def\zb{\overline{z}}

\def\eps{\epsilon}

\titleformat*{\section}{\bfseries\Large }
\titleformat*{\subsection}{\bfseries \large}
\titleformat*{\subsubsection}{\bfseries}

\begin{document}

\title{\LARGE Bounding the Space of Holographic CFTs with Chaos}

\authors{Eric Perlmutter}
\institution{PU}{Department of Physics, Princeton University, Princeton, NJ 08544, USA}

\abstract{Thermal states of quantum systems with many degrees of freedom are subject to a bound on the rate of onset of chaos, including a bound on the Lyapunov exponent, $\lambda_L\leq 2\pi /\beta$. We harness this bound to constrain the space of putative holographic CFTs and their would-be dual theories of AdS gravity. First, by studying out-of-time-order four-point functions, we discuss how $\lambda_L=2\pi/\beta$ in ordinary two-dimensional holographic CFTs is related to properties of the OPE at strong coupling. We then rule out the existence of unitary, sparse two-dimensional CFTs with large central charge and a set of higher spin currents of bounded spin; this implies the inconsistency of weakly coupled AdS$_3$ higher spin gravities without infinite towers of gauge fields, such as the $SL(N)$ theories. This fits naturally with the structure of higher-dimensional gravity, where finite towers of higher spin fields lead to acausality. On the other hand, unitary CFTs with classical $W_{\infty}[\lambda]$ symmetry, dual to 3D Vasiliev or hs[$\lambda$] higher spin gravities, do not violate the chaos bound, instead exhibiting no chaos: $\lambda_L=0$. Independently, we show that such theories violate unitarity for $|\lambda|>2$. These results encourage a tensionless string theory interpretation of the 3D Vasiliev theory. We also perform some CFT calculations of chaos in Rindler space in various dimensions.}

\date{}

\maketitle
\setcounter{tocdepth}{2}
\tableofcontents

\section{Introduction and summary}
The study of quantum chaos has lent new perspectives on thermal physics of conformal field theories and gravity \cite{Shenker:2013pqa, Maldacena:2015waa}. Geometric structure in the bulk may be destroyed by small perturbations whose effects grow in time and spread in space, otherwise known as the butterfly effect. This accounts for scrambling by black holes, destroys entanglement, and, via holography, gives a view into Lorentzian dynamics of conformal field theories at large central charge \cite{Shenker:2013pqa, Shenker:2013yza, Leichenauer:2014nxa, kitaev, Roberts:2014isa, Roberts:2014ifa, Jackson:2014nla, Shenker:2014cwa, Maldacena:2015waa, Polchinski:2015cea, Caputa:2015waa, Hosur:2015ylk, Stanford:2015owe, Gur-Ari:2015rcq, Berkowitz:2016znt, Polchinski:2016xgd, Fitzpatrick:2016thx, Michel:2016kwn}. 
  
Inspired by the classical picture \cite{larkin}, a quantity that has been identified as a sharp diagnostic of quantum chaos is the out-of-time-order (OTO) four-point correlation function between pairs of local operators,
\e{i1}{\la V W(t)VW(t)\ra_\b~.}
We use a common notation: $V$ sits at $t=0$, and the operators are separated in space. The onset of chaos is seen as an exponential decay in time of this correlator, controlled by $\exp(\l_L t)$. The rate of onset is set by $\l_L$, the Lyapunov exponent. Under certain conditions that are easily satisfied by many reasonable thermal systems, $\l_L$ is bounded above by \cite{Maldacena:2015waa}
\e{bound}{\l_L\leq {2\pi\over\b}~.}
The bound is saturated by Einstein gravity, nature's fastest scrambler \cite{Sekino:2008he}. Understanding how exactly this bound fits into the broader picture of CFT constraints and their relation to the emergence of bulk spacetime, and studying the range of chaotic behaviors of CFTs more generally, are the general goals of this paper.

This work touches on various themes in recent study of conformal field theory and the AdS/CFT correspondence. The first is the delineation of the space of CFTs. An abstract CFT is (perturbatively) specified by the spectrum of local operators and their OPE coefficients: $\lbrace \D_i, C_{ijk}\rbrace$. As evidenced by the conformal bootstrap, imposing crossing symmetry and unitarity leads to powerful constraints on this data. It is not yet known what the precise relation is between $\lbrace \D_i, C_{ijk}\rbrace$ and the chaotic properties of a generic CFT, say, $\l_L$. One would like to use OTO correlators to constrain the CFT landscape: given the existence of a bound on chaos, a natural goal is to exclude certain putative CFTs which violate it. This tack would provide a Lorentzian approach to the classification of CFTs.

The strong form of the AdS/CFT correspondence posits that every CFT is dual to a theory of quantum gravity in AdS. At the least, a subspace of all CFTs can be mapped via holography to the space of weakly coupled theories of gravity or string/M-theory in AdS. Given that string and M-theory are tightly constrained by their symmetries, this suggests that any consistent CFT possesses a level of substructure over and above the manifest requirements of conformal symmetry. One might hope to enlist chaos in the quest to ``see'' the structure of AdS string or M-theory compactifications from CFT. 

At large central charge $c$ and with a sufficiently sparse spectrum of light operators $\D_i\ll c$, a universality emerges: such CFTs appear to be dual to weakly coupled theories of AdS gravity, that in the simplest cases contain Einstein gravity. These CFTs obey certain other unobvious constraints: for example, corrections to $a-c$ in four-dimensional CFTs are controlled by the higher spin spectrum \cite{Camanho:2014apa}. Identifying the set of sufficient conditions for the emergence of a local bulk dual is an open problem. There is already evidence that $\l_L=2\pi/\b$ is at least a necessary criterion, but one would like to make a sharper statement. Explicitly connecting the value of $\l_L$ with the strong coupling OPE data would permit a direct derivation of $\l_L=2\pi/\b$ from CFT, which is presently lacking in $d>2$ and has been done under certain conditions on the operators $V$ and $W$ in $d=2$ \cite{Roberts:2014ifa}.

Not all weakly coupled theories of gravity are local: one can, for instance, add higher spin fields. In AdS$_{D>3}$, there are no-go results: namely, one cannot add a finite number of either massive or massless higher spin fields, for reasons of causality \cite{Camanho:2014apa} and -- in the case of massless fields -- symmetry \cite{Bekaert:2010hw, Maldacena:2011jn}. In AdS$_3$, the constraints are less strict. For one, the graviton is non-propagating. Moreover, higher spin algebras, i.e. $\W$-algebras, with a finite number of currents do exist. 

Consider theories which augment the metric with an infinite tower of higher spin gauge fields. Other than string theory, these include the Vasiliev theories \cite{Vasiliev:1990en, vasstar, Prokushkin:1998bq}; see \cite{Giombi:2012ms, Didenko:2014dwa} for recent reviews. These are famously dual to $O(N)$ vector models in $d\geq 3$ CFT dimensions (and, in $d=3$, Chern-Simons deformations thereof \cite{Giombi:2011kc,Aharony:2011jz}). One widely held motivation for studying the Vasiliev theories in $d$ dimensions is that they morally capture the leading Regge trajectory of tensionless strings in AdS \cite{Sundborg:2000wp}. For the supersymmetric AdS$_3$ Vasiliev theory with so-called shs$_2[\l]$ symmetry, this is now shown to be literally true \cite{Gaberdiel:2014cha, Gaberdiel:2015mra, Gaberdiel:2015wpo}: CFT arguments imply that this super-Vasiliev theory forms a closed subsector of type IIB string theory on AdS$_3\times S^3 \times T^4$ in the tensionless limit, $\a'\rar\i$. More generally, it is unclear whether other, e.g. non-supersymmetric, Vasiliev theories are UV complete, or whether they can always be viewed as a consistent subsector of a bona fide string theory. 
  
In AdS$_3$, there seem to be other consistent theories of higher spin gravity: to every $\W$-algebra arising as the Drinfeld-Sokolov construction of a Lie algebra $G$, one can associate a pure higher spin gravity in AdS$_3$ cast as a $G\times G$ Chern-Simons theory. This builds on the original observation that general relativity in AdS$_3$ can be written in this fashion with $G=SL(2,\R)$ \cite{Achucarro:1987vz, Witten:1988hc}. Such pure Chern-Simons theories have been studied in the context of AdS/CFT, especially for $G=SL(N,\R)$ and $G=$ \hsl. The former contains a single higher spin gauge field at every integer spin $2\leq s \leq N$ which generate an asymptotic $W_N$ symmetry \cite{Campoleoni:2010zq}. The latter, a one-parameter family labeled by $\l$, contains one higher spin gauge field at every integer spin $s\geq 2$ which generate an asymptotic $\winf$ symmetry \cite{Pope:1989sr, Henneaux:2010xg, Gaberdiel:2011wb}. The 3D Vasiliev theory \cite{Prokushkin:1998bq} contains the \hsl\ theory as a closed subsector. All of these theories should be viewed as capturing the universal dynamics of their respective $\W$-algebras at large central charge. These theories have been studied on the level of the construction of higher spin black holes and their partition functions (e.g. \cite{Gutperle:2011kf, Ammon:2012wc, Bunster:2014mua, deBoer:2014fra}), entanglement and R\'enyi entropies and Wilson line probes (e.g. \cite{Ammon:2013hba,  deBoer:2013vca, Chen:2013dxa, Perlmutter:2013paa,Datta:2014ska, Long:2014oxa, deBoer:2014sna}), conformal blocks \cite{deBoer:2014sna,Hegde:2015dqh}, and flat space limits \cite{Afshar:2013vka}, among other things. 

To introduce dynamics, one would like to consistently couple these pure higher spin theories to matter, or embed them into string theory. However, it is far from clear that these $SL(N)$-type theories, with finite towers of higher spin gauge fields, are not pathological. The notion of a finite tower of higher spin fields feels quite unnatural, and is highly unlikely to descend from string theory. A heuristic argument is that in a tensionless limit $\a'\rar\i$, {\it all} operators on the lowest Regge trajectory would become massless, not only a finite set; then if we are guided by the principle that every CFT is dual to (some limit of) a string theory in AdS, or by some milder notion of string universality \cite{Haehl:2014yla,Belin:2014fna}, the notion of a holographic, unitary 2d CFT with a finite number of higher spin currents seems suspicious. As an empirical matter, the only known construction of a fully nonlinear AdS$_3$ higher spin gravity coupled to matter that is consistent with unitarity is the Vasiliev theory, which has an infinite tower of higher spin currents; likewise, there are no known $W_N$ CFTs with the aforementioned properties.

In this paper, we will initiate a systematic treatment of chaotic OTO correlators in CFTs with weakly coupled holographic duals. We will realize some of the goals mentioned above. Our results are in the spirit of the conformal bootstrap program: we exclude regions of the CFT landscape by imposing consistency properties on correlation functions. In our setting, we are working with Lorentzian, out-of-time-order correlators, relating dynamical statements about the development of quantum chaos and scrambling in thermal systems \cite{Shenker:2013pqa, Maldacena:2015waa} to the question of UV completeness. Our work has a similar flavor to \cite{Hartman:2015lfa, Hartman:2016dxc}, which uses the Lorentzian bootstrap to enforce causality in shock wave backgrounds. 

\ssec{Summary of results}
Our basic philosophy is, following \cite{Roberts:2014ifa}, to study OTO four-point functions of the form \eqr{i1} in $d$-dimensional CFTs by computing vacuum four-point functions, and performing a conformal transformation to a thermal state. In $d>2$, this yields the Rindler thermal state. In $d=2$, this yields the thermal state of the CFT on a line with arbitrary $\b$. In the large $c$ limit, we diagnose chaos by looking at planar correlators; in particular, we study their Regge limits. A conformal transformation leads to an OTO correlator of the form
\e{i3}{\vwvw \approx 1+{e^{\l_L t}\over c}{f(x)+\ldots}}
It follows that in these thermal states, the chaotic properties of the CFT can in principle be inferred from OPE data at $\O(1/c)$. In this paper, we make this concrete. (This last statement assumes that $V$ and $W$ are light operators, with conformal dimensions parametrically less than $c$, but we will also treat the case of $V$ and $W$ being heavy in $d=2$, with similar results.)

\sssec*{Chaotic correlators in holographic CFTs}

In Section \ref{s3}, we consider chaos in CFTs with weakly coupled local gravity duals, with no higher spin currents.\footnote{Throughout the paper, the phrase ``weakly coupled gravity'' means $G_N\ll1$ and generically includes a coupling to a finite number of matter fields; if we wish to denote the gravitational sector alone, we refer to ``pure gravity.''} We take $V$ and $W$ to be arbitrary light scalar primaries. By recalling properties of the strongly coupled OPE at $\O(1/c)$, we argue that for times $\b\ll t \ll {\b\over 2\pi}\log c$,
\e{i4}{\vwvw \sim 1-{i\over c \eps_{12}^*\eps_{34}}e^{{2\pi\over \b}(t-x)}f(x)+\ldots}
where the $\eps_{ij}$ parameterize the Euclidean times of the operators, in a notation explained below and borrowed from \cite{Roberts:2014ifa}. $f(x)$ is a function of the spatial separation whose general form we determine; see equation \eqr{feta12}, where $\eta = \exp(-4\pi x/\b)$. \eqr{i4} implies $\l_L=2\pi/\b$ and $t_* = \l_L^{-1}\log c$, matching the Einstein gravity behavior. 

We emphasize that our goal is not to prove this from CFT, but to understand how it relates to properties of the gravity duals. Accordingly our arguments are not made from first principles in CFT; for example, we do not rigorously apply conformal Regge theory techniques \cite{Costa:2012cb} to derive sufficient conditions for \eqr{i4} using CFT arguments alone, which is itself an interesting open problem. Rather, our arguments are based on necessary conditions a prototypical CFT must satisfy for the existence of an emergent local bulk theory, in the spirit of \cite{Heemskerk:2009pn, ElShowk:2011ag}, and known facts about correlators computed from the bulk. This analysis is analogous to Section 6 of \cite{Maldacena:2015iua}, where the emergence of bulk-point singularities is derived from properties of the OPE at strong coupling.  We also introduce a simplified toy model in which the results above may be derived.

One way to phrase \eqr{i4} is that $\l_L$ may be read off from the stress tensor exchange alone. That is, the spin of the highest-spin current in a holographic CFT determines $\lambda_L$. This is a key principle in what follows. A corollary of our result is a derivation of the butterfly velocity in Rindler space, 
\e{i5}{v_B={1\over d-1}\quad\quad\quad (\text{Rindler})}
This is determined by the exchange of the lowest-twist spin-2 operator, which is the stress tensor. 

In Appendix \ref{appb}, we give an example of \eqr{i4} in strongly coupled $\N=4$ super-Yang-Mills (SYM), where we take $V=W=\O_\2$. Our analysis also clarifies the relationship between the sparseness condition and $\l_L=2\pi/\b$: in particular, this result is somewhat insensitive to the density of scalar and vector primary operators, and does not require the strictest definition of sparseness.

\sssec*{Chaotic destruction of higher spin theories}
In Section \ref{s4} we focus on $d=2$, and upgrade the previous analysis to include higher spin currents of bounded spin $s\leq N$, where $N>2$ is finite. The same principles imply that for generic $V$ and $W$,
\e{i6}{\vwvw \sim 1-{i\over c \eps_{12}^*\eps_{34}}e^{{2\pi\over \b}(N-1)t}f(x)+\ldots}
The Lyapunov exponent is
\e{i7}{\l_L={2\pi\over\b}(N-1)}
This violates the chaos bound. It follows from our assumptions that unitary, holographic 2d CFTs with finite towers of higher spin currents do not exist. Not only would such CFTs violate the chaos bound, but as we review, the results of \cite{Hartman:2015lfa} imply that they would be acausal: that is, these higher spin CFTs would be too-fast scramblers. 
 \begin{figure}[t!]
   \begin{center}
 \includegraphics[width = .4\textwidth]{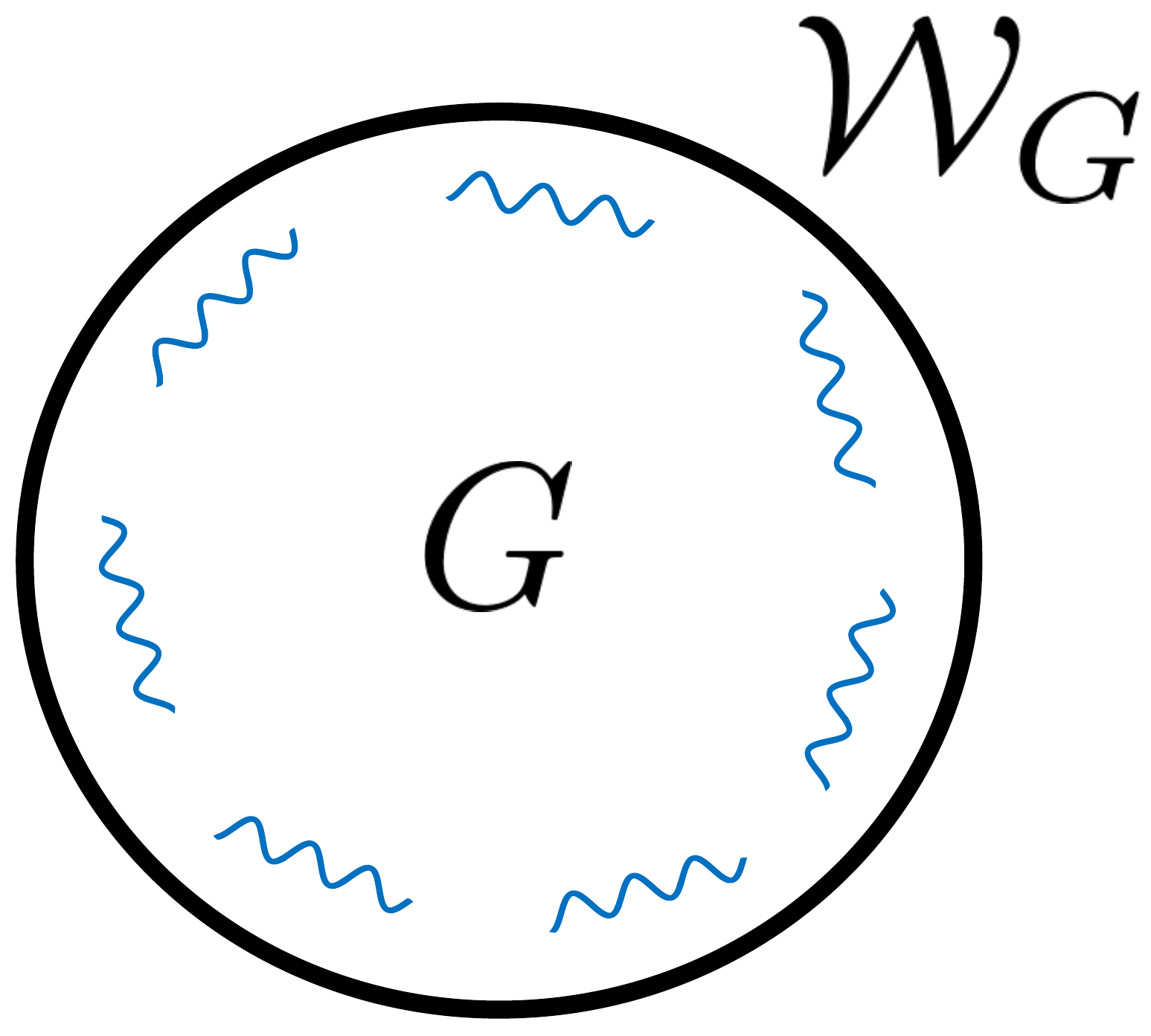}
 \caption{A weakly coupled theory of higher spin gravity in AdS$_3$ may be viewed as matter coupled to a $G\times G$ Chern-Simons theory for some Lie algebra $G$. The boundary gravitons of $G$ generate an asymptotic $\W$-symmetry, $\W_G$. When rank($G$) is finite, such theories are inconsistent.}
 \label{wg}
 \end{center}
 \end{figure}

To bolster these claims, we rigorously study the case where $V$ and $W$ have higher spin charges that scale with large $c$. This was inspired by the analogous calculation in the Virasoro context \cite{Roberts:2014ifa}. In sparse CFTs, correlators of such operators are computed exactly, to leading order in large $c$, by the semiclassical vacuum conformal block of the $\W$-algebra \cite{Hartman:2013mia, Hartman:2014oaa,Asplund:2014coa, deBoer:2014sna, Hegde:2015dqh}. Taking $\W=W_N$, the semiclassical $W_N$ vacuum block is now known in closed form for any $N$ \cite{Hegde:2015dqh}, so we can compute its Regge limit explicitly. The resulting function again violates the chaos bound. (See e.g. equation \eqr{cie} for $N=3$.)

The bulk dual statement is that weakly coupled higher spin gravities with finite towers of higher spin gauge fields are inconsistent. This rules out the $SL(N)$ higher spin gravities. As we explain in Section \ref{s43}, our CFT calculations of the OTO correlators map directly to the way one would calculate the same quantity in the bulk, via certain bulk Wilson line operators studied in \cite{Ammon:2013hba,  deBoer:2013vca, Castro:2014mza, deBoer:2014sna,Hegde:2015dqh}; they therefore constitute a direct bulk calculation as well. In purely bulk language, the problem can be equivalently phrased as an acausality of a ``higher spin shock wave'' induced by a higher spin-charged perturbation of the planar BTZ black hole. Alternatively, the Regge limit of AdS$_3$ Mellin amplitudes grows too fast. A corollary of our result is that $SL(N)$-type higher spin gravities cannot be coupled to string or M-theory.

This result fits very nicely with known features of higher-dimensional gravity. Weakly coupled theories of gravity in AdS$_{D>3}$ with a finite number of higher spin fields suffer from violations of causality \cite{Camanho:2014apa}. In AdS$_3$, this conclusion does not hold, as the bulk graviton is non-propagating. However, upon introducing matter, the physics is the same. It is useful to compare the status of $SL(N)$-type theories in AdS$_3$ with Gauss-Bonnet theory in AdS$_{D>3}$. Whereas both require infinite towers of higher spin degrees of freedom to be completed (though see \cite{Papallo:2015rna}), Gauss-Bonnet comes with a coupling $\l_{\rm GB}\sim M_{\rm GB}^2$, which determines the energy scale $E\sim M_{\rm GB}$ at which massive higher spin fields must appear to restore causality; on the other hand, $SL(N)$ gravity has no scale besides $L_{\rm AdS}$, and cannot be viewed as an effective field theory. 

Altogether, this reduces various computations in $SL(N)$ higher spin theories -- including entanglement entropies as Wilson lines, efforts to find gauge-invariant causal structure, and models of higher spin black hole formation -- to algebraic statements about $\W$-algebra representation theory, rather than dynamical statements about actual unitary, causal CFTs.\footnote{A special case not covered by the above is pure higher spin gravity. We are skeptical that such theories exist for small $G_N$; if they do, they are devoid of dynamics.} For non-dynamical questions, the $SL(N)$ theories may still be useful: for example, they remain approachable toy models of more complicated theories of higher spin gravity and of stringy geometry; some of their observables may be analytically continued   to derive results in \hsl\ higher spin gravity (e.g. \cite{Gaberdiel:2011zw, Gaberdiel:2012ku, Gaberdiel:2013jca, Campoleoni:2013lma, Hegde:2015dqh}); and $\W$-algebras do appear widely in CFT.\footnote{Non-unitarity also has its place in the AdS/CFT dictionary, e.g. \cite{Grumiller:2008qz, Anninos:2011ui, Gaberdiel:2012ku, Perlmutter:2012ds, Vafa:2014iua}, although ideally not packaged with acausality.} 

This begs the question of what happens when an infinite tower of massless higher spin fields is introduced.

\sssec*{Regge behavior in $\winf$ CFTs and 3D Vasiliev theory}

In Section \ref{s5}, we consider chaos in 2d CFTs with $\winf$ symmetry. We continue to apply the principle that the $\winf$ vacuum block at $\O(1/c)$ can be used to derive $\l_L$. Doing so requires deriving its Regge limit. This in turn requires performing the infinite sum over single higher spin current exchanges; see Figure \ref{wdecomp}. The result is highly sensitive to the relations among OPE coefficients, i.e. the higher spin charges of $V$ and $W$. Taking both $V$ and $W$ to sit in the simplest representation of $\winf$, i.e. the fundamental representation (which obeys unitarity for $\l\geq -1$), we find a remarkably simple result for the $\winf$ vacuum block:
\e{i8}{\cF_{{\rm vac},\infty}(z|\l)  = 1+{(1-\l^2)\over c}\big(z\,{}_2F_1(1,1,1-\l;z)+\log(1-z)\big)+\O\left({1\over c^2}\right)~.}
In the Regge limit, the $\O(1/c)$ term goes like a constant, which implies 
\e{i9}{\l_L=0~.}
There is no chaos. The result $\l_L=0$ is non-trivial, unlike higher spin CFTs in $d>2$, because CFTs with $W_{\infty}[\l]$ symmetry are not necessarily free. We expect that \eqr{i8} will find other applications. 

This has intriguing implications for the status of non-supersymmetric Vasiliev theory. The quantum numbers of $V$ and $W$ chosen above are those of the Vasiliev scalar field. We believe that our result encourages the tensionless string theory interpretation described earlier. One especially relevant feature of string theory for our purposes is the phenomenon of Regge-ization of amplitudes, in which infinite towers of massive string states sum up to give soft high-energy behavior \cite{Amati:1987wq, Amati:1987uf,Amati:1988tn}. It is sometimes said that string theory is the unique theory with a consistent sum over higher spin states. Our calculation suggests that this is not strictly true: the non-supersymmetric 3D Vasiliev theory provides another, simpler example. This suggests that the non-supersymmetric Vasiliev theory may be shown to be a limit or subsector of string theory, as in the supersymmetric AdS$_3\times S^3\times T^4$ case described above.

As an aside, using arguments independent of chaos, we show that unitary CFTs with a classical (that is, large $c$) $W_{\i}[\l]$ chiral algebra with $\l>2$ do not exist. Correspondingly, 3D Vasiliev and pure \hsl\ higher spin gravities with $\l>2$ have imaginary gauge field scattering amplitudes. This follows from the fact that, as we show, the classical $\winf$ algebra is actually complex for $\l>2$.

\sssec*{AdS/CFT sans chaos}
In Section \ref{s6}, we give a selection of chaotic computations in familiar CFTs that are relevant to AdS/CFT: namely, chiral 2d CFTs, symmetric orbifold CFTs, and CFTs with slightly broken higher spin symmetry. Chiral CFTs are non-chaotic. We perform an explicit computation of an OTO correlator in the D1-D5 CFT at its orbifold point, Sym$^N(T^4)$, again finding an absence of chaos. Finally, we argue that in slightly broken higher spin CFTs \cite{Maldacena:2012sf} in arbitrary dimension, the Lyapunov exponent in thermal states on $S^1\times \mathbb{R}^{d-1}$ should vanish to leading order in $1/c$. This gives a physical motivation to study $\l_L$ to higher orders in $1/c$. 

\centerline{\noindent\rule{7cm}{0.4pt}}
\vs
\vs
 
The sections outlined above are bookended by a short Section \ref{s2}, in which we set up the calculations and briefly review the Regge limit and the chaos bound; and by a discussion in Section \ref{s7}. Finally, we include a handful of appendices with supplementary calculations. 

\section{Chaotic Correlators}\label{s2}
We will study OTO four-point functions of pairs of local primary operators in thermal states of $d$-dimensional CFTs,
\e{e21}{{\la V^\dag W^\dag(t) V W(t)\ra_\b\over \la V^\dag V\ra_\b \la W^\dag (t)W(t)\ra_\b}}
where operators are time-ordered as written. We achieve this by a conformal transformation from the vacuum. We focus mostly on $d=2$ CFTs on the cylinder with inverse temperature $\b$, so we set up the problem in those variables; the result for $d$-dimensional Rindler space can be read off at the end by setting $\b=2\pi$.

Consider local scalar primary operators $V,W$ with respective conformal weights $h_v=\hb_v$ and $h_w=\hb_w$, where more generally,
\e{e2}{(h,\hb) = \left({\D+s\over 2},{\D-s\over 2}\right)}
Conformal invariance constrains the vacuum four-point function of $V$ and $W$ to take the form
\e{cra}{{\la  V^{\dag}(z_1,\zb_1)V(z_2,\zb_2) W^{\dag}(z_3,\zb_3)W(z_4,\zb_4)\ra\over \la  V^{\dag}(z_1,\zb_1)V(z_2,\zb_2) \ra \la W^{\dag}(z_3,\zb_3)W(z_4,\zb_4)\ra} = {\A(z,\zb)}}
for some function $\A(z,\zb)$ of the conformally invariant cross-ratios,
\e{e3}{z={z_{12}z_{34}\over z_{13}z_{24}}~, \quad \zb={\zb_{12}\zb_{34}\over \zb_{13}\zb_{24}}}
where $z_{ij}\equiv z_i-z_j$ as usual. We refer to $\A(z,\zb)$ as a reduced amplitude. It is invariant under the conformal map to the cylinder,
\e{e4}{z = e^{{2\pi\over\b}(t-x)}~, \quad \zb = e^{{2\pi\over\b}(-t-x)}}
with thermal periodicity $t \sim t+i \b$.
For Euclidean correlators, $\zb_i=z_i^*$, and time is imaginary. In Lorentzian configurations, $\zb_i\neq z_i^*$, and time is complex. 

As explained in \cite{Roberts:2014ifa}, the OTO correlator \eqr{e21} can be obtained by a particular analytic continuation of \eqr{cra} from the Euclidean regime. When $\zb\neq z^*$, correlators have a branch cut running from $z\in(1,\i)$, and the analytic continuation requires crossing this cut. This leads to new singularities. The relevant limit for chaos has been studied before, known as the Regge limit, and
boils down to the following manipulations of $\A(z,\zb)$. First, take $z$ clockwise around the branch point at $z=1$, leaving $\zb$ alone:\footnote{We occasionally will use the following notation: $\rar$ means we pass to the second sheet, and $\sim$ means that we keep the leading behavior near $z=0$ on the second sheet, i.e. we take the Regge limit.}
\es{pres}{(1-z) \rar e^{-2\pi i}(1-z)}
This defines a Lorentzian amplitude in which $z$, but not $\zb$, lives on the second sheet of the function $\A(z,\zb)$. Then take $z$ and $\zb$ to zero, holding their ratio fixed:
\e{}{z\rar 0~,\quad \zb\rar 0~, \quad {\zb\over  z}\equiv\eta~\text{fixed}}
This defines $\A^\Reg(z,\eta)$. 

\begin{figure}
\centering
\begin{subfigure}{.5\textwidth}
  \centering
  \includegraphics[width=1.\linewidth]{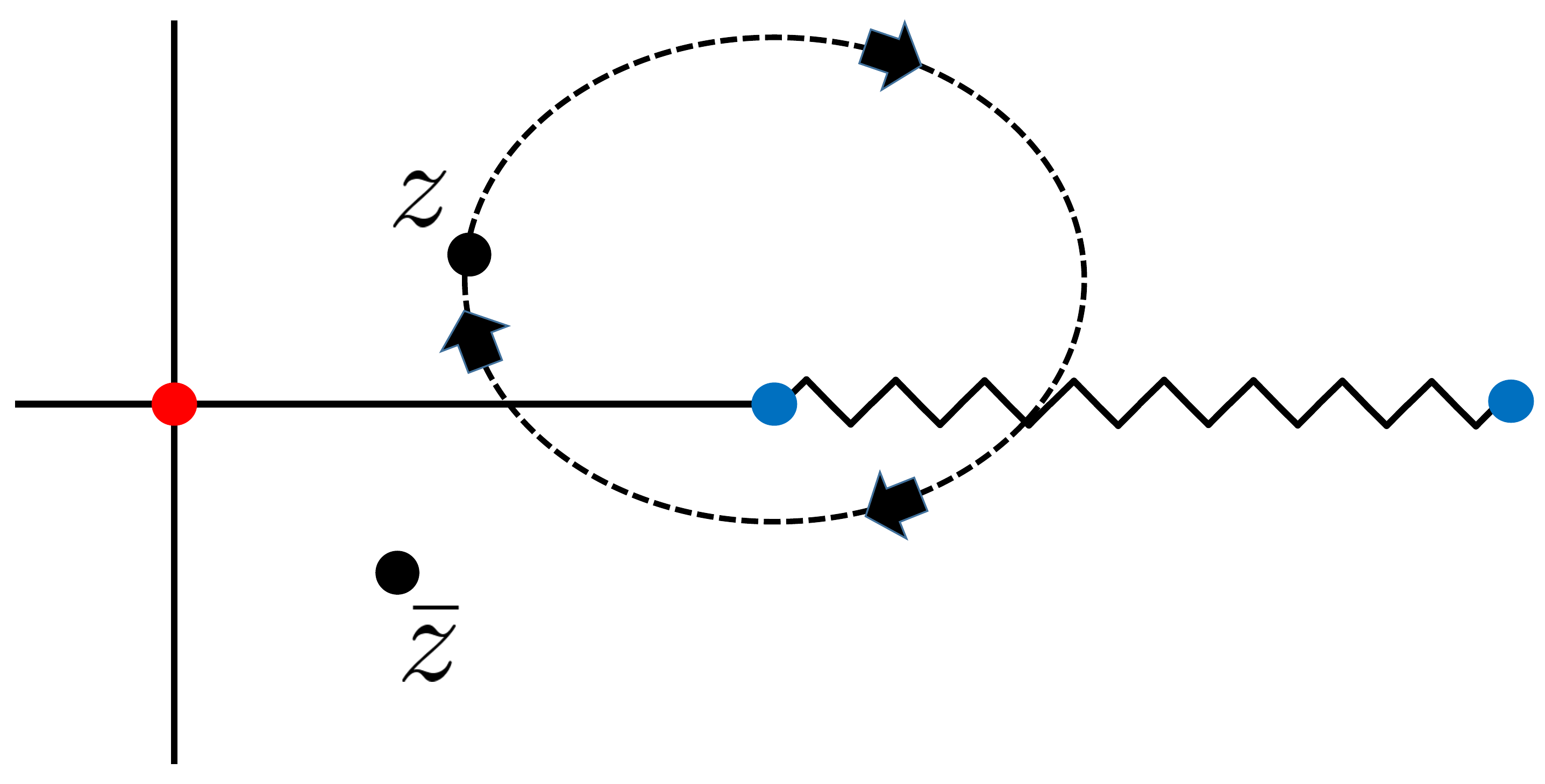}
  \label{fig:sub1}
\end{subfigure}%
\begin{subfigure}{.5\textwidth}
  \centering
  \includegraphics[width=.48\linewidth]{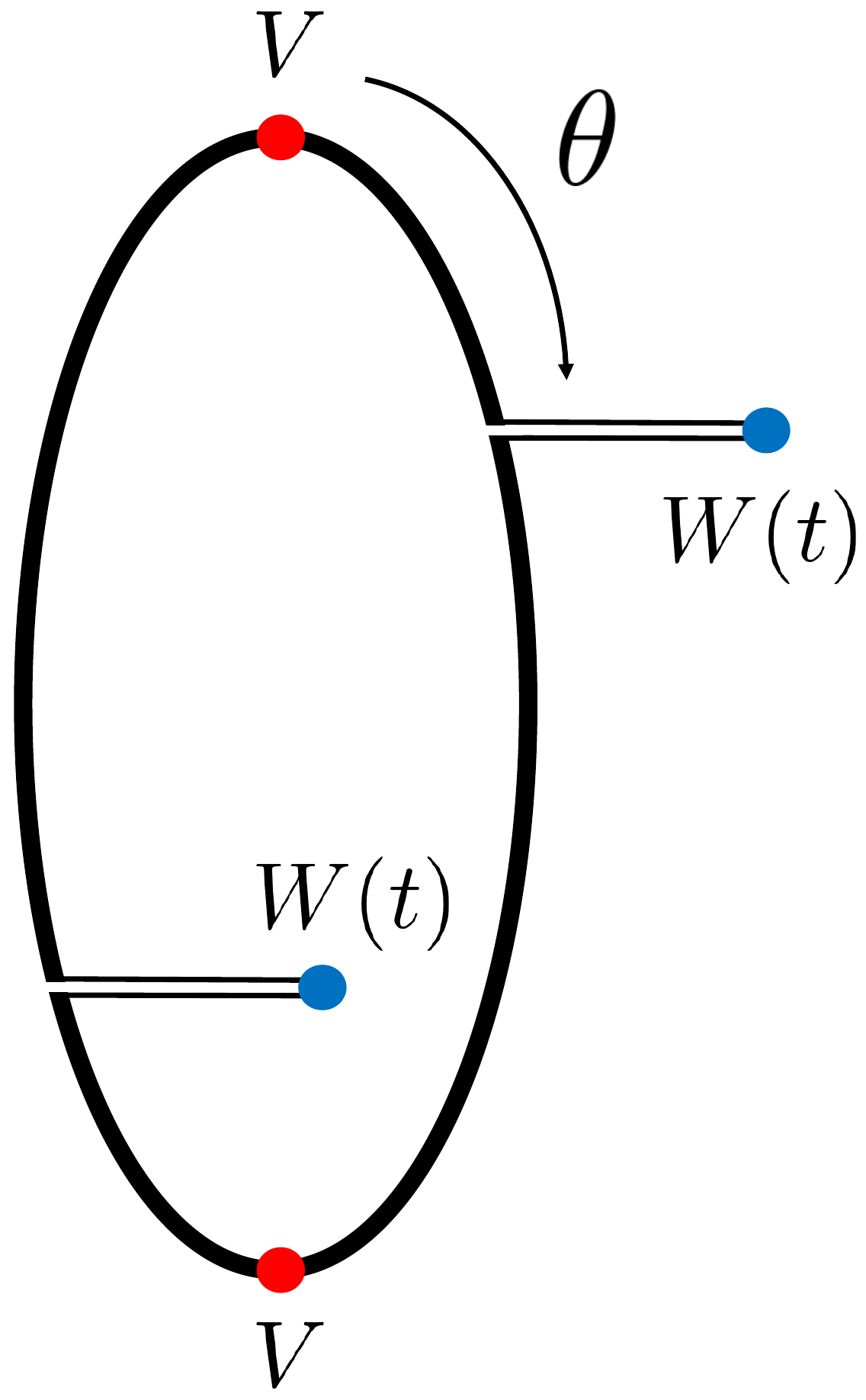}
  \label{fig:sub2}
\end{subfigure}
\caption{On the left, the analytic continuation of $z$ onto the second sheet, passing clockwise around $z=1$. On the right, the canonical \cite{Shenker:2014cwa} diagram of the out-of-time-order arrangement of operators along the Euclidean time circle. $W(t)$ undergoes Lorentzian time evolution orthogonal to the circle. We have placed the operators diametrically opposite from one another in pairs, as in \eqr{e26}, with the two pairs separated by an angle $\theta$. In terms of the imaginary time $\tau$, $\theta = {\pi\over 2}-{2\pi\over\beta}\tau$.}
\label{Lor}
\end{figure}

The appearance of the Regge limit is explained as follows. To obtain the OTO correlator of interest from the Euclidean amplitude $\A(z,\zb)$, one assigns complex time $t_i+i\eps_i$ to each operator, then evolves the $t_i$ to the desired kinematic configuration. We take $V$ and $V^{\dag}$ to sit at $t=x=0$, and $W$ and $W^\dag$ to sit at $t>x>0$:
\es{}{z_1 &= e^{{2\pi\over\b}i \eps_1}~, \quad\quad\quad~ \zb_1 = e^{-{2\pi\over\b}i \eps_1}\\
z_2 &= e^{{2\pi\over\b}i \eps_2}~, \quad\quad\quad~ \zb_2 = e^{-{2\pi\over\b}i \eps_2}\\
z_3 &= e^{{2\pi\over\b}(t+i \eps_3-x)}~, \quad \zb_3 = e^{{2\pi\over\b}(-t-i \eps_3-x)}\\
z_4 &= e^{{2\pi\over\b}(t+i \eps_4-x)}~, \quad \zb_4 = e^{{2\pi\over\b}(-t-i \eps_4-x)}}
The cross-ratio can be read off from \eqr{e3}.  At $t=x$, $W$ crosses the lightcone of $V$, and $z$ passes through the cut. See Figure \ref{Lor}. 
Evolving to later times $t-x\gg\b$ puts us in the Regge regime, with the desired time-ordering $\la V^\dag W^\dag VW\ra$. In this regime,\footnote{In $d=2$, one could opt to use Zamolodchikov's $q$-variable, $q = \exp(i K(1-z)/K(z))$, where $K(z) = {\pi\over 2}\hyp(\half,\half,1,z)$ \cite{zamo2}. This maps {\it all} sheets of the cut plane $\mathbb{C}\backslash( 1,\i)$ to the unit disk $|q|\leq 1$. The Regge limit corresponds to taking $q \rar i$. While we will not find occasion to use this variable anymore in this paper, it is undoubtedly useful for studying Lorentzian correlators in generic CFTs \cite{Maldacena:2015iua}.}
\e{zzb}{z \approx -e^{{2\pi\over \b}(x-t)}\eps_{12}^*\eps_{34}~, \quad \zb \approx -e^{{2\pi\over \b}(-x-t)}\eps_{12}^*\eps_{34}}
and
\e{e5}{\eps_{ij} \equiv i\left(e^{{2\pi\over \b}i\eps_i}-e^{{2\pi\over \b}i\eps_j}\right)}
To preserve the operator ordering $\la V^\dag W^\dag VW\ra$, the $\eps_i$ obey $\eps_1<\eps_3<\eps_2<\eps_4$; when all $\eps_i$ are distinct, all operators are separated in imaginary time. 

To summarize, the thermal, Lorentzian correlator relevant for chaos is
\e{regform}{{\la V^\dag W^\dag(t) V W(t)\ra_\b\over \la V^\dag V\ra_\b \la W^\dag (t)W(t)\ra_\b} = \A^\Reg(z,\eta)~, \quad\text{where}~~ z = -e^{{2\pi\over \b}(x-t)}\eps_{12}^*\eps_{34}~,~~ \eta= e^{-{4\pi\over \b}x}} 
$V$ sits at $t=0$. We will often refer to $\A^\Reg(z,\eta)$ when making statements about chaos, with the identifications in \eqr{regform} understood. Precisely the same formula applies for chaos in $d>2$ CFTs in Rindler space, setting $\b=2\pi$. The relation of $z$ and $\zb$ to the familiar higher-dimensional cross ratios $u$ and $v$ is
\e{uvdef}{u = {x_{12}^2x_{34}^2\over x_{13}^2x_{24}^2} = z\zb~, \quad v={x_{14}^2x_{23}^2\over x_{13}^2x_{24}^2} = (1-z)(1-\zb)}
In the rest of this section and the next, we leave the $\dag$ implicit.

\subsec{A bound on chaos}\label{s21}

An especially useful, and fairly general, choice for operator positions around the thermal circle is to place them diametrically opposite in pairs: $\eps_2=\eps_1+\b/2$ and $\eps_4=\eps_3+\b/2$. Fixing $\eps_1=0$ without loss of generality, we define the angular displacement between the pairs as 
\e{nathan}{\theta \equiv {2\pi\over \b}\eps_3~.}
In this arrangement, 
\e{e26}{\eps_{12}^*\eps_{34} = 4e^{i\theta}~, \quad \text{where}\quad 0\leq\theta\leq \pi}
which implies the behavior
\e{e27}{z\approx-4e^{i\theta}e^{{2\pi\over \b}(x-t)}}%
for $t- x\gg\b$. The range of $\theta$ is bounded as indicated in order to preserve the ordering $VWVW$. 
Note that $\Im(z)\leq 0$. When $\theta=\pi/2$, the operators are spaced equally.\footnote{One-sided correlators have $\eps_{12}=\eps_{34}=0$, which leads to divergences in $\A^\Reg(z,\eta)$; these can be regulated, most crudely by just not taking the $\eps_i$ strictly to zero.} We also note the imaginary time parameterization used in \cite{Maldacena:2015waa},
\e{e28}{\theta = {\pi\over 2}-{2\pi\over\b}\t}

This pairwise arrangement of operators leads to a bound on the rate of chaotic time evolution \cite{Maldacena:2015waa}. The authors prove a general statement about analytic functions bounded on the half strip, which they then apply to OTO correlators. Consider a function of complex time, $f(t+i\t)$, which obeys the following conditions: {\bf i)} $f(t+i\t)$ is analytic in the half-strip $|\t|\leq \b/4$, i.e. for $0\leq \theta\leq \pi$, {\bf ii)} $f(t)$ is real, and {\bf iii)} $|f(t+i\t)|\leq 1$ throughout the strip. Then $f(t)$ obeys
\e{e29}{{1\over 1-f}\left|{df\over dt}\right|\leq {2\pi\over \b}+\O(e^{-{4\pi\over \b}t})}
We are ignoring possible sources of error in this bound that are carefully discussed in \cite{Maldacena:2015waa} and reviewed in \cite{Fitzpatrick:2016thx}; these are not important for the large $c$ theories we will discuss. Actually, $f(t)$ need not be real; the generalization is
\e{e210}{{1\over 1-|f|}\left|{df\over dt}\right|\leq {2\pi\over \b}+\O(e^{-{4\pi\over \b}t})}
This is necessary when $f(t+i\t)$ is an OTO correlator of non-Hermitian operators. 

The bound  may be applied to functions of the form
\e{e211}{f(t) \approx 1-\eps \,e^{\l_L t}+\ldots}
where $0<\eps\ll1$ is a small parameter. Its sign ensures that $f(t)$ decays rather than grows as $t$ increases. \eqr{e210} implies $\l_L\leq {2\pi/\b}$. This expression is valid for $\l_L^{-1}\ll t \ll \l_L^{-1} \log 1/\eps$. The upper bound defines the scrambling time $t_*$, at which the $\eps$-expansion breaks down; resummation of higher-order effects in $\eps$ ensure a smooth descent towards zero. For complex $\eps=\eps_1+i\eps_2$,
\e{}{|f| = 1-{\eps_1}\,e^{\l_Lt}+\ldots}
The bound requires $\eps_1>0$ and $\l_L\leq {2\pi/\b}$, while $\eps_2$ is unconstrained.

In a large $c$ CFT, $\eps \propto 1/c$. A corollary of the chaos bound in large $c$ theories is a bound on $t_*$,
\e{}{t_* \geq {\b\over 2\pi}\log c}
This is the (updated version of the) fast scrambling conjecture \cite{Sekino:2008he}. 

\sssec*{The origins of the bound}

It is important to emphasize the fundamentality of the physical inputs leading to the chaos bound. The analyticity requirement is the statement that for operators separated along the thermal circle -- that is, non-coincident in Euclidean time -- the correlator must not have any singularities. The boundedness requirement is equivalent to the statement that the OTO correlator decays, rather than grows, due to chaos. %

Moreover, there is a close connection between chaos and causality bounds \cite{Hartman:2015lfa,Hartman:2016dxc}.  In the language of those papers, the correlator $\la VVWW\ra$ is the two-point function $VV$ in the ``shockwave state,'' $|W\ra \equiv W|0\ra$. ``Causal'' means that for any choice of $V$ and $W$, the commutator $\la W|[V,V]|W\ra$ vanishes for spacelike separated $V$ operators:
\e{}{\la W|[V(x_1),V(x_2)]|W\ra = 0\quad \text{for}~(x_1-x_2)^2>0}
This happens if and only if the second-sheet correlator is analytic and bounded above by 1 in the half-strip, which are the same inputs as for the chaos bound.\footnote{To get from \cite{Hartman:2015lfa} to here, take $z_{\rm there} = (1-z)_{\rm here}$. Then $\sigma_{\rm there} = -z_{\rm here} = 4e^{i\theta}e^{{2\pi\over\b}(x-t)}$, and the region ${\rm Im}(\sigma_{\rm there})\geq 0$ is ${\rm Im}(z_{\rm here})\leq 0$, which is the half-strip. The semicircle of \cite{Hartman:2015lfa} has radius $R=4e^{{{2\pi\over\b}}(x-t)}$; we keep this small by going to $(t-x)/\b\gg1$, approaching $z=0$ on the second sheet.}

\ssec{A toy model for violation of the chaos bound}\label{s22}
A simple function that illustrates what the chaos bound is all about, and will be central to our later analysis, is the following:
\e{testf}{f(t+i\t) = 1+ {i\eps\over z^{n}}+\ldots}
where $|\eps|\ll1$ is a real small parameter of either sign, and $n\in\Z_+$ for simplicity. The $\ldots$ can denote terms of $\O(\eps^2)$, and/or higher powers of $z$. Viewing this function as a chaotic correlator, the exponential map \eqr{e27} implies a Lyapunov exponent
\e{}{\l_L={2\pi n\over \b}}
 \begin{figure}[t!]
   \begin{center}
 \includegraphics[width = .5\textwidth]{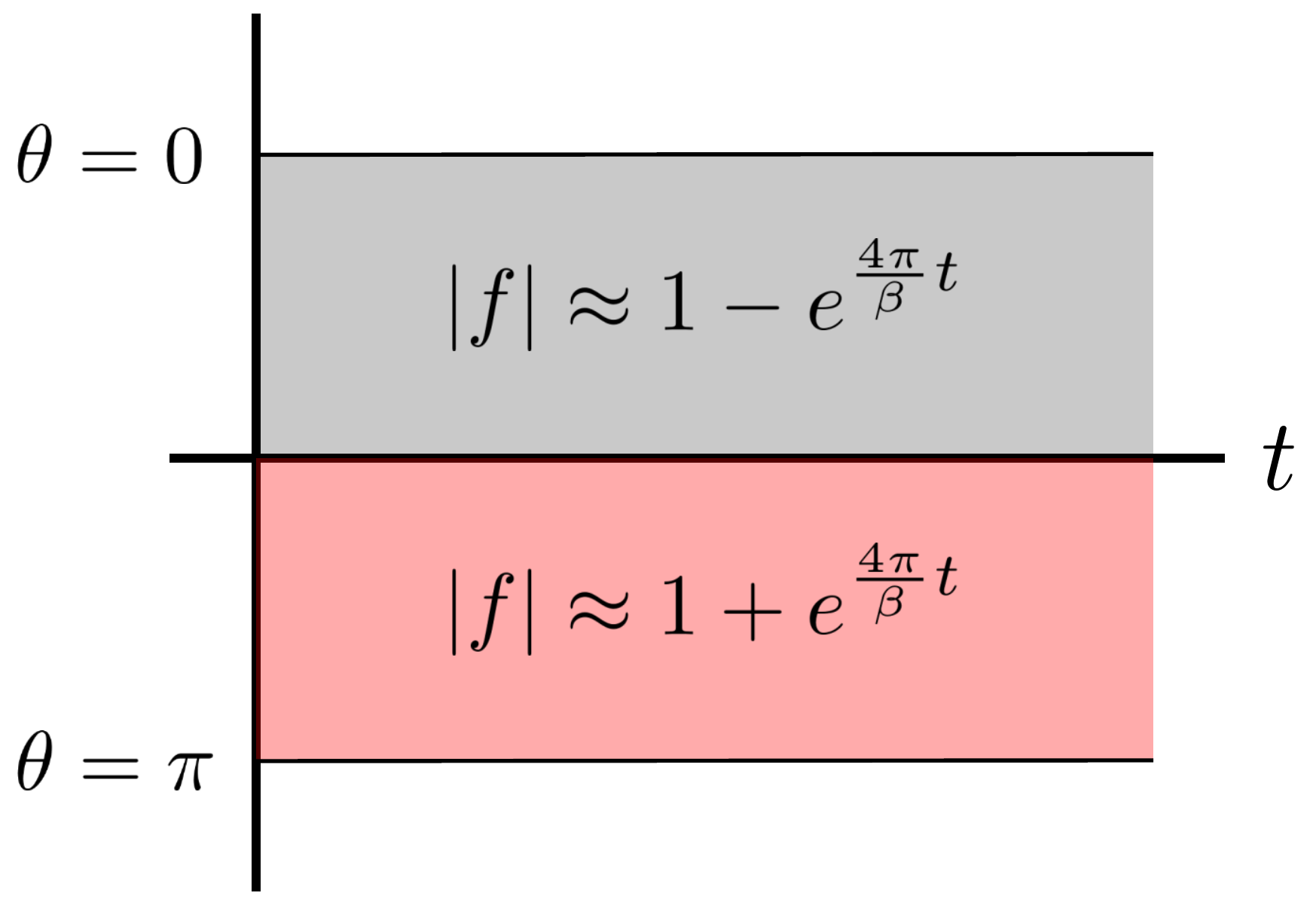}
 \caption{Functions of the form \eqr{testf} violate the chaos bound in portions of the half-strip $0\leq \theta\leq\pi$. Here we take $n=2$ (and $\eps<0$), suppressing coefficients. The modulus of $f$ grows with time in the lower (red) sub-strip.}
 \label{strip}
 \end{center}
 \end{figure}
To get a feel for this function, take the relation between $z$ and complex time to be that in \eqr{e27}. Rescaling $\eps$ by positive $t$-independent coefficients, 
\e{}{f(t+i\t) = 1+i\eps \,e^{-in\theta}e^{{2\pi n\over\b}t}+\ldots}
Its magnitude is
\e{}{|f(t+i\t)| = 1+\eps(-1)^n \sin(n\theta) e^{{2\pi n\over\b}t}+\ldots}
Recall that $0\leq\theta\leq \pi$. 

When does this grow with $t$? Equivalently, when is $|f(i\t)|\geq 1$? For $n=1$, we have
\e{}{|f(i\t)| = 1-\eps \,\sin\theta}
When $\eps>0$, this is bounded from above by 1 for all admissible $\theta$. But for general $n>1$, there are $\lfloor {n\over 2}\rfloor$ sub-strips within the full strip $0\leq \theta \leq \pi$ in which the correlator grows exponentially with $t$. This is true for either sign of $\eps$. See Figure \ref{strip}.

One concludes that the function \eqr{testf} cannot describe the OTO correlator in a consistent chaotic system. This is equivalent to saying that $\l_L=2\pi n/\b$ violates the bound on chaos: for functions of the form \eqr{testf}, analyticity and boundedness of $f(t+i\t)$ follow from $\l_L\leq 2\pi/\b$, and vice versa.\footnote{Note that within the sub-strips in which the correlator does decay, one can derive a bound on $\l_L$. Indeed, for a strip of vertical width $\b/2n$, the bound on $\l_L$ is $\l_L \leq 2\pi n/\b$. This follows from identical arguments as in \cite{Maldacena:2015waa}, again mapping the sub-strips each to a unit disc and using the Schwarz-Pick theorem. However, the point is that the bound must apply over the full strip.} The same function was recently discussed in \cite{Hartman:2015lfa} in the context of causality violation in CFT, where $f(t+i\t)$ was a correlator in the lightcone limit, $\eps=\eta\rar 0$ for fixed $z$.

\sec{Chaotic Correlators in Holographic CFTs}\label{s3}
In general, the chaotic behavior of correlation functions is sensitive to the OPE data of the CFT, the choice of thermal state, and to some extent on the choice of operators $W,V$. However, calculations in classical Einstein gravity and in $d=2$ CFT, performed for heavy operators with $\D_w\gg\D_v\gg1$, suggest that in typical holographic CFTs, chaotic correlators of arbitrary local operators take a universal form, including a Lyapunov exponent $\l_L=2\pi/\b$, and scrambling time $t_*=\l_L^{-1}\log c$. 

The goal of this section is to connect this to knowledge of the OPE data of general holographic CFTs. For general scalar primaries $V$ and $W$ in the light spectrum -- that is, with $\D_w,\D_v\sim \Oc(1)$ -- one would like to show not only that $\l_L=2\pi/\b$, but that the dependence on the spatial separation of $V$ and $W$ takes a universal form. 

This is, in general, a difficult problem to analyze purely in CFT: it amounts to understanding sufficient conditions, currently unknown, on strongly coupled OPE data that give rise to Regge scaling $\A^{\Reg}(z,\eta)\sim z^{-1}$. A framework for this problem was put forth in \cite{Costa:2012cb}. Our goal is different, and more modest: we instead want to understand what the properties of a weakly coupled bulk theory imply about chaos in the dual CFT. Accordingly we discuss the Regge behavior of holographic correlators, translated into the language of chaos. This will serve as a stepping stone to the next sections, where we apply this analysis to putative AdS$_3$/CFT$_2$ pairs with higher spin currents added. 

We will introduce a simple toy model of a prototypical bulk theory, whose Regge behavior is soluble and which shares features with more general holographic chaos. This statement is supported by a calculation, in Appendix \ref{appb}, of chaos in strongly coupled $\N=4$ SYM: in particular, its Regge scaling and position dependence match those of the toy model.

\subsec{The general picture}
Known properties of holographic CFTs with Einstein gravity duals support the following picture of OTO correlators:

\vs
{\it Take $V$ and $W$ to be arbitrary local primary operators in a CFT$_d$ with large central charge $c$, a sparse spectrum of light operators, and no parametrically light single-trace operators of spin $s>2$. This is the characteristic spectrum of a CFT$_d$ with a weakly coupled Einstein gravity dual. Then for such a CFT$_d$ in Rindler space, or a CFT$_2$ on $\R\times S^1$ with any $\b$, 
\e{form1}{\A^\Reg(z,\eta) \sim 1+{i\over cz}f(\eta)+\ldots}
where $\ldots$ includes terms subleading in the Regge limit and in the $1/c$ expansion. In terms of $x$ and $t$,
\e{form2}{\vwvw \sim 1-{i\over c \eps_{12}^*\eps_{34}}e^{{2\pi\over \b}(t-x)}f(x)+\ldots}
It follows that $\l_L=2\pi/\b$. }

\vs
\noindent If we further define $t_*$ as the time at which the $1/c$ expansion  breaks down, then \eqr{form2} also implies $t_* = \l_L^{-1}\log c$. To obey the chaos bound, we must also have $f(\eta)>0$ for $0\leq \eta<1$. 

In what follows, we will be able to constrain the functional form of $f(\eta)$: see \eqr{feta12}. Moreover, \eqr{form2} makes a prediction for the evaluation of $\la VWVW\ra$ as a bulk wave function overlap integral \cite{Shenker:2014cwa}, for {\it light} fields $V$ and $W$. We will say more about this in the Discussion.
\vs


The main physical point of the result \eqr{form1} is that $z^{-1}=z^{1-2}$, where 2 is the spin of the stress tensor, which is the highest-spin current in the theory. Essentially all CFTs have a local stress tensor, but {\it not} all CFTs have Regge scaling $z^{-1}$; so this simple relation between the spectrum of currents and the Regge scaling must follow from non-trivial features of holographic CFTs. This is sometimes called ``graviton dominance'' of bulk amplitudes \cite{Cornalba:2006xm}. We now phrase this non-triviality in CFT language.

\vs


We consider the Regge limit of vacuum four-point functions $\la VVWW\ra$. We take $V$ and $W$ to be scalar operators. In a general CFT, the reduced amplitude $\A(z,\zb)$ can be expanded in $s$-channel conformal blocks of $SO(d+1,1)$ for symmetric tensor exchange, $G_{\D,s}(z,\zb)$:
\e{globalde}{\A(z,\zb) = \sum_{p}a_p G_{\D_p,s_p}(z,\zb)}
where
\e{}{a_p \equiv C_{VVp}C^p_{~WW}}
is the product of OPE coefficients for exchange of the symmetric tensor primary $\O_p$ with conformal dimension $\D_p$ and spin $s_p$. In a unitary CFT, $a_p$ is real, but can have either sign. This sum is infinite, but convergent for $|z|<1, |\zb|<1$ independently \cite{Pappadopulo:2012jk, Hartman:2015lfa}. We consider CFTs in which both $a_p$ and $\D_p$, and hence the amplitude $\A(z,\zb)$, admit expansions in $1/c$,
\es{}{a_p &= a^{(0)}_p +{a^{(1)}_p\over c}+\ldots\\
\D_p &= \D_p^{(0)}+{\gamma_p\over c}+\ldots\\
\A(z,\zb)&= \A^{(0)}(z,\zb) + {\A^{(1)}(z,\zb)\over c}+\ldots}
where zeroth order quantities may be computed in mean field theory. $G_{\D,s}$ is independent of $c$. To diagnose chaos, we focus on the planar connected correlator, $\A^{(1)}(z,\zb)$, which takes the form
\e{adecomp}{\A^{(1)}(z,\zb) = \sum_{p}a^{(1)}_p G_{\D^{(0)}_p,\,s_p}(z,\zb) + a^{(0)}_p \g_p \,\p_{\D}G_{\D_p^{(0)},\,s_p}(z,\zb)}
We have not yet specified the spectrum of the theory.

In general, one can only take the Regge limit of $\A(z,\zb)$ if it is known in closed form, which is rarely the case at strong coupling. Passage to the second sheet term-by-term in the conformal block expansion of $\A(z,\zb)$ generically requires a resummation: in particular, higher spin operators contribute more strongly in the Regge limit, and all CFTs contain operators with arbitrarily large spin.\footnote{Surmounting this challenge with incomplete information about $\A(z,\zb)$ is the essential goal of \cite{Costa:2012cb}.} These include descendant operators, like $\p_{\mu_1}\ldots \p_{\mu_s} V$, or the ``double-trace'' primaries appearing in the lightcone bootstrap. On the other hand, a result on the Regge scaling of conformal blocks, $G^{\Reg}_{\Delta,s}(z,\eta) \propto z^{1-s}$, implies that whenever the conformal block sum is restricted to a sum over {\it primaries} of bounded spin, its Regge limit can be taken block-by-block. (We give the precise form for $G^{\Reg}_{\Delta,s}(z,\eta)$ in \eqr{regblok}.)

Now we note that holographic CFT spectra have a generalized free field structure. Recalling that we are interested in the $\O(1/c)$ part of $\A(z,\zb)$, let us enumerate the light primaries\footnote{We ignore heavy exchanges, which are suppressed as $a_{\rm Heavy} \sim e^{-\D}$ and decouple at large $c\sim \D$.} appearing in \eqr{adecomp} in a holographic CFT:
\vs

{\bf i)} A sparse spectrum of light single-trace operators of $\D\sim \O(c^0)$ and $s\leq 2$. This includes the stress tensor, with $\D=d$ and $s=2$.\footnote{In \cite{Hartman:2014oaa}, a precise notion of sparseness has been proposed for CFT$_2$: for all $\D\lesssim c/12$ (where the vacuum has $\D=0$), the density of states obeys $d(\D)\lesssim e^{2\pi\D}$.}
\vs
{\bf ii)} Double-trace operators $[VV]_{n,s}$ and $[WW]_{n,s}$. These take the schematic form
\e{}{[VV]_{n,s} \approx V\p^{2n}\p_{\mu_1}\ldots \p_{\mu_s}V}
and likewise for $[WW]_{n,s}$. These are indexed by a spin $s=0,1,\ldots L$, where $L$ is a maximum spin, and $n=0,1,\ldots,\i$. Their dimensions are
\e{}{\D(n,s) \approx \D^{(0)}(n,s)+{\gamma(n,s)\over c}+\ldots~, \quad\text{where}\quad \D^{(0)}(n,s) = 2\D_v + 2n+s}
 \begin{figure}[t!]
   \begin{center}
 \includegraphics[width = \textwidth]{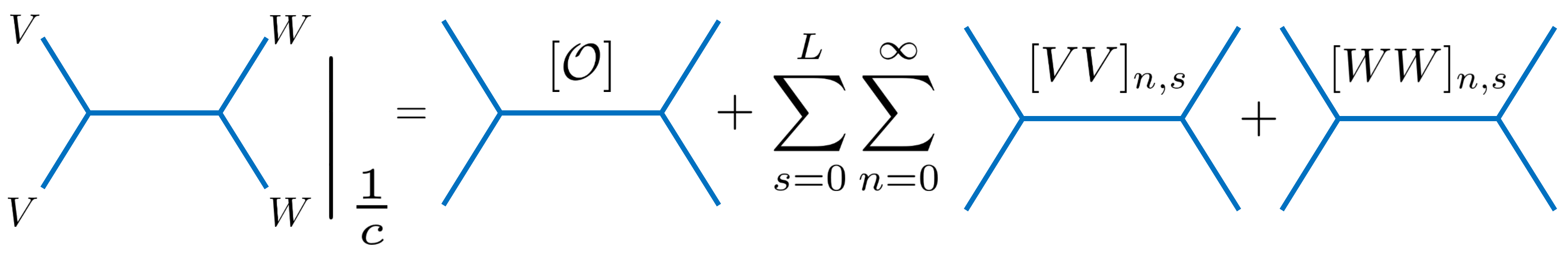}
 \caption{The decomposition of the connected piece of $\la VVWW\ra$ into $s$-channel conformal blocks, to leading order in $1/c$. The first term stands for the exchanges of single-trace operators $\O$. In a typical holographic CFT, $L=\i$, but the Regge limit corresponds to an exchange of effective spin $L_{\rm eff}=2$.}
 \label{holcorr}
 \end{center}
 \end{figure}
See Figure \ref{holcorr}.

In a CFT with an Einstein-type gravity dual, $L=\i$. There is no spin truncation. The $s>2$ double-trace exchanges in, say, the $S$-channel conformal block decomposition come from $T$- or $U$- channel exchange diagrams in the bulk. Nevertheless, the ``effective spin'' $L_{\rm eff}$ that determines the Regge scaling of $\A(z,\zb)$ equals the spin of the graviton, $L_{\rm eff}=2$: that is, $\A^{\Reg}(z,\eta) \sim z^{1-2}$. This resummation effect is the CFT image of the ``graviton dominance'' of bulk tree-level amplitudes \cite{Cornalba:2006xm}. Moreover, causality in the bulk requires $f(\eta)>0$ for $0\leq \eta\leq 1$. A more general weakly coupled bulk theory -- including stringy effects, say \cite{Shenker:2014cwa} -- has Reggeon spin $L_{\rm eff}\leq 2$, as required by causality \cite{Camanho:2014apa}. 

%

\subsubsec{A simple toy model}

It is perhaps useful to introduce a toy model for a local bulk theory, inspired by \cite{Heemskerk:2009pn}, in which one can derive $\lambda_L=2\pi/\beta$ directly. This model is simple -- in particular, it has $L=2$, so it does not require Regge resummation techniques. However, it has two main virtues: its Regge scaling is soluble, and $\A^{\Reg}(z,\eta)$ exhibits the same position dependence as the amplitudes of more complicated cases. The result is equations \eqr{atoy} and \eqr{feta12}.

The toy model is of the sort studied in \cite{Heemskerk:2009pn}: in particular, consider a bulk theory of a graviton interacting with two scalar fields $\phi_v$ and $\phi_w$, dual to $V$ and $W$, respectively, with quartic interactions of the form $\phi_v^2\phi_w^2$, $(\p\phi_v)^2\phi_w^2$, and $(\p\phi_v)^2(\p\phi_w)^2$. This theory is local and causal \cite{Camanho:2014apa, Hartman:2015lfa}. Its spectrum obeys the generalized free field structure, but the only single-trace operator appearing in the OPE is the stress tensor. 

To derive $\A^{\Reg}(z,\eta)$, we need to establish some facts about conformal blocks. Specifically, the Regge limit of $G_{\D,s}$ behaves like a spin-$s$ exchange, even though $G_{\D,s}$ includes descendant contributions of unbounded spin. In general $d$, the conformal Casimir equation for $G_{\D,s}$ simplifies, admitting a closed-form hypergeometric solution \cite{Cornalba:2006xm}
\e{regblok}{G_{\D,s}^\Reg(z,\eta) = i z^{1-s}\Gc_{\D,s}(\eta)}
where
\e{regblok2}{\Gc_{\D,s}(\eta)\equiv C(\D,s)\eta^{\D-s\over 2}\hyp\left({d-2\over 2},\D-1,\D-{d-2\over 2},\eta\right)  }
with $C(\D,s)$ a positive prefactor given in Appendix \ref{appa}. 
Note the $z^{1-s}$ behavior as advertised. This can be easily checked against the Regge limit of the closed-form blocks in even $d$. See Appendix \ref{appa} for details.  

The following two additional properties of $\Gc_{\D,s}(\eta)$ are significant. One, 
\e{e38}{\Gc_{\D,s}(\eta)>0\quad\quad (0\leq\eta <1)}
for all $d\geq 2$ and $s>0$, assuming $\D$ satisfies the unitarity bound $\D\geq d-2+s$. And two, the expansion around $\eta=0$ is organized by the twists of the operators living in the conformal family $(\D,s)$.

%


These properties allow us to compute $\A^{\Reg}$ from \eqr{adecomp}. First, let us temporarily ignore the graviton. Computing the tree-level four-point amplitude holographically as a sum of quartic contact Witten diagrams, its connected piece has only double-trace exchanges; this is a known fact about contact diagrams. By construction, $L=2$: thus, the total spin sum in \eqr{adecomp} is bounded from above, and is dominated in the Regge limit by spin-2 exchanges. Using the form of $G_{\D,2}^{\Reg}$, we see that
\e{atoy}{\A^\Reg(z,\eta)= 1+{i\over cz}f(\eta)+\ldots}
where $f(\eta)$ is determined by the sum over single- and double-trace spin-2 primary exchanges, 
\e{feta}{f(\eta) \equiv \sum_{p}a^{(1)}_p \Gc_{\D_p,2}(\eta) + a^{(0)}_p \g_p \,\p_{\D}\Gc_{\D_p,2}(\eta)}
and $\ldots$ includes the subleading spin-0,1 exchanges, and terms suppressed by powers of $1/c$.

What changes when we turn on gravity? For distinct operators $V\neq W$ in our four-point function, only a single new bulk diagram contributes to $\A(z,\zb)$ at tree-level, namely, the graviton exchange in the $\phi_v\phi_v-\phi_w\phi_w$ channel.\footnote{There are also crossed-channel graviton exchanges contributing to the correlators $\la VVVV\ra$ and $\la WWWW\ra$. But resorting to the observation that $\lambda_L$ should be independent of one's choice of operators in a holographic CFT, these cases should give the same result as $\la VVWW\ra$, although a direct calculation would be harder as explained earlier.} In the dual CFT conformal block decomposition in the $VV-WW$ channel, this is known to add the exchange of the stress tensor, and to contribute to the exchange of the double-trace operators $[VV]_{n,s}$ and $[WW]_{n,s}$ of spins $s\leq 2$ only (e.g. \cite{Cornalba:2006xm, Hijano:2015zsa}). Therefore, at $\Oc(1/c)$ we still exchange only operators of spin $s\leq 2$, and $\A^\Reg(z,\eta)$ is dominated by the spin-2 exchanges whose contributions we can compute block-by-block. Said another way, the graviton exchange gives the universal dominant contribution. 

\sssec*{\it The form of $f(\eta)$}
We can say more about $f(\eta)$. The first term in \eqr{feta} runs over single- and double-trace primaries. This includes the stress tensor. For single-trace primaries, $a^{(0)}_p=0$ due to large $c$ factorization. Whether the anomalous dimension terms turn on depends on the relative values of $\D_v,\D_w$ \cite{Liu:1998th}:
\e{2xcond}{a^{(0)}_{[VV]_{n,2}}, a^{(0)}_{[WW]_{n,2}}\neq 0 \quad \text{iff}\quad \D_v-\D_w\in\Z}
The coefficients $a^{(0)}_{[VV]_{n,2}}$ were determined in \cite{Fitzpatrick:2011dm}. Noting that
\e{}{\p_{\D}\Gc_{\D,2}(\eta) = \half \Gc_{\D,2}(\eta)\log\eta+\text{(non-log terms)}~,}
the double-trace anomalous dimensions lead to a $\log\eta$ term in $f(\eta)$ if \eqr{2xcond} is satisfied.

Altogether, then, $f(\eta)$ can be written in the form
\e{feta12}{f(\eta) =  \eta^{d-2\over 2}(f_1(\eta)+f_2(\eta)\log\eta)}
We have pulled out the leading twist stress tensor contribution. Both $f_{1}(\eta)$ and $f_2(\eta)$ are analytic near $\eta=0$, obeying $f_1(0)\neq 0$ and $f_2(0)=0$. $f_2(\eta)$ reflects the presence of double-trace anomalous dimensions, and vanishes unless \eqr{2xcond} holds.

While we have derived \eqr{feta12} within our toy model, it also applies in more generic cases where $L=\infty$ but $L_{\rm eff}=2$. This follows from graviton dominance, together with the fact that the conformal block decomposition of direct-channel graviton exchange includes both the stress tensor as well as spin $s\leq 2$ double-trace exchanges. 
\vs
As for positivity of $f(\eta)$, proving this for our toy model would take us somewhat astray from the main thread of this paper. However, \eqr{e38} implies that any single-trace operator with $a_p^{(1)}>0$ contributes positively to \eqr{feta12}, and one can show that double-trace operators also contribute positively.\footnote{The argument is as follows. The double-trace contributions come from both the contact and graviton exchange diagrams. Contact diagrams are given by $D$-functions. 
%
%
%
%
%
%
For example, one can easily show using the method of \cite{Heemskerk:2009pn} that a $\l(\p\phi_v)^2(\p\phi_w)^2$ term in the bulk Lagrangian contributes to the reduced amplitude as 
\e{}{\A(z,\zb) \supset \l u^{\D_v}(1+u+v) \Db_{\D_v+1,\D_v+1,\D_w+1,\D_w+1}(z,\zb)}
where $\Db_{\D_v+1,\D_v+1,\D_w+1,\D_w+1}$ is the reduced $D$-function. For integer $\D_v,\D_w$, one may check the sign-definiteness of the Regge limit of this object using the expression for $\Db_{1111}(z,\zb)$ given in Appendix \ref{appb}, together with $D$-function identities; see e.g. the appendix of \cite{Arutyunov:2002fh}. (Essentially this same point was made in \cite{Maldacena:2015iua}; see also \cite{Komargodski:2016gci} for a similar constraint on the sign of $\l$.) 
Showing that the contribution from graviton exchange is also positive takes some more work; nevertheless, this follows from the results of Section 5.7 of \cite{Hijano:2015zsa}. In particular, equation 4.18 of  \cite{Hijano:2015zsa} also holds for arbitrary spin exchanges, and for $s>0$ exchanges, the right-hand side of 4.18 is positive due to the unitarity bound.}

\sssec{A top-down calculation: $\N=4$ SYM}

In Appendix \ref{appb}, we perform an explicit computation of $\A^\Reg(z,\eta)$ in $\N=4$ SYM at large $\l$. We take $V=W$ both to be the 1/2-BPS scalar operator in the \2 of the $SU(4)$ R-symmetry. The features described above are all visible there, and may be cleanly interpreted in terms of the $\2\times \2$ OPE at large $\l$. In particular, we note that $f(\eta)$ does indeed take the form \eqr{feta12}.

\ssec{Lessons and implications}\label{s32}
Before moving on, let us extract some key points from the above.

\sssec{$\l_L$ from the vacuum block alone}
A main message of \eqr{form1} is that the stress tensor exchange is sufficient to read off $\l_L=2\pi/\b$ and $t_*=\l_L^{-1}\log c$, while the remaining exchanges simply modify the $x$-dependence of $\la VWVW\ra$. 

It is important to note that, in general, the lightcone limit of $\A(z,\zb)$ cannot be used to read off $\l_L$. In a CFT dual to string theory, for example, where one must sum over infinite towers of higher spin operators dual to massive string states in the bulk, $\l_L<2\pi/\b$ \cite{Shenker:2014cwa}. We have shown that the $\eta\ll1$ expansion of $\A^\Reg(z,\eta)$ at $\O(1/c)$ is a lightcone expansion, a feature which is special to holographic CFTs with local bulk duals. 

\sssec{Butterfly velocity in Rindler space}
Taking $V$ and $W$ to have large spatial separation $x\gg1$, but still obeying $x\ll t$, defines the butterfly velocity, $v_B$:
\e{}{\vwvw\Bigg|_{x\gg1} \sim 1-{i\over c\eps_{12}^*\eps_{34}}e^{{{2\pi\over\b}\left(t-{x\over v_B}\right)}}+\ldots}
$v_B$ parameterizes the spatial growth of chaotic effects under time evolution. Since $\eta=\exp(-4\pi x/\b)$, this is the $\eta\rar0$ limit of $\A^{\Reg}(z,\eta)$. So in any holographic CFT, $v_B$ is determined by the spin-2 operator of lowest twist, which is of course the stress tensor. Its contribution is, ignoring constants,
\e{}{\A^\Reg(z,\eta\rar0) \approx 1+{i\over cz}(\eta^{d-2\over 2}+\ldots)}
Trading $(z,\eta)$ for $(x,t)$ yields
\e{}{v_B={1\over d-1}\quad\quad\quad (\text{Rindler})}
The result can be derived by an Einstein gravity calculation using shock wave techniques in a hyperbolic black hole background \cite{douglasdan}.

In CFTs with Einstein gravity duals, the butterfly velocity on the plane is \cite{Shenker:2013pqa}
\e{}{v_B = \sqrt{{d\over 2(d-1)}}\quad\quad\quad (\text{Planar, any }\b)}
The planar and Rindler velocities need not, and do not, agree. 

The Rindler result is robust under local higher-derivative corrections to the Einstein action. Rindler space is conformal to the hyperbolic cylinder $\mathbb{H}^{d-1}\times S^1$ with $\b=2\pi$. If we consider chaos in hyperbolic space for $\b\neq2\pi$, we have no right to use vacuum correlators, and our derivation does not apply. In planar geometries, $v_B$ {\it does} change as a function of higher derivative couplings \cite{Roberts:2014isa}. Thus we expect that $v_B$ in hyperbolic space is actually temperature-dependent in higher-derivative gravity. It should be possible to check this using shock waves in hyperbolic black hole backgrounds; to verify this in CFT, one would need to compute OTO correlators not in the vacuum, but on $\mathbb{H}^{d-1}\times S^1$ with generic $\b$. This is similar to the difference between computing entanglement entropy and R\'enyi entropy across a sphere. 

\sssec{On sparseness}
The value of $\l_L$ is sensitive to the spectrum of spins, not conformal dimensions, present in the CFT. Since sparseness refers to the latter, it is not directly related to the value of $\l_L$.

First,
\e{}{\l_L={2\pi\over\b} \quad \nRightarrow \quad \text{Sparseness}}
Consider adding, say, $10^{100}$ scalar operators of fixed $\D\ll c$ to an otherwise sparse CFT. This leaves $\l_L=2\pi/\b$ intact, but spoils sparseness.\footnote{On the other hand, it is logically possible that imposing a gap for $s>2$ operators implies sparseness; in other words, that all non-sparse CFTs must have an infinite tower of light higher spin operators. This is the operating assumption in \cite{Heemskerk:2009pn}.} Admittedly, this is a weak violation of sparseness: near $\D\approx c$, the density of states is still sub-Hagedorn, so this modification does not ruin the validity of the $1/c$ expansion, and remains consistent with the existence of a bulk dual with Einstein gravity thermodynamics.\footnote{We thank Ethan Dyer for discussions on this point.} A recent example of a non-sparse theory with maximal chaos is the SYK model \cite{Polchinski:2016xgd, Maldacena:2016hyu}.

Conversely, and more definitively,
\e{}{ \text{Sparseness}\quad \nRightarrow \quad\l_L={2\pi\over\b} }
For example, the presence of higher-spin operators in the light spectrum will change the value of $\l_L$. We will see explicit examples of sparse CFTs with $\l_L\neq 2\pi/\b$ in Section \ref{s4}, where we consider symmetric orbifold CFTs, and in the next section, where we consider sparse 2d CFTs with higher spin currents.

\sssec{Are pure theories of AdS$_3$ gravity chaotic?}

$V$ and $W$ are primary operators dual to bulk fields carrying local degrees of freedom. In $D>3$ bulk dimensions, such processes do not require the introduction of matter: gravitons can create geometry, and destroy entanglement. In $D=3$, unlike in higher dimensions \cite{Abbott:2016blz}, there are no gravitational waves: gravitons in AdS$_3$ live at the boundary, and the dynamics of the stress tensor alone are not chaotic. This is to say that we should not consider $\l_L=2\pi/\b$ as a feature of {\it pure} AdS$_3$ gravity: only when mediating interactions between matter fields do the gravitons behave chaotically. 

It may well be that pure semiclassical AdS$_3$ gravity does not exist \cite{Maloney:2007ud}. In any case, coupling the theory to matter, as in string or M-theory embeddings, is the only way to introduce non-trivial dynamics, including chaos. This is the sense in which we consider $\l_L=2\pi/\b$ to be a property of weakly coupled theories of 3D gravity.

\section{Chaotic Destruction of Higher Spin Theories}\label{s4}

We now add higher spin currents to the CFT. In $d>2$, higher spin CFTs have correlation functions that coincide with those of free theories \cite{Maldacena:2011jn} so we take $d=2$. We will first consider correlators $\la V^\dag W^\dag V W\ra_\b$ of generic $V$ and $W$, generalizing the analysis of the previous section, and then take $V$ and $W$ to have charges scaling like $c$ in a semiclassical large $c$ limit. The latter will be more rigorous, as it allows us to compute some correlators {\it exactly}, at leading order in large $c$. The upshot is simple to state: in putative large $c$ 2d CFTs with currents of spins $s\leq N$ for some finite integer $N>2$, OTO correlators of local scalar primary operators violate the chaos bound. 

The holographic dual of our conclusion is that would-be dual theories of AdS$_3$ higher spin gravity with finite towers of higher spin currents are pathological.

\ssec{Chaotic correlators in higher spin 2d CFTs}\label{s41}

Consider a set of holomorphic single-trace currents $\lbrace J_s(z) \rbrace$, where $s\leq N$ for some $N\in\Z$. We normalize our currents as 
\e{jnorm}{\la J_s(z) J_{s'}(0)\ra  = {c N_s\delta_{s,s'}\over z^{2s}}}
for some $N_s$. Altogether, the currents generate a $\W$-algebra. There may be multiple currents of a given spin, but we leave this implicit. All but $J_2(z) = T(z)$ are Virasoro primaries.

As in previous section, we will use vacuum four-point functions to diagnose chaos in the thermal state on the cylinder with arbitrary $\b$. $V$ and $W$ are $\W$-primaries carrying charges $q^{(s)}$ under the higher spin zero modes,
\e{qs}{J_{s,0}|W\ra = q_w^{(s)}|W\ra~,\quad J_{s,0}|V\ra = q_v^{(s)}|V\ra}
In this notation, $q^{(2)}=h$, the holomorphic conformal weight. 
Generically, $q^{(s)}\neq 0$ for all $s$.

What is the spectrum of a putative holographic higher spin 2d CFT? Like all holographic CFTs, the spectrum is inferred from the properties of a weakly coupled theory of gravity in AdS. In the present case, the bulk theory would be a higher spin gravity in AdS$_3$ which has $G_N\ll 1$, a set of higher spin gauge fields $\lbrace \varphi_s\rbrace$ whose boundary modes give rise to an asymptotic $\W$ symmetry algebra generated by $\lbrace J_s\rbrace$, and some perturbative matter fields whose density is fixed as a function of $G_N$. 

At $\O(1/c)$, we may apply the lesson of Section \ref{s3}: in a CFT dual to a weakly coupled theory of gravity, the spectrum of currents determines $\l_L$ as $\l_L=(2\pi/\beta)\times(s_{\rm max}-1)$. This captures the fact that, generalizing the ``graviton dominance'' of non-higher spin theories, the Regge scaling is determined by the exchange Witten diagram of $\varphi_{s_{\rm max}}$. 
%
%
%

%


Given this spectral data, it follows that for generic $V$ and $W$,
\e{hsnreg}{\A^\Reg(z,\eta) = 1+{i\over cz^{N-1}}f(\eta)+\ldots}
The function $f(\eta)$ is real and smooth, but need not be positive. In terms of $x$ and $t$,
\e{}{{\la V^\dag W^\dag(t) V W(t)\ra_\b\over \la V^\dag V\ra_\b \la W^\dag (t)W(t)\ra_\b} = 1-{i\over c \eps_{12}^*\eps_{34}}e^{{2\pi(N-1)\over \b}(t-x)}f(x)+\ldots}
This function is precisely of the form of our toy function \eqr{testf}. Therefore, it violates the chaos bound:
\e{}{\l_L={2\pi\over\b}(N-1)}
Likewise, the correlator implies a scrambling time
\e{}{t_* ={\b\over 2\pi(N-1)}\log c}
As $N$ increases, the correlator grows in an increasing number of sub-strips of the half-strip, arrayed in regular intervals with spacing linear in $N$. For $N=3$, we drew this behavior in Figure \ref{strip}.
\vs
It follows that:
\vs
\centerline{\it Unitary, holographic 2d CFTs with finite towers of higher spin currents do not exist.}
\vs
\noindent In other words, the only finitely generated $\W$-algebra consistent with unitarity in a large $c$ CFT is the Virasoro algebra. This also rules out non-sparse CFTs of the sort discussed in Section \ref{s32}, which violate the sparseness condition with only low-spin operators.\footnote{Note that when $\W$ is finitely generated (i.e. the set $\lbrace J_s\rbrace$ is finite), a primary under $\W$ branches into $\lesssim \exp(2\pi \sqrt{\text{rank}(\W)\D/6})$ Virasoro primaries in the Cardy regime $\D\gg c$. This gives an upper bound on the scaling near $\D\approx c$. Since the $d=2$ sparseness condition on the density of light states is exponential in $\D$, the distinction between Virasoro primary and $\W$-primary is irrelevant for diagnosing sparseness.}  If infinitely-generated $\W$-algebras with currents of bounded spin exist, then CFTs with these symmetries, too, are ruled out. 
Recalling the discussion in Section \ref{s21}, we may phrase this chaos bound violation in another way: holographic higher spin CFTs are non-unitary and acausal.

\sssec*{Are there exceptions?}

 One family of large $c$ CFTs with $W_N$ symmetry is the $W_N$ minimal models at negative level $k=-N-1$, the so-called ``semiclassical limit'' \cite{Gaberdiel:2012ku,Perlmutter:2012ds}; but these are non-unitary. 

A large $N$ symmetric orbifold Sym$^N(X)$, where the seed CFT $X$ has $\W$-symmetry $\W_X$ and finite central charge, has a much larger chiral algebra, $(\W_X)^N/S_N$, which becomes infinitely generated in the large $N$ limit; it also has ``massive'' higher spin operators besides the currents. 

There is one hypothetical class of CFTs that escape our conclusion: a CFT with no non-chiral light operators $V$ and $W$. This would be dual to pure higher spin gravity, for example, which has only $\W$-gravitons and black hole states. Given the difficulty in constructing duals of pure AdS$_3$ Einstein gravity, the viability of these theories is nevertheless dubious; we will return to this in the AdS/CFT context in Section \ref{s43}.

\ssec{Chaos for heavy operators in $W_N$ CFTs}\label{s42}

\begin{figure}[t!]
   \begin{center}
 \includegraphics[width = \textwidth]{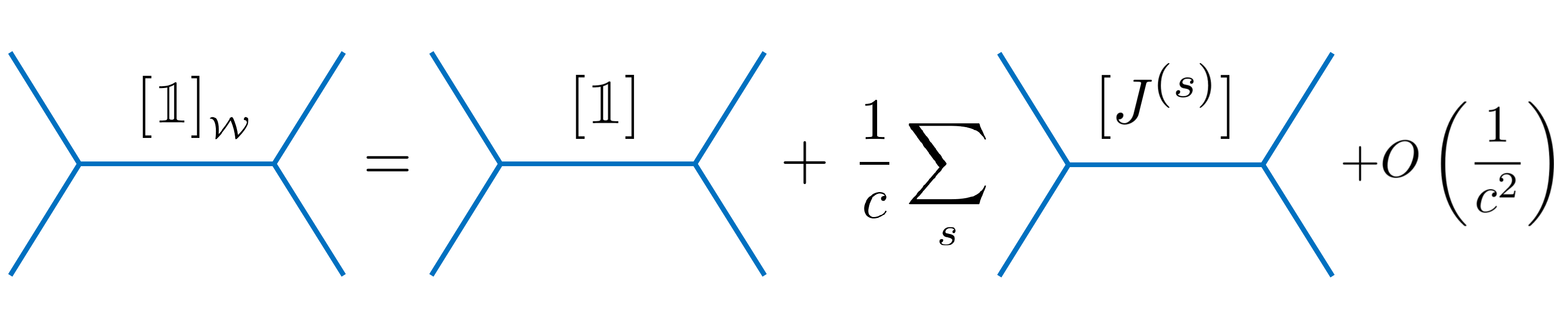}
 \caption{At $\O(1/c)$, a $\W$-algebra vacuum block (left side) branches into a sum of {\it global} blocks for simple current exchanges. All composite operator exchanges are suppressed by higher powers of $1/c$.}
 \label{wdecomp}
 \end{center}
 \end{figure}

Given the extended conformal symmetry $\W$, one may form conformal blocks with respect to $\W$, rather than $SO(3,1)$. Instead of \eqr{globalde}, we could have expanded in the $s$-channel as
\e{wde}{\A(z,\zb) = \sum_{p} a_p \F_{p,\W}(z)\overline{\F}_{p,\W}(\zb)}
where $\F_{p,\W}(z)$ are the holomorphic blocks for exchange of a $\W$-primary operator $\O_p$. A nice feature of $\F_{\vac,\W}(z)$ is that at $\O(1/c)$, the only exchanges are the simple current exchanges:
\e{fvaci}{\F_{\vac, \W}(z) = 1+{1\over c}\sum_{s} {q^{(s)}_vq^{(s)}_w\over N_s} g_s(z)+\O(c^{-2})}
where
\e{}{g_s(z) \equiv z^s \hyp(s,s,2s,z)}
is the holomorphic global block for dimension-$s$ exchange. (See Figure \ref{wdecomp}.) We are assuming that the OPE coefficients $q^{(s)}$ do not scale with $c$. 
\eqr{fvaci} follows from the fact that $\la J_s|J_s\ra \sim c$, and that all composite operators in the vacuum module of $\W$ (e.g. :$J_sJ_s$:) have norms of $\O(c^2)$ or greater. This explains why, for our chaotic correlators involving light operators $V$ and $W$, none of the complication of the $\W$-algebra blocks $\F_{\vac,\W}$ was necessary. 

To gather more data on what goes wrong in holographic higher spin CFTs, we now consider operators $V$ and $W$ whose charges scale with $c$ in the large $c$ limit:
\e{sclim}{q^{(s)}_{v,w} \rar\infty~,\quad  c\rar\infty~,\quad  {q^{(s)}_{v,w}\over c}~\text{fixed}~,\quad  {q_v^{(s)}\over c}\ll 1}
In this limit, \eqr{fvaci} is insufficient, because higher orders in $1/c$ come with positive powers of the charges. However, the point of computing these heavy correlators is that the sum over blocks simplifies: in particular, \eqr{wde} is dominated by the semiclassical vacuum block, $\F_{\vac,\W}$, up to exponential corrections in $c$:
\e{}{\A(z,\zb) = \F_{\vac,\W}(z)\overline{\F}_{\vac,\W}(\zb)}
The vacuum dominance follows from the definition of a sparse CFT, and has been supported by many computations \cite{Hartman:2013mia, Asplund:2014coa, deBoer:2014sna, Hegde:2015dqh}.

To be concrete, we now take $\W=W_N$. The semiclassical vacuum block, which we call $\F_{\vac,N}$, is known in closed-form for any $N$ \cite{Hegde:2015dqh}.  Then in the Regge limit,
\e{presn}{\A^\Reg(z,\eta) = \F^\Reg_{\vac,N}(z)}
Compared to the case where $V$ and $W$ are light, this ``re-sums'' an infinite set of global blocks for multi-$J_s$ exchange. The resulting expressions for $\A^\Reg(z,\eta)$, exact to leading order in large $c$, are more intricate than our result \eqr{hsnreg}. Nevertheless, they still violate the chaos bound.

\vs
Our calculations are directly inspired by those of Roberts and Stanford \cite{Roberts:2014ifa} in the Virasoro case. We begin by reproducing, then reinterpreting, their calculation.\footnote{In the remainder of this section, $\F_{\vac,\W}(z)$ refers to the blocks in the semiclassical limit \eqr{sclim}, and we use the common convention  $\F_{\vac,\W}(z) \approx z^{-2h_v}(1+\ldots)$ for ease of comparison to previous works.}

\sssec{Warmup: Virasoro}

We want to compute $\F_{\vac}^\Reg$, the Regge limit of the semiclassical Virasoro vacuum block $\F_{\vac}\equiv \F_{\vac,2}$. More precisely, we choose operators $V$ and $W$ whose holomorphic conformal dimensions scale as in \eqr{sclim}:
\e{sclimv}{h_{v,w} \rar\infty~,\quad  c\rar\infty~,\quad  {h_{v,w}\over c}~\text{fixed}~,\quad  {h_v\over c}\ll 1}
In this semiclassical limit, the Virasoro vacuum block is \cite{Fitzpatrick:2014vua}
\e{fvacvir}{\F_{\vac}(z) = \left({ \a (1-z)^{-1+\a\over 2}\over1-(1-z)^{\a}}\right)^{2h_v} }
where 
\e{alpha}{\a = \sqrt{1-24{h_w\over c}}}
Actually, the same result holds even in the ``heavy-light limit'' of \cite{Fitzpatrick:2015zha}, in which $h_v$ is held fixed as $c$ becomes large. (And even in this regime, the vacuum dominance of the correlator is believed to hold \cite{Asplund:2014coa}.)

\cite{Roberts:2014ifa} studies this object at small $h_w/c$, whereby
\e{}{\a \approx 1-2\eps~, \quad \text{where} \quad \eps \equiv {6 h_w\over c}\ll 1}
Imposing this limit on $\F_{\vac}$ and keeping only the terms inside the parenthesis that would contribute to linear order in $\eps$, \cite{Roberts:2014ifa} write
\e{}{\F_{\vac}(z) \approx \left({  (1-z)^{-{\eps}}\over1-(1-z)^{1-2\eps}}\right)^{2 h_v}}
In the Regge limit,
\e{fregz}{\F^\Reg_{\vac}(z) = \left({1\over z-4\pi i \eps}\right)^{2h_v}}
In the Lorentzian variables, this reads
\e{RS}{{\la V^\dag W^\dag(t) V W(t)\ra_\b\over \la V^\dag V\ra_\b \la W^\dag (t)W(t)\ra_\b} \sim \left({1\over 1+ {h_w\over c}{24 \pi i\over  \eps_{12}^*\eps_{34}} e^{{2\pi\over \b}(t-x)}}\right)^{2h_v}}
The initial decrease in time is exponential, with $\l_L=2\pi/\b$. Note that the sign in the denominator is crucial: it ensures that the magnitude of the correlator decreases in time, for any choice of the $\eps_{ij}$. At even later times, the correlator has lost an order one fraction of its original value; this happens at $t\approx t_*+x$, where $t_* = {\b\over 2\pi}\log (c/h_w)$. 

As noted in \cite{Roberts:2014ifa}, there is some subtlety in this interpretation. One point regards the scrambling time. In using the semiclassical conformal block, one holds $h_w/c$ fixed. Then strictly speaking, $t_*$ as defined above is parametrically smaller than the scrambling time one expects from Einstein gravity, $t_* = {\b\over 2\pi}\log c$. This is presumably an artifact of the semiclassical limit. To wit, if one analytically continues \eqr{RS} to a regime in which $h_w$ is held fixed but large in the large $c$ limit, it exactly matches a shock wave calculation in 3D gravity in the same regime of dimensions. This suggests that \eqr{RS} captures the correct physics even when $W$ and $V$ are not parametrically heavy, and that we may extrapolate \eqr{RS} to
\e{}{{\la V^\dag W^\dag(t) V W(t)\ra_\b\over \la V^\dag V\ra_\b \la W^\dag (t)W(t)\ra_\b} \sim \left({1\over 1+ {24 \pi i h_w\over  \eps_{12}^*\eps_{34}} e^{{2\pi\over \b}(t-t_*-x)}}\right)^{2h_v}}
with $t_* = {\b\over 2\pi}\log c$. Then at early times $t\ll t_*+x$, the correlator decreases as $\exp({{2\pi t/\b}})$, and at late times $t\approx t_*+x$ decays to zero as $\exp({-{4\pi h_vt/ \b}})$.

A more obvious point is that \cite{Roberts:2014ifa} only expanded $\a$ to linear order in $\eps\propto h_w/c$, but don't fully expand the block to linear order. Doing so, one finds
\e{}{\F^\Reg_{\vac}(z) =z^{-2h_v}\left(1+{8\pi i h_v \eps\over z} + \O(\eps^2)\right)}
Expanding near $\eps\ll1$ commutes with going to the second sheet, so the term of $\O(\eps)$ should simply be the Regge limit of the stress tensor exchange. Indeed,
\e{}{a_T\,g_2(z) = {2 h_vh_w\over c}z^2\hyp(2,2,4,z) \sim {48 \pi i  h_vh_w\over cz}+\ldots}
where we use the hypergeometric monodromy around $z=1$,
\e{monod}{\hyp(s,s,2s,z) ~\rar ~\hyp(s,s,2s,z)+{2\pi i\over (2s-1)}{\Gamma^2(2s)\over \Gamma^4(s)}z^{1-2s}\hyp(1-s,1-s,2(1-s),z)}
where $s\in\Z$. We also note that expanding $\F_\vac(z)$ to all orders in $\eps$ and keeping only the leading term in small $z$ at each order, the result re-sums to \eqr{fregz}; this gives a partial justification for the method of \cite{Roberts:2014ifa}. 

\sssec{$W_N$}
We now perform the calculation of \eqr{presn} for $W_N$ with charges \eqr{sclim}.
$\F_{\vac,N}$ was derived in \cite{Castro:2014mza,deBoer:2014sna} for $N=3$, and for arbitrary $N$ in \cite{Hegde:2015dqh}. Its bulk interpretation is of a ``heavy'' field $W$ generating a classical background with higher spin charge, in which the ``light'' operator $V$ moves. Moreover, there is evidence that the $W_N$ vacuum blocks given below are also valid for $q_v^{(s)}$ held fixed in the large $c$ limit, and that even in that case, $\F_{\vac,N}$ is still the dominant saddle point of the correlation function \cite{Hegde:2015dqh}. As we discussed earlier, this is the case for Virasoro.

We will first do the computation for $N=3$, where a single spin-3 current is added to the CFT. This case demonstrates all of the essential physics present at general $N$, an assertion we support with computations at $N=4$ and at arbitrary $N$ in Appendix \ref{appc}.
\vs
In $W_3$, there is only one higher spin charge, so we drop the superscript on $q_{v}^{(3)}, q_w^{(3)}$. The semiclassical $W_3$ vacuum block is \cite{deBoer:2014sna, Hegde:2015dqh}\footnote{We ignore irrelevant factors of the UV cutoff of the CFT. Also, in this subsection, we use the normalization of \cite{deBoer:2014sna}, in which $N_3 = 5/6$.}
\e{fvac3}{\F_{\vac,3}(z) = ((1-z)^2 m_1m_2)^{- h_v/2}\left({m_2\over m_1}\right)^{3q_v/ 2}}
where
\es{}{m_1 & = {2\over n_{12}n_{23}n_{31}} \big( n_{12}(1-z)^{n_3}+\text{cyclic}\big)~, \\
m_2& ={2\over n_{12}n_{23}n_{31}} \big( n_{12}(1-z)^{-n_3}+\text{cyclic}\big)}
with $n_{ij}\equiv n_i-n_j$. 
The $n_i$ are roots of the cubic equation
\be\label{pn}
n^3-\alpha^2n -4q =0~,
\ee
where we have defined a rescaled charge,
\e{}{q \equiv {6\over c}q_w}

Note the absence of a quadratic term in \eqr{pn}, which implies $\sum_i n_i=0$. Note also that under $q\rar -q$, one root has odd parity while the product of the other two roots has even parity. At small $q$,
\es{}{n_1 &\approx -{4\over \a^2}q - {64\over \a^8} q^3 + \O(q^5)\\
n_2 &\approx \a + {2\over \a^2} q - {6\over \a^5} q^2 + {32\over \a^8} q^3 -{210\over \a^{11}}q^4 + \O(q^5)\\
n_3 &\approx -\a + {2\over \a^2} q + {6\over \a^5} q^2 + {32\over \a^8} q^3 +{210\over \a^{11}}q^4 + \O(q^5)}
When $q=0$, one recovers the Virasoro block \eqr{fvacvir}. 

We want to take \eqr{fvac3} to the chaos regime. We study each piece of \eqr{fvac3} in turn, starting with the term $((1-z)^2 m_1m_2)^{- h_v/2}$. This is the only surviving term for an uncharged probe, $q_v=0$. To first non-trivial order in $q$, 
\es{e429}{((1-z)^2&m_1m_2)^{-1} = \frac{\alpha ^4 z^{2 \alpha -2}}{\left(z^{\alpha }-1\right)^4}\Bigg[1+\frac{12 q^2 }{\a^6(z^{\a}-1)^4} \Big(6 \alpha ^2 \left(z^{2 \alpha }+1\right) z^{\alpha } \log ^2 z\\&+\alpha  \left(z^{4 \alpha }-14 z^{3 \alpha }+14 z^{\alpha }-1\right) \log z-\left(z^{\alpha }-1\right)^2 \left(5 z^{2 \alpha }-22 z^{\alpha }+5\right)\Big)+\O(q^4)\Bigg]_{z\rar 1-z}}
Taking the Regge limit, we find
\e{chwa}{((1-z)^2m_1m_2)^{-1}\sim \left({1\over z-4\pi i \eps}\right)^4\left[1 - (24\pi q)^2\left({1\over z-4\pi i \eps}\right)^4 + \O(q^4)\right]}
It is straightforward to proceed to higher orders. Perturbation theory through $\O(q^{16})$ is consistent with the following result:
\e{}{((1-z)^2m_1m_2)^{-1}\sim \left({1\over z-4\pi i \eps}\right)^4{1\over 1-Y}}
with
\e{}{Y \equiv (24\pi i q)^2\left({1\over z-4\pi i \eps}\right)^4}
Turning now to the $(m_2/m_1)^{3q_v/2}$ term in \eqr{fvac3}, the first few orders read
\es{chy}{{m_2\over m_1}\approx 1+2\sqrt{Y}+2Y+2Y^{3/2}+\O(Y^2)}
Perturbation theory through $\O(q^{16})$ is consistent with the following result:
\e{}{{m_2\over m_1} \sim {1+\sqrt{Y}\over 1-\sqrt{Y}}}

Putting this all together, restoring the $c$-dependence and plugging the block into \eqr{presn}, we find the OTO correlator to be
\es{cie}{\A^\Reg(z,\eta)={\left((1-{24\pi i h_w\over cz})^2+{144\pi i q_w\over cz^2}\right)^{3q_v}\over \left((1-{24\pi i h_w\over c z})^4+\left({144\pi  q_w\over c z^2}\right)^2\right)^{{h_v+3q_v\over 2}}}}
where $z$ is given in \eqr{zzb}. 

Not unexpectedly, this violates the bound on chaos. The essential point is that every $q_w$ appears with a $1/cz^2$. This implies that the decay rate is exponential, controlled by a ``spin-3 Lyapunov exponent''
\e{}{\l_L^{(3)}={4\pi\over \b}}
As in the Virasoro case, the calculation breaks down at late enough times, signifying the decrease of the correlator. Introducing the ``spin-3 scrambling time'' $t_*^{(3)}$,
\e{}{t_*^{(3)} \equiv {\b\over 4\pi}\log c}
the OTO correlator is thus
\es{cif}{{\la V^\dag W^\dag(t)VW(t) \ra_{\b}\over \la V^\dag V \ra_{\b} \la W^\dag(t)W(t) \ra_{\b}}  = {\left((1+ {24\pi ih_w\over \eps_{12}^*\eps_{34}}e^{{2\pi\over\b}(t-2t_*^{(3)}-x)})^2+{144\pi i q_w\over(\eps_{12}^*\eps_{34})^2}e^{{4\pi\over\b}(t-{t_*^{(3)}}-x)}\right)^{3q_v}\over \left(\left(1+ {24\pi ih_w\over \eps_{12}^*\eps_{34}}e^{{2\pi\over\b}(t-2t_*^{(3)}-x)}\right)^4 + {(144\pi q_w)^2\over (\eps_{12}^*\eps_{34})^4}e^{{8\pi\over \b}(t-{t_*^{(3)}}-x)}\right)^{{h_v+3q_v\over 2}}}}
Scrambling sets in when $t\approx t_*^{(3)}+x$ -- long before $t= t_*+x$ -- at which point the correlator decays to zero as $\exp\left(-{4\pi h_vt/\b}\right)$. 

As we reviewed in Section \ref{s2}, because $\l_L^{(3)}>2\pi/\b$, either the correlator is not analytic in the entire half-strip, and/or it grows in time rather than decaying. Either is a fatal outcome for a theory. To put the problem in sharpest relief, we take $t\approx t_*^{(3)}+x$ (or $h_w\rar 0$) and $q_v\rar0$, and place the operators in the arrangement \eqr{e27}, with a displacement angle $\theta=\pi/4$. Then $(\eps_{12}^*\eps_{34})^4=-4^4$, and the correlator reads
\es{cif}{{\la V^\dag W^\dag(t)VW(t) \ra_{\b}\over \la V^\dag V \ra_{\b} \la W^\dag(t)W(t) \ra_{\b}}   = {1\over \left(1 - \left({3\pi q_w\over 4}\right)^2e^{{8\pi\over \b}(t-{t_*^{(3)}}-x)}\right)^{{h_v\over 2}}}}
The correlator grows in time. Indeed, it {\it diverges} for $t\approx x+t_*^{(3)}$. More generally, for operators diametrically opposite on the thermal circle, the correlator will diverge for any $\theta\in[0,\pi]$ such that $\Re(e^{4i\theta})<0$. This carves out two substrips, $\theta\in[{\pi\over 8}, {3\pi\over 8}]$ and $\theta\in[{5\pi\over 8}, {7\pi\over 8}]$, of the full strip in which the correlator is non-analytic. Turning on $q_v\neq 0$ does not evade this conclusion. 

One can also check that to linear order in small $q_w/c$, the result\footnote{Here we are assuming that these formulas hold even for $q_v$ of order $c^0$, as motivated above, so that we can treat $q_wq_v/ c$ as small.}%
\es{}{\A^\Reg(z,\eta) \approx 1+{432\pi i q_wq_v\over cz^2}+\ldots}
matches the Regge limit of the spin-3 current exchange block, $g_3(z) = z^3\hyp(3,3,6,z)$, using \eqr{monod} and $N_3 = 5/6$. 

\sssec{General spins}

These pathologies of the $W_3$ result only get worse as we add higher spins. For explicit calculations at $N=4$ and at arbitrary $N$, see Appendix \ref{appc}. The results are consistent with general expectations:  when $V$ and $W$ are charged under a spin-$s$ primary, its contribution to the onset of chaos is characterized by the ``higher spin Lyapunov exponent''
\e{lams}{\l_L^{(s)}={2\pi(s-1)\over \b}}
and to the onset of scrambling by the ``higher spin scrambling time''
\e{}{t_*^{(s)} = {\b\over 2\pi(s-1)}\log c}
When $V$ and $W$ carry charges of spins $s\leq s_{\rm max}$, the leading chaotic behavior is controlled by $s_{\rm max}$.

\ssec{In the bulk: Ruling out AdS$_3$ higher spin gravities}\label{s43}
Let us now invoke AdS/CFT. The dual higher spin gravities would contain the gauge fields $\lbrace \varphi_s\rbrace$ coupled to some matter. The gauge sector may be succinctly packaged as a $G\times G$ Chern-Simons theory for some Lie group $G\supset SL(2,\R)$. With AdS$_3$ boundary conditions, the boundary ``gravitons'' of $G$ generate a $\W$-algebra, call it $\W_G$, which is the Drinfeld-Sokolov reduction of $G$ \cite{Campoleoni:2011hg}. See Figure \ref{wg}.

The bulk dual of our CFT conclusion is:
\vs
\centerline{\it Weakly coupled higher spin gravities with finite towers of higher spin fields are inconsistent.} 
\vs
\noindent This includes the oft-studied $G=SL(N,\R)$ theories \cite{Campoleoni:2010zq}, where $\W_{SL(N,\R)}=W_N$. 

In fact, our $W_N$ calculations of Section \ref{s42} double as direct bulk calculations. This follows from a series of recent works \cite{Ammon:2013hba, deBoer:2013vca, Castro:2014mza, deBoer:2014sna, Hegde:2015dqh}. It has been firmly established that the semiclassical vacuum block $\F_{\vac,N}$ is  computed as a certain bulk Wilson line operator constructed from the $SL(N)$ connections. On the other hand, this Wilson line also computes holographic four-point functions of two heavy and two light operators in precisely the semiclassical limit \eqr{sclim}. We may schematically write these relations, valid at leading order in large $c$, as
\e{}{SL(N)~\text{Wilson line} ~ \approx ~\langle \O_H|\O_L\O_L|\O_H\rangle ~ \approx ~ |\F_{\vac,N}|^2}
Importantly for us, this relation is believed to hold for {\it any} potential theory of $SL(N)$ gauge fields coupled to matter, as a matter of gauge invariance \cite{Ammon:2013hba}. Given this, the CFT and bulk calculations are identical, and the inconsistency of $SL(N)$ higher spin gravity is explicitly established by our calculations. An analogous statement holds for any bulk algebra $G$. 

\sssec*{Higher spin shock waves}

An equivalent way of computing the OTO correlator in the bulk is not via analytic continuation of a Euclidean correlator, but by directly probing a backreacted shock wave solution \cite{Shenker:2013pqa}. 

At early times, consider perturbing a planar BTZ black hole by an operator ($W$) carrying higher spin charge. Our calculation shows that the infalling higher spin quanta generate a ``higher spin shock wave'' that, in the absence of an infinite tower of higher spin gauge fields, acts acausally. For the two-sided BTZ black hole, the shock wave destroys the entanglement of the thermofield double state too fast. Said another way, these higher spin gravities are too-fast scramblers. 

Note that, like a higher spin black hole \cite{Ammon:2011nk}, a higher spin shock wave is no longer purely geometric: the higher spin fields $\varphi_s$ are sourced by $W$. Just as the shock wave line element picks up a $g_{uu}$ component along the null direction $u$, the spin-$s$ tensor fields $\varphi_s =\varphi_{\mu_1\mu_2\ldots\mu_s}$ should acquire components of schematic form
\e{}{\varphi_{uu\ldots u} = q_w^{(s)}h^{(s)}(u,x)}
with spin-dependent profiles $h^{(s)}(u,x)$ determined by the field equations. These couple to null spin-$s$ currents $J_{uu\ldots u} \propto q_w^{(s)}$ induced by the motion of $W$. In a higher spin shock of a two-sided BTZ black hole, acausality will be manifest as a causal connectivity between the left and right CFTs in the perturbed thermofield double state, similar to the effect of a time-advance in higher derivative gravity.\footnote{See e.g. \cite{Camanho:2014apa}, Section 6. We thank Aitor Lewkowycz for discussions on this point.} This would be visible in the analytic structure of a two-sided correlator of a probe dual to $V$, in the shock wave background.\footnote{A more tractable construction of the higher spin shock wave would use the Chern-Simons description. In that language, a two-sided correlator would be computed as a two-sided Wilson line. In carrying out such a calculation, one must choose an appropriate gauge for the shock wave connections; \cite{Castro:2016ehj} motivates a specific gauge choice for the pure thermofield double state, which would also be useful in the shock wave context.}

The properties of scattering through shock waves may also be phrased using amplitudes.\footnote{Note also the recent work in massive 3D gravity, \cite{Edelstein:2016nml}.} The effect of a particle traveling through a shock wave is determined by the high-energy behavior of four-point, tree-level, high-energy scattering amplitude, call it $\A_{\rm tree}$ (e.g. \cite{Cornalba:2006xm}). In flat space or AdS gravity in $d>3$ dimensions, finite numbers of higher spin fields, massive or massless, lead to unacceptably fast growth of $\A_{\rm tree}$ in the high energy regime of large $s$ and fixed $t$ \cite{Camanho:2014apa}. The tree-level amplitude for exchange of a spin-$J$ field grows at large $s$ like 
\e{}{\A^{(J)}_{\rm tree}(s,t) \sim G_N s^{J}}
whereas the total amplitude $\A_{\rm tree}$ must not grow faster than
\e{treebound}{\A_{\rm tree}(s,t)\lesssim G_N s^2}

In AdS$_{d>3}$, the $s^J$ behavior holds for spin-$J$ exchange between external fields of any spin, including pure graviton scattering. In AdS$_3$, pure gauge field scattering is trivial. However, spin-$J$ exchange between external matter fields still behaves as $\A_{\rm tree} \sim G_N s^{J}$. In other words, AdS$_3$ Witten diagrammatics with external matter have the same high-energy scaling as in higher dimensions. In this sense our result may have been anticipated from \cite{Camanho:2014apa}.  In the Regge limit, the Mellin amplitude for massless spin-$J$ exchange between pairs of scalars $V$ and $W$ in AdS$_3$ may be read off from the results of \cite{Costa:2014kfa}:
\e{}{M^{(J)}_{\rm tree}(s,t) \sim - G_N s^J\left({2q^{(J)}_vq^{(J)}_w\over 3 N_J}{2^{1-2J}\Gamma(2J)(J-1)_J\over \Gamma(\D_v)\Gamma(\D_w)\Gamma^4(J)}\right){{}_3F_2\left(1-\D_v,1-\D_w,-{t\over 2}; J,1-{t\over 2}; 1\right)\over t}}
where $s$ and $t$ are Mellin variables in the conventions of \cite{Costa:2014kfa}, and the quantities $q^{(J)}_{v,w}$ and $N_J$ were defined in \eqr{jnorm}--\eqr{qs}. (Note that for $J=2$, the coefficient of $G_N s^2$ is positive when $t<0$ for all $\D_v,\D_w\geq 0$, consistent with causality.) 
 \begin{figure}[t!]
   \begin{center}
 \includegraphics[width = .75\textwidth]{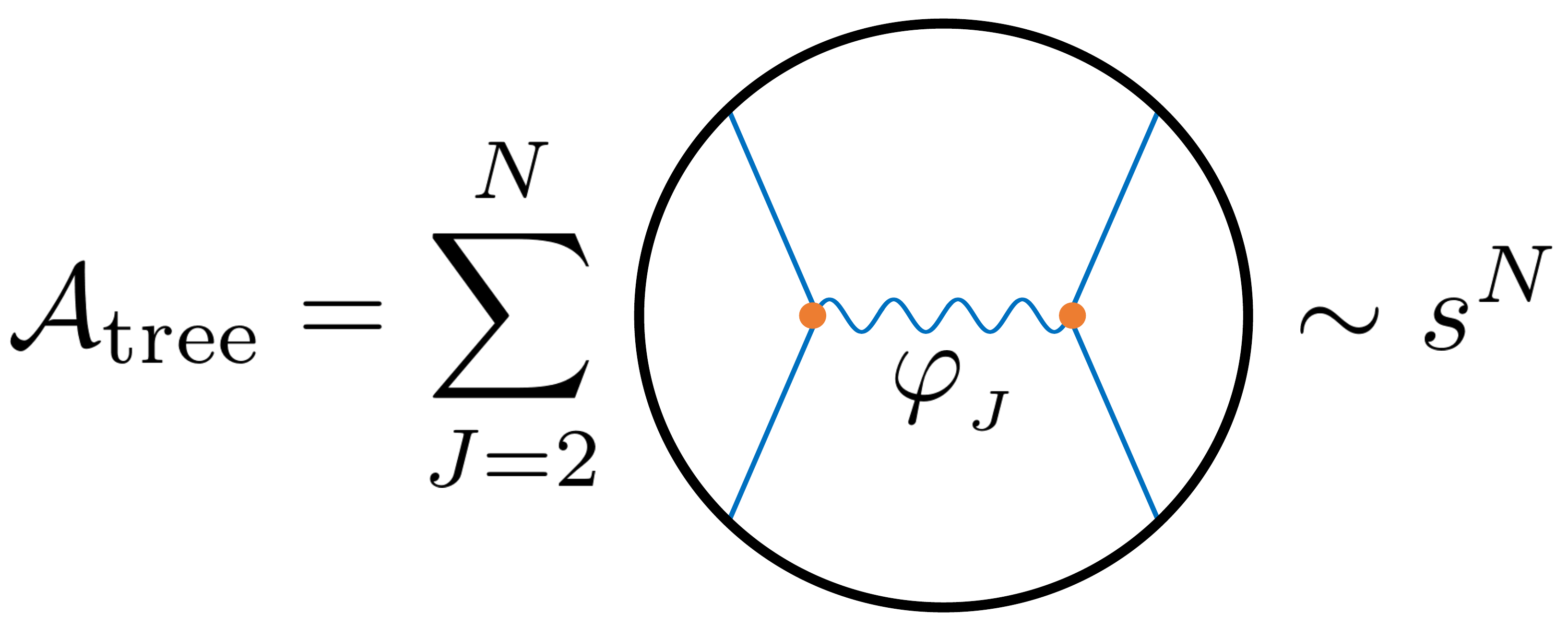}
 \caption{Four-point, tree-level Mellin amplitudes for scalar scattering in AdS$_3$ higher spin gravity grow like $s^N$, where $N$ is the spin of the highest-spin field that couples to the scalars. This is the bulk dual of \eqr{hsnreg}.}
 \label{atree}
 \end{center}
 \end{figure}
A theory with fields of spin $J\leq N$ with finite $N>2$ violates \eqr{treebound}, see Figure \ref{atree}. This growth with $s$ is the AdS$_3$ manifestation of the CFT violation of the chaos bound in \eqr{hsnreg}.

\sssec*{Higher spin gravitational actions from CFT}

An open question in the higher spin community has been whether one can consistently couple matter to $SL(N)$ higher spin gravity. The only example we know is $SL(N)$ Vasiliev theory -- that is, Vasiliev theory at $\l=\pm N$ with the gauge fields of spin $s>N$ truncated -- which contains a scalar field. This theory is holographically dual to the ``semiclassical limit'' of the $W_N$ minimal models, as recently reviewed in \cite{Hegde:2015dqh}; this is a non-unitary limit. Our result shows that it is impossible to construct other, non-Vasiliev theories of $SL(N)$ gauge fields coupled to matter that are actually consistent with CFT unitarity.

\sssec*{Pure higher spin gravity}
Pure AdS$_3$ higher spin gravity, with no matter, is not ruled out by our arguments. The discussion of Section \ref{s32} could be applied essentially verbatim to this case. While pure higher spin gravity, like pure Einstein gravity, is a conceptually interesting theory to consider, it has no dynamics. Based partly on interpretational difficulties in pure gravity, we suspect that weakly coupled pure higher spin gravities, and their would-be CFT duals, do not exist.\footnote{For a somewhat different perspective, see \cite{Honda:2015mel}, which claims to compute the exact $SL(N,\mathbb{C})$ higher spin gravitational path integral over manifolds with solid torus topology. We only note that the result is strongly constrained by several assumptions about the allowed saddle points in the path integration; also, the result is only determined up to an additive constant which is critical for distinguishing among possible dual CFTs.}

An obvious question is what happens when there is an infinite tower of massless higher spin fields. This will be the subject of the next section.

\sec{Regge Behavior in $\winf$ CFTs and 3D Vasiliev Theory}\label{s5}

We turn now to the most interesting case of a higher spin theory: one with an infinite tower of currents. In particular, we consider a 2d CFT with a current at each spin $s=2,3,\ldots,\i$, that altogether furnish a classical $W_{\infty}[\l]$ symmetry. This is the asymptotic symmetry algebra of \hsl\ Chern-Simons theory in AdS$_3$, which also forms the gauge sector of 3D Vasiliev theory \cite{Henneaux:2010xg, Gaberdiel:2011wb}.

With the same justification as in earlier sections, we assume that the $W_{\infty}[\l]$ vacuum block at $\O(1/c)$ in the Regge limit is sufficient to read off $\l_L$. In fact, we will be able to derive the vacuum block for all $z$, not only in the Regge limit. Due to the infinite tower of spins, we need to perform a resummation. The result will therefore be highly sensitive to the interrelations among the coefficients of the different terms in the sum, which are fixed by the higher spin charges of $V$ and $W$. These are constrained to furnish a representation of $W_{\i}[\l]$. Happily, we will find that the sum over spins ``Regge-izes'' to give a result consistent with the chaos bound: $\l_L=0$. 

\ssec{Resumming higher spins in $W_{\i}[\l]$}\label{s51}

With normalization \eqr{jnorm}, the block reads
\e{}{\cF_{\rm vac,\infty}(z|\l) = 1+{1\over c}{\F_{\vac,\i}^{(1)}(z|\l)}+O\left({1\over c^2}\right)}
where
\e{fwb}{\F_{\vac,\i}^{(1)}(z|\l)\equiv \sum_{s=2}^{\infty}{ \qvs q_w^{(s)}\over N_s}z^s {}_2F_1(s,s,2s,z)}
To evaluate this, we need to fix $\qvs, \qws$. $V$ and $W$ are primary operators, which means that they furnish a highest-weight representation of $\winf$. Highest-weight representations of $\winf$ can be specified by Young tableaux; these may be thought of as $SU(N)$ Young tableaux, analytically continued to non-integer $N$. The simplest choice is to take $V=W$ in the so-called ``minimal'' representation, or fundamental representation, which we denote $V=W={\bf f}$. This is an especially pertinent choice: the single-particle states of the scalar field in the Vasiliev theory carry these quantum numbers.\footnote{
There are also conjugate representations with charges obtained by taking $\l\rar-\l$. In Vasiliev language, this is the scalar in alternate quantization.} For the representation ${\bf f}$, the higher spin charges for arbitrary $s$ were derived in Section 5 of \cite{Ammon:2011ua}. The ratio $\qvs \qws/N_s$, which is invariant under rescaling $J_s$, is
\e{qqn}{{\qvs({\bf f}) \qws({\bf f})\over N_s} =  {(1-\l^2)\Gamma(1-\l)\over \Gamma(1+\l)}{\Gamma^2(s)\Gamma(s+\l)\over \Gamma(2s-1)\Gamma(s-\l)}}
Note that in a normalization in which $N_2=1/2$, as is typical for the Virasoro algebra, the conformal dimension is $h({\bf f}) = q^{(2)}({\bf f})=(1+\l)/2$. So it only makes sense to consider $\l> -1$ in this calculation, otherwise $V$ and $W$ have negative norm.\footnote{Note, though, that setting $\l=-N$ for $N\in\Z$, all charges $q^{(s>N)}$ vanish and we recover our previous result $\l_L=2\pi(N-1)/\b$ for $W_N$. This is to be expected from the fact that $W_{\i}[\pm N]\cong W_N$ after modding out generators of spins $s>N$.}

We now plug \eqr{qqn} into \eqr{fwb} and perform the sum. We do this by using the integral representation of the hypergeometric function,
\e{hypint}{\hyp(s,s,2s,z) = {\Gamma(2s)\over \Gamma^2(s)}\int_0^1 dt (t(1-t))^{s-1}(1-z t)^{-s}~,}
then exchanging the order of the sum and integral:
\e{}{\F_{\vac,\i}^{(1)}(z|\l) = {(1-\l^2)\Gamma(1-\l)\over \Gamma(1+\l)} \int_0^1 dt(t(1-t))^{-1}\left[\sum_{s=2}^{\infty}{(2s-1)\Gamma(s+\l)\over \Gamma(s-\l)}\left({z t(1-t)\over 1-z t}\right)^{s}\right]}
For various rational values of $\l$, the sum can be done. Upon integrating, we infer the following elegant result for general $\l$:
\e{fvacinf}{\cF_{{\rm vac},\infty}^{(1)}(z|\l)  = (1-\l^2)\big(z\,{}_2F_1(1,1,1-\l,z)+\log(1-z)\big)}
This formula admits a nifty proof. Labeling our two representations of $\finf(z|\l)$ as
\es{}{A&= \sum_{s=2}^{\infty}\left({(1-\l^2)\Gamma(1-\l)\over \Gamma(1+\l)}{\Gamma^2(s)\Gamma(s+\l)\over \Gamma(2s-1)\Gamma(s-\l)}\right)z^s {}_2F_1(s,s,2s,z)~, \\
B&= (1-\l^2)\big(z\,{}_2F_1(1,1,1-\l,z)+\log(1-z)\big)}
where $A$ is \eqr{fwb} and $B$ is \eqr{fvacinf}, we want to prove that $A=B$. Writing out the series expansion of the hypergeometric functions in $A$, collecting terms of a given power of $z$, and performing the sum over $s$, one finds
\e{A4f3}{A =\sum_{p=2}^{\infty}  A_pz^p~,\quad \text{where}\quad A_p={3(p-1)\over p(p+1)}(1+\l)^2\,{}_4 F_3\Farg{
1,\,{5\over 2},\,2-p,\,2+\l}{{3\over 2},\, 2+p, \,2-\l}{-1}}
On the other hand, the series expansion of $B$ reads
\e{}{B =\sum_{p=2}^{\infty}  B_pz^p~,\quad \text{where}\quad B_p={(1-\l^2)\over p}\left({\Gamma(p+1)\Gamma(1-\l)\over \Gamma(p-\l)}-1\right)}
To show that $A_p=B_p$, we use the following welcome identity for a closely related ${}_4F_3$ (see e.g. \cite{opac-b1078126}, p.561):
\e{}{{}_4 F_3\Farg{
1,\,{3\over 2},\,1-p,\,1+\l}{{1\over 2},\, 1+p, \,1-\l}{-1} = {\Gamma(1+p)\Gamma(1-\l)\over \Gamma(p-\l)}}
Note that all of its parameters besides the 1 are shifted by 1 relative to the ${}_4F_3$ of interest in \eqr{A4f3}. In fact, the two ${}_4F_3$'s share a simple relation, which is clear by series expansion:
\es{}{{}_4 F_3\Farg{
1,\,{3\over 2},\,1-p,\,1+\l}{{1\over 2},\, 1+p, \,1-\l}{-1}&= 1-3{(1-p)(1+\l)\over (1+p)(1-\l)}{}_4 F_3\Farg{
1,\,{5\over 2},\,2-p,\,2+\l}{{3\over 2},\, 2+p, \,2-\l}{-1}}
$A_p=B_p$ follows.

With $\finf(z|\l)$ in hand, we can now take its Regge limit with ease. For $\l\notin \Z$, the monodromies under $(1-z) \rar e^{-2\pi i}(1-z)$ are
\es{}{z\,{}_2F_1(1,1,1-\l,z) &~\rar~ z\,{}_2F_1(1,1,1-\l,z)-2\pi i \l e^{-\pi i \l}\left({z\over 1-z}\right)^{1+\l}\\
\log(1-z) &~\rar~ \log(1-z)-2\pi i}
Taking the Regge limit $z\rar 0$, and recalling that we restrict to $\l>-1$, the leading term is the constant coming from the log:
\e{}{\cF_{{\rm vac},\infty}^{(1)}(z|\l)  \sim -2\pi i(1-\l^2)+\ldots}
That is, the Regge limit of the $\winf$ vacuum block for charges \eqr{qqn} reads
\e{finalinf}{\F^\Reg_{\vac,\i}(z|\l)  = 1-{2\pi i(1-\l^2)\over c}+\ldots}
The sum over spins gives a softer behavior than any term in the sum; indeed, the result does not grow at all!

Thus, we conclude that the OTO correlator $\la V^\dag W^\dag V W\ra_\b$ is characterized by a vanishing Lyapunov exponent:
\e{}{\l_L=0~.}

\subsubsec{Redux: Conformal Regge theory}
It was somewhat remarkable that the infinite sum over higher spin exchanges defining $\finf(z|\l)$ could be performed, yielding the simple expression \eqr{fvacinf} whose Regge limit was trivial to extract. We now perform a different computation: we instead take the Regge limit of each global conformal block appearing in the sum \eqr{fwb}, {\it before} performing the sum, then keep the leading term near $z=0$. This is essentially a realization of conformal Regge theory \cite{Costa:2012cb}.

We start from \eqr{fwb} with charges \eqr{qqn}. The monodromy of $\hyp(s,s,2s,z)$ around $z=1$ is given in \eqr{monod}. 
We have $s\geq 2$, so the second term dominates as $z\rar 0$. Keeping only the leading order term at each spin and plugging into \eqr{fwb}, the sum we want to perform is
\e{regf}{\F_{\vac,\i}^{(1),{\rm Regge}}(z|\l) \stackrel{?}{=} 2\pi i{(1-\l^2)\Gamma(1-\l)\over \Gamma(1+\l)}\left[\sum_{s=2}^{\i}{\Gamma(s+\l)\over \Gamma(s-\l)}{\Gamma(2s)\over \Gamma^2(s)}z^{1-s}\right]_{z\rar0}}
Each term is more divergent than the next. Performing the sum, the right-hand side of \eqr{regf} becomes
\es{}{&-2\pi i(1-\l^2)\left(1-\hyp\left({3\over 2},1+\l,1-\l,{4\over z}\right)\right)}
At small $z$,
\es{}{1-\hyp\left({3\over 2},1+\l,1-\l,{4\over z}\right)&\approx 1-z^{3/2}\frac{i  \Gamma (1-\lambda ) \Gamma \left(\lambda -\frac{1}{2}\right)}{8 \Gamma \left(-\lambda -\frac{1}{2}\right) \Gamma (1+\lambda)}\\&+z^{1+\l}\frac{(-1)^{-\lambda } 2^{-2 \lambda -1} \Gamma (1-\lambda ) \Gamma \left(\frac{1}{2}-\lambda \right)}{\sqrt{\pi } \Gamma (-2 \lambda )}+\ldots}
Recalling that $\l>-1$, the leading term precisely agrees with \eqr{finalinf}.

In conformal Regge theory, knowledge of the spectrum of exchanges and their couplings to the external operators is sufficient to compute the Regge limit of the correlation function, in the form of an effective ``Regge pole'' of spin $j$ living in the complex spin plane. We have traded an infinite sum over higher spin current exchanges for a single effective exchange of $j=1$. Presumably the same result could have been reached by directly employing the techniques developed in \cite{Costa:2012cb}, although we did not use them here. 

The analogous computation in CFTs with string theory duals -- a sum over higher spin states dual to Regge trajectories of the closed string -- is a hallmark of their UV finiteness. It is tantalizing to see a similar structure operating here.

\sssec{Comments}

\sssec*{Holographic interpretation}

Unlike the case of $SL(N)$-type higher spin gravities, a weakly coupled \hsl\ higher spin theory is causal. It obeys the chaos bound, with $\l_L=0$ for all $\l$. Despite the fact that a dual CFT need not be free, these bulk theories behave similarly to Vasiliev theories in $d>2$: their dynamics is non-chaotic, and thus, in a specific sense, integrable.

The 3D Vasiliev theory is the only known theory of \hsl\ higher spins that couples to matter. Our results suggest that the higher spin black holes of \cite{Kraus:2011ds}, with an infinite tower of higher spin charges, cannot be formed in Vasiliev theory by throwing higher spin quanta into a BTZ black hole. Perhaps they cannot form at all.

In our calculation, an infinite sum over higher spin exchanges yields a result with a causal Regge limit. The Vasiliev theory has far fewer fields than string theory -- indeed, it has no massive higher spin states at all -- but nevertheless exhibits a stringy structure, as discussed in the introduction. It would be fascinating to try to find a specific string theory in which the non-supersymmetric Vasiliev theory embeds.

\sssec*{Other representations of $\winf$}

$\l_L$ is supposed to be independent of the choice of $V,W$. However, since our calculation is sensitive to the precise choice of charges, we repeat the derivation for a different choice. In Appendix \ref{appd}, we take $V$ and $W$ to be distinct operators, with $V= {\bf f}$ in the minimal representation as before, and $W$ now in the antisymmetric two-box representation, {\bf asym}$_2$. With these charges, the result for $\finf(z|\l)$ on the first sheet is inferred to be
\es{}{\finf(z|\l) &= \sum_{s=2}^{\infty}{q_v^{(s)}({\bf f}) q_w^{(s)}({\bf asym}_2)\over N_s} z^s \hyp(s,s,2s,z)\\
&= 2(1-\l^2) \left(z\,{}_3F_2(3,1,1;2,1-\l;z)+\log(1-z)\right)}%
Taking its Regge limit, the constant term from the log again dominates, giving $\l_L=0$. %

\sssec*{$\l=1$ and free bosons}

The case $\l=1$ is special:
\e{}{\finf(z|1) = {z^2\over (1-z)^2}}
It has trivial monodromy, going like $z^2$, not $z^0$, in the Regge limit. This is related to the fact that at $\l=1$, the algebra linearizes. The linear algebra, often known as $W_{\i}^{\rm PRS}$ \cite{Pope:1989ew}, may be realized by a theory of $c$ free bosons. The operator with charges \eqr{qqn} is the marginal bilinear $\p\phi \bar\p\phi$. In this way, the $\l=1$ case is reminiscent of the free $O(N)$ vector models in $d>2$, where $V$ and $W$ are the $d=2$ analogs of the bilinear $O(N)$ singlet operator $J_0 = \phi^i \phi^i$. Indeed, the four-point function of $J_0$ in the $d$-dimensional free $O(N)$ model \cite{Dolan:2003hv},
\es{4pfon}{{\la J_0(x_1)J_0(x_2)J_0(x_3)J_0(x_4)\ra\over \la J_0(x_1)J_0(x_2)\ra\la J_0(x_3)J_0(x_4)\ra} &= 1+u^{d-2}+\left({u\over v}\right)^{d-2} \\&+ {4\over N}\left(u^{d/2-1}+\left({u\over v}\right)^{d/2-1}+u^{d/2-1}\left({u\over v}\right)^{d/2-1}\right)}
has a simple monodromy, and the connected part leads to a negative Lyapunov exponent, $\l_L = -2\lfloor{d-1\over 2}\rfloor$.

Note also that if one performed the Regge summation in \eqr{regf} at fixed $\l=1$, one would find a leading term of $\O(z^{3/2})$, as opposed to the correct scaling $\O(z^2)$. This is secretly because the constant term of \eqr{finalinf} vanishes at $\l=1$, and in the Regge analysis, the subleading terms are not to be trusted. This example highlights the fact that the Regge technique is not always applicable: in particular, the same mismatch happens for the free $O(N)$ bosons in all $d$ \cite{douglas}. This may be true of free theories in general.

\sssec*{An upper bound on $\l$?}

Note that the sign of \eqr{finalinf} depends on whether $\l>1$. Because there is no $z$-dependence, and because it is imaginary, this term is not constrained by the chaos bound to be sign-definite. Nevertheless, in Section \ref{s53} we {\it do} derive a bound on $\l$, without using chaos. Namely, we prove that unitary, large $c$ CFTs with $\winf$ symmetry can only exist for $\l\leq 2$.

\sssec*{$W_N$ minimal models}
One family of known, unitary CFTs with large $c$ and $\winf$ symmetry is the `t Hooft limit,  introduced by Gaberdiel and Gopakumar, of the $W_N$ minimal models \cite{Gaberdiel:2010pz,Gaberdiel:2012uj} . The limit CFTs have $\winf$ symmetry with $0\leq \l\leq 1$. As this is a large $N$ limit of a soluble CFT, it is unsurprising that it would have $\l_L=0$. What we have shown is that this is a feature of the $\winf$ algebra, independent of any particular CFT realization. 

\ssec{$\l_L=0$ at finite $c$}\label{s52}

One might worry that in the presence of an infinite tower of higher spin currents, using the vacuum block alone to diagnose chaos misses something. We now provide evidence to the contrary, in the specific context of the  `t Hooft limit of the $W_N$ minimal models. These may be defined via the coset construction
\e{}{{SU(N)_k\oplus SU(N)_1\over SU(N)_{k+1}}}
where 
\e{}{c=(N-1)\left(1-{N(N+1)\over (N+k)(N+k+1)}\right)}
Correlation functions in the 3D Vasiliev theory may be computed using the $W_N$ minimal models in the `t Hooft limit,%
\e{}{N,k\rar\i~, \quad \l\equiv{N\over N+k+1}~\text{fixed}}
where $\l$ is identified with the $\l$ of the bulk. The bulk scalar field, which has $m^2=-1+\l^2$, is taken in standard quantization, and is dual to the minimal model primary $(f,0)$.\footnote{That $\l$ is defined this way, and not as $\l=N/(N+k)$, is required by the choice of standard quantization. This ensures the consistency of the bulk 1-loop free energy with the $\O(N^0)$ CFT central charge \cite{Giombi:2013fka}.} Note that $0\leq \l\leq 1$. Euclidean correlators in the $W_N$ minimal models are known for all values of $N$ and $k$ (e.g. \cite{Papadodimas:2011pf,Chang:2011vka, Hijano:2013fja}). We can analytically continue these to compute the OTO correlators of interest, which contain {\it all} exchanges, vacuum and otherwise. 

The calculation to follow supports many of the statements in this paper: namely, that we can use the $1/c$ vacuum block alone to diagnose $\l_L$; and that higher orders in $1/c$ do not compete with $\l_L$ so derived.%

As in Section \ref{s51}, we take $V$ and $W$ both in the minimal representation. This identifies them with the minimal model primaries
\e{}{V=(f,0)~, \quad W=(f,0)}
The $(f,0)$ operator, a scalar, has conformal dimension
\e{}{\D = {(N-1)(2N+k+1)\over N(N+k)}}
which becomes $\D\approx 1+\l$ in the limit. In \cite{Papadodimas:2011pf}, the four-point function $\la V V^{\dag} W W^{\dag}\ra$ was computed to be
\es{PR}{\langle V(\infty) V^{\dag}(1)W(z,\zb)  W^{\dag}(0)\rangle
= |z(1-z)|^{-2\D}\Big(h_1(z)h_1(\zb)+N_1h_2(z)h_2(\zb)\Big)}
where we define
\es{}{h_1(z) &\equiv (1-z)^{k+2N\over k+N}\hyp\left({k+N+1\over k+N}, -{1\over k+N}, -{N\over k+N}, z\right)\\
h_2(z) &\equiv z^{k+2N\over k+N}\hyp\left({k+N+1\over k+N}, -{1\over k+N}, {2k+3N\over k+N}, z\right)}
and
\e{}{N_1 \equiv -{\G{k+2N-1\over k+N}\G{-N\over k+N}^2\G{2k+3N+1\over k+N}\over \G{-k-2N-1\over k+N}\G{1-N\over k+N}\G{2k+3N\over k+N}^2}}

We want to take the Regge limit of \eqr{PR} normalized by the two-point functions, i.e. of the reduced amplitude\footnote{We have accounted for both possible Wick contractions, since $V=W$. With this convention, $\A(z,\zb) \rar 1$ in the `t Hooft limit.}
\es{}{\A(z,\zb) &= \langle V(\infty) V^{\dag}(1)W(z,\zb)  W^{\dag}(0)\rangle{|z(1-z)|^{2\D}\over |z|^{2\D}+|1-z|^{2\D}}\\
&={1\over |z|^{2\D}+|1-z|^{2\D}}(h_1(z)h_1(\zb) + N_1h_2(z)h_2(\zb))}

To begin, consider the $h_2$ terms. In the Regge limit, this will vanish: up to monodromy coefficients,
\es{2xtr}{{1\over |z|^{2\D}+|1-z|^{2\D}}h_2(z)h_2(\zb) &\sim \zb^{k+2N\over k+N}z^{{k+2N\over k+N}}\left(z^{1-{2k+3N\over k+N}} +\ldots\right) \\&= \zb^{k+2N\over k+N}+\ldots}
The exponent is positive for all physical $N,k$. In the `t Hooft limit, ${k+2N\over k+N} \rar 1+\l+\O({1\over N})$. Turning now to the $h_1$ terms
\e{PRout}{{1\over |z|^{2\D}+|1-z|^{2\D}}h_1(z)h_1(\zb)  \sim e^{-2\pi i \left({k+2N\over k+N}-\D\right)}(1+\O(z,\zb))}
Using ${k+2N\over k+N}-\D = 1/(N-\l)$, in the large $N$ limit, this becomes
\e{}{ \A^\Reg(z,\eta) = 1-{2\pi i \over N}+\ldots}
where $\ldots$ includes higher order terms in $1/N$ and in $z,\zb$. Noting that $c = N(1-\l^2)+\O(N^0)$ in the `t Hooft limit, the $1/c$ expansion of the chaotic correlator is therefore
\e{regagain}{\A^\Reg(z,\eta) = 1-{2\pi i (1-\l^2)\over c}+\ldots}
which precisely matches our result from $\finf$ alone. Moreover, we learn something important: the leading term \eqr{PRout} has an expansion to all orders in $1/c$. This strongly supports the notion that $\l_L=0$ can indeed be read off by taking large $c$ first.

Furthermore, if we expand in $s$-channel conformal blocks, we can trace \eqr{regagain} back to the $\winf$ vacuum block. Consider the reduced amplitude
\es{}{\A(z,\zb)&=|z|^{2\D}\langle V(\infty) V^{\dag}(1)W(z,\zb)  W^{\dag}(0)\rangle\\
&= |1-z|^{-2\D}(h_1(z)h_1(\zb)+N_1 h_2(z)h_2(\zb))}
In the $VV$--$WW$ channel, this decomposes into $\winf$ conformal blocks $\F_{p,\i}(z|\l)$, as in \eqr{wde}, in the `t Hooft limit. Focusing on the holomorphic pieces to save space, we find
\es{e536}{(1-z)^{-\D} h_1(z) &\approx 1+{1-\l^2\over c}\big(\log(1-z)-\l^2\p^{(3)}\hyp(0,1,-\l,z)\\&\quad\quad\quad\quad\quad\quad\quad\quad+\l(\p^{(2)}-\p^{(1)})\hyp(0,1,-\l,z)\big)+\O(1/c^2)\\
(1-z)^{-\D} h_2(z)&\approx z^{1+\l}\left(1+\O(1/c)\right)}
where $\p^{(i)}$ acts on the $i$'th parameter. The last line of \eqr{e536} encodes the contribution of double-trace operators $[VV]_{n,s}$ and $[WW]_{n,s}$, which have holomorphic weights $h=1+\l+n+s$. Together with its anti-holomorphic part, this is subleading in the Regge limit, cf. \eqr{2xtr}, so we focus on the first lines. The derivatives can be simplified easily, using the series representation of the hypergeometric function. One finds%
%
\es{}{\p^{(3)}\hyp(0,1,-\l,z)&=0\\
\p^{(2)}\hyp(0,1,-\l,z)&=0\\
\p^{(1)}\hyp(0,1,-\l,z) &=-{z\over \l}\,\hyp(1,1,1-\l,z) }
Plugging in above, we get
\e{}{(1-z)^{-\D}h_1(z)\approx 1 + {1-\l^2\over c}\big(z \,\hyp(1,1,1-\l,z)+\log(1-z)\big)+\ldots}
The $1/c$ piece is precisely $\finf(z|\l)$. This example provides evidence that the $1/c$ vacuum block is sufficient to determine $\l_L$ in general holographic CFTs.

\ssec{No unitary $\winf$ CFTs for $\l>2$}\label{s53}
This section has nothing to do with chaos. However, we include it because it provides a complementary constraint on the space of large $c$ higher spin theories. 

We now prove the following claim: 
\vs
\centerline{ {\it Unitary, large $c$ 2d CFTs with a $W_{\i}[\l]$ chiral algebra with $\l>2$ do not exist.}}
\vs

\noindent The proof is simple. We observe that while the classical (i.e. large $c$) $\winf$ algebra is defined for any $\l$, it is actually {\it complex} for $\l>2$.\footnote{For $\l=\pm N$, we can truncate to $W_{\i}^{\rm cl}[\pm N]/\chi_N\cong W^{\rm cl}_N$, where $\chi_N$ is the ideal consisting of generators of spins $s>N$. The latter algebra obviously exists, and is real. Our  statement applies to the $W^{\rm cl}_{\infty}[\pm N]$ algebra {\it before} truncating, when it is still an infinite-dimensional algebra.} Mathematically, this is perfectly acceptable. However, if a chiral algebra describes the current sector of a CFT, its structure constants are identified with three-point coefficients of currents. In a diagonal basis of two-point functions with real and positive norms, unitarity forces these coefficients to be real. 

CFT three-point functions are holographically computed as cubic scattering amplitudes in AdS. So the dual claim is that  
\vs
\centerline{{\it 3D Vasiliev and pure \hsl\ higher spin gravities with $\l>2$ have}}
\centerline{{\it imaginary gauge field scattering amplitudes.}}
\vs
\noindent This statement is independent of the matter sector, and extends to any bulk theory, known or unknown, containing a \hsl\ subsector of higher spin gauge fields.
 \begin{figure}[t!]
   \begin{center}
 \includegraphics[width = .53\textwidth]{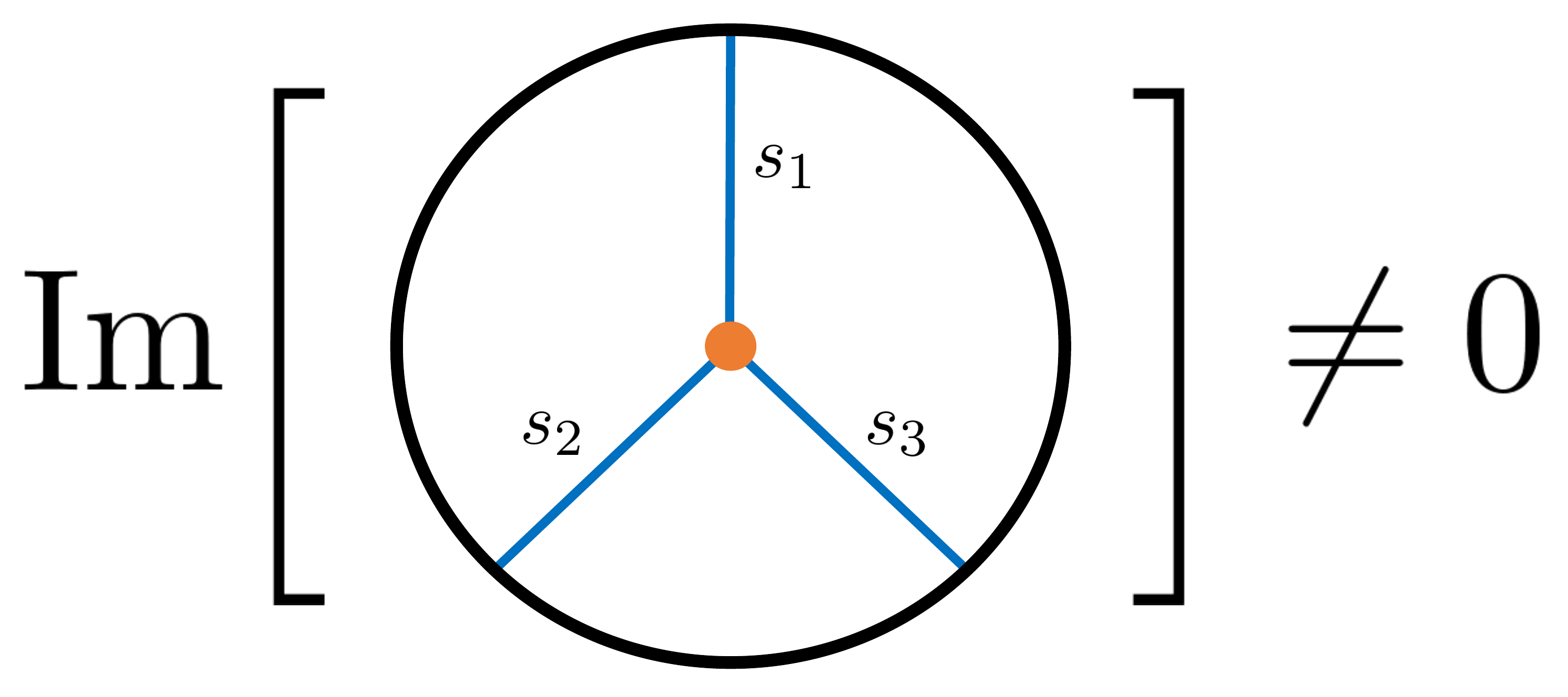}
 \caption{In \hsl\ higher spin gravity, which also forms the pure gauge sector of 3D Vasiliev theory, an infinite set of three-point scattering amplitudes is imaginary for any value $\l>2$. This follows from the existence of a classical $W_\i[\l>2]$ asymptotic symmetry algebra, which is complex.}
 \label{f7}
 \end{center}
 \end{figure}

Onto the proof. In \cite{Gaberdiel:2012ku}, the structure of the quantum (i.e. finite $c$) $W_{\infty}[\l]$ algebra was studied, and  several structure constants $C_{ij}^k$ -- that is, the three-point coefficients of higher spin currents of spins $i,j$ and $k$ -- were determined. It was convincingly argued that all $C_{ij}^k$ are determined in terms of two free parameters, which can be taken to be $c$ and $C_{33}^4$. In the diagonal basis \eqr{jnorm}, normalizing the currents as $N_s=1/s$,\footnote{Note that the choice $N_s=1/s$ is the same normalization as in \cite{Gaberdiel:2012ku}, but their use of the symbol $N_s$ is different.} $C_{33}^4$ is related to $\l$ and $c$ as \cite{Gaberdiel:2012ku}%
\e{gsq}{\g^2 \equiv (C_{33}^4)^2 = {64(c+2)(\l-3)(c(\l+3)+2(4\l+3)(\l-1))\over (5c+22)(\l-2)(c(\l+2)+(3\l+2)(\l-1))}}
For fixed $c$, there is a range of values for which $\g^2<0$. Taking $c\rar\infty$ yields the value of $\g$ in the classical algebra, which we denote $W_{\i}^{\rm cl}[\l]$:
\e{}{\lim_{c\rar\infty}\g^2 = {64\over 5}{\l^2-9\over \l^2-4}}
Clearly, 
\e{}{\g^2<0~\text{when}~2<\l<3}
As explained above, this rules out $W_{\infty}^{\rm cl}[2<\l<3]$ as the chiral algebra of a unitary CFT. 

We can now starting climbing our way up the spin ladder. \cite{Gaberdiel:2012ku} also derive the structure constant $C_{34}^5$. At large $c$, 
\e{}{\lim_{c\rar\infty}(C_{34}^5)^2 = {{375\over 112}\lim_{c\rar\infty}\g^2-25} = {125\over 7}{\l^2-16\over \l^2-4}}
We now have
\e{}{(C_{34}^5)^2 <0~\text{when}~2<\l<4}
This rules out $W_{\infty}^{\rm cl}[2<\l<4]$ as the chiral algebra of a unitary CFT. 

The pattern at higher spins is clear. In Appendix \ref{appe} we exclude the entire range $\l>2$, as OPE coefficients involving successively higher spins become imaginary for successively higher $\l$. In particular, we show, using only general properties of the algebra, that 
\e{3s1}{(C_{3s}^{s+1})^2 = {\l^2-s^2\over \l^2-4}\times (\text{Positive})}
Taking $s\rar\infty$ proves our claim.

When $\l\rar\i$, one should be more careful. For example, if one takes the limit $\l\rar\i, c\rar\i$ with $c/\l^3$ fixed, the structure constants are all real (see e.g. section 4.2 of \cite{Beem:2014kka}). Indeed, there is a well-known unitary theory which realizes these structure constants as three-point couplings: namely, the $d=6$, (2,0) superconformal field theories with Lie algebra su$(N)$, where $N=\l\rar\i$. The aforementioned limit of $W^{\rm cl}_{\infty}[\l]$ is the chiral algebra of the protected sector of the (2,0) theory at large $N$, and thus determines its OPE. It seems likely that CFTs with vector-like growth $c\propto\l\rar\i$ cannot exist.

\sec{AdS/CFT Sans Chaos}\label{s6}

So far, we have studied theories with varying amounts of chaos. Vasiliev theory aside, the landscape of AdS/CFT contains other interesting and more familiar examples of theories with $\l_L$ far below its upper bound. We present some cases here. 

\subsec{Chiral CFTs}

Chiral CFTs are $d=2$ CFTs with $c_R=0$ and $c_L=24k$ where $k\in\Z$. These are perhaps the most symmetric of all CFTs: every operator is either a current or a descendant of a current. In other words, the CFT consists solely of the vacuum module of an exotic $\W$-algebra with $c=24k$. Due to the high degree of symmetry, such theories should not be chaotic, for any value of $k$. Indeed, given some holomorphic primary current $J_s(z)$ with conformal weight $h=s\in\Z$, its four-point function is constrained to take the form \cite{Bouwknegt:1988sv, Headrick:2015gba}
\e{}{\la J_s(\i)J_s(1)J_s(z)J_s(0)\ra =z^{-2s} \sum_{n=0}^{\lfloor2s/3\rfloor} c_n {z^{2n} (1-z+z^2)^{2s-3 n}\over (1-z)^{2s-2n} }}
where the constants $c_n$ are determined by the OPE. This has trivial monodromy around $z=1$, hence $\l_L=0$. The same conclusion obviously holds for holomorphically factorized CFTs. 

It has been suggested that the CFT dual to pure AdS$_3$ quantum gravity holomorphically factorizes \cite{Witten:2007kt}. This would imply a hidden infinite-dimensional symmetry among the tower of BTZ black hole states. The result $\l_L=0$ for factorized CFTs would seem to be in tension with a value $\l_L=2\pi/\b$ associated to chaotic evolution in Einstein gravity. Likewise for the behavior of two-interval mutual information after a global quench \cite{Asplund:2015eha}: whereas chiral and factorized CFTs exhibit a ``dip'' in their entanglement entropy after the quench, classical AdS$_3$ gravity shows no such effect, as the entanglement ``scrambles'' maximally.\footnote{The presence of strong chaos and entanglement scrambling have recently been argued to be different manifestations of the same underlying physics \cite{Hosur:2015ylk}.} As explained in Section \ref{s32}, this tension is illusory, for the same reason that factorization of the dual CFT is not patently false: pure quantum gravity is topological. Analogous comments apply to chiral gravity.

\subsec{The D1-D5 CFT}
Consider a symmetric orbifold CFT, Sym$^N(X)$ CFT, for some seed CFT $X$. All such CFTs have a large chiral algebra that becomes infinitely generated at large $N$. We again expect these theories to be non-chaotic for generic choices of $V$ and $W$. We note that all Sym$^N$ CFTs are sparse in the precise sense of \cite{Hartman:2014oaa}.

Technology for correlators in Sym$^N$ CFTs has been developed in e.g. \cite{Lunin:2000yv, Pakman:2009zz,Burrington:2012yn}. We now perform an OTO correlator calculation in a CFT especially relevant for holography, namely, the D1-D5 CFT at the symmetric orbifold point, Sym$^N(T^4)$. This is a $\N=(4,4)$ SCFT with a global $SO(4)_I\cong  SU(2)_{I_1}\times SU(2)_{I_2}$ 
symmetry, and central charge $c=6N$. In terms of the dual string description, $N=N_1N_5$, where $N_1$ and $N_5$ count D1- and D5-branes, respectively. Each of the $N$ copies is a $c=6$ theory of four real bosons $X^i$ and their fermionic superpartners, where $i=1\ldots 4$.  

In \cite{Burrington:2012yn}, a four-point function of two twist operators and two non-twist operators (among others) was computed in this theory. In particular, consider the following two Virasoro primary operators:
\es{}{\O_d(z,\zb) &= \eps_{AB}G_{-\half}^{-A}\tilde G_{-\half}^{\dot -B}\sigma_2^{+\dot +}\\
\Phi_{\rm dil}(z,\zb) &= \sum_{\kappa=1}^N \p X^i_{(\kappa)}\pb X^i_{(\kappa)}}
$\O_d$ is an exactly marginal scalar primary, which is a superconformal descendant of a certain twist operator. $A,B=1,2$ are indices of the $SU(2)_{I_1}$. $\Phi_{\rm dil}$ is the BPS operator dual to the dilaton in the bulk; the index $\kappa$ denotes the copy of $(T^4)^N$. See \cite{Burrington:2012yn} for further details. 

\cite{Burrington:2012yn} viewed the four-point function $\la \O_d\O_d\Phi_{\rm dil}\Phi_{\rm dil}\ra$ as a second-order perturbation of the two-point function $\la \Phi_{\rm dil}\Phi_{\rm dil}\ra$ under the exactly marginal deformation away from the orbifold point. Thinking of $V=\O_d$ and $W=\Phi_{\rm dil}$, we are simply interested in this correlator as a quantity in Sym$^N(T^4)$ itself. Including combinatoric factors of $N$, the $S_N$-invariant correlator, normalized by the disconnected piece, is
\es{}{&{\la  \O_d(0)\O_d(z,\zb)\Phi_{\rm dil}(1)\Phi_{\rm dil}(\infty)\rangle\over \la  \O_d(0)\O_d(z,\zb)\rangle \la \Phi_{\rm dil}(1)\Phi_{\rm dil}(\infty)\rangle} = {N-2\over N} + {1\over N}\Bigg[1+2^{-2} \left|{2-z\over \sqrt{1-z}}\right|^2\\&+2^{-4}\left({z^2(2-z)\over (1-z)^{3/2}}\right)\left(\sqrt{1-\zb}(2-\zb)\right) + 2^{-4}\left({\zb^2(2-\zb)\over (1-\zb)^{3/2}}\right)\left(\sqrt{1-z}(2-z)\right) \\&+2^{-4}\left|{z^2\over (1-z)^{3/2}}\right|^2+2^{-4}\left|{z^2\over \sqrt{1-z}}\right|^2\\&-2^{-5}\left({z^2\over (1-z)^{3/2}}\right)\left({\zb^2\over \sqrt{1-\zb}}\right)-2^{-5}\left({\zb^2\over (1-\zb)^{3/2}}\right)\left({z^2\over \sqrt{1-z}}\right)\Bigg]}

We now take its Regge limit. Except for the constant terms, every term picks up a minus sign as we cross the cut. Expanding near $z,\zb=0$ on the second sheet, 
there is no divergence, and hence no chaos. This is consistent on general grounds with the existence of a higher spin symmetry enhancement of the tensionless type IIB string in AdS$_3 \times S^3\times T^4$.

\ssec{Slightly broken higher spin theories and $1/c$ corrections}
An obvious question is whether $\l_L$ receives $1/c$ corrections, and whether this idea is even sensible. (A first pass in $d=2$ CFT was recently taken in \cite{Fitzpatrick:2016thx}.) One affirmative argument comes from considering slightly broken higher spin CFTs. We use this term in the original sense of \cite{Maldacena:2012sf}: these are CFTs with some large parameter $\tilde N\sim c$ and a coupling $\tilde\l$, with higher spin symmetry breaking of the schematic form
\e{}{\p\cdot J = {\tilde\l\over \sqrt{\tilde N}}JJ}
 This maps to a quantum breaking of bulk higher spin gauge symmetry. The canonical family of such theories is the set of $d=3$ bosonic and fermionic $O(N)$ models at large $N$ and their Chern-Simons deformations \cite{Giombi:2011kc,Aharony:2011jz}, where $\tilde N\propto N$ and $\tilde \l$ is fixed by the `t Hooft coupling $\l \equiv N/k$. 
 
How does chaos develop in these CFTs? We assert that, for finite temperature states on the cylinder $S^1\times \mathbb{R}^{d-1}$, such theories have 
\e{sbhs}{\l_L(\tilde \l)\approx {f(\tilde \l)\over \tilde N}+O\left({1\over \tilde N^2}\right)}
This motivates an extension of the usual definition of $\l_L$ to higher orders in $1/N$. In contrast, CFTs with ``classical'' higher spin symmetry breaking, like planar $\N=4$ SYM, have $\l_L \approx f(\l)+\O(1/N)$. One way to understand \eqr{sbhs} for the $O(N)$ models is from the bulk 4D Vasiliev description. The scalar and higher spin gauge fields pick up mass shifts only through loops. But a shock wave calculation is classical.

\eqr{sbhs} may be generalized to other spatial manifolds, such as $\mathbb{H}^{d-1}$, where $\l_L\neq 0$ in the free theory at $\tilde \l=0$:
\e{sbhs2}{\l_L(\tilde\l) - \l_L(0) \approx {f(\tilde \l)\over \tilde N}+O\left({1\over \tilde N^2}\right)}
This allows us to perform an explicit check: we consider the $d=3$ critical $O(N)$ model in Rindler space, reading off $\l_L$ for the vacuum four-point function of the scalar operator $J_0$ using the techniques of this paper. This correlator has been computed in both the free (cf. \eqr{4pfon}) and critical models \cite{Leonhardt:2003du}; in the latter,
\es{4pfcon}{&{\la J_0(x_1)J_0(x_2)J_0(x_3)J_0(x_4)\ra\over \la J_0(x_1)J_0(x_2)\ra\la J_0(x_3)J_0(x_4)\ra} = 1+u^{2}+\left({u\over v}\right)^{2} \\&+ {1\over N}\Big(u^2\left(u^{-3/2}(1+u-v)+v^{-3/2}(1+v-u)-(uv)^{-3/2}(1-u-v)\right)\Big)}
Under the Lorentzian continuation \eqr{pres}, the half-integer powers of $v$ pick up a minus sign; in the Regge limit, both \eqr{4pfon} and \eqr{4pfcon} give $\la J_0J_0J_0J_0\ra \sim z^{2}$, i.e. $\l_L=-2$, confirming \eqr{sbhs2} for this particular case.

A related statement pertains to any two holographic large $c$ CFTs related by a double-trace flow, as obtained by swapping standard ($\D_+$) for alternate ($\D_-$) boundary conditions of a bulk field in AdS. If we denote these CFTs' respective Lyapunov exponents as $\l_L^{\D_{\pm}}$, then a natural claim is that
\e{}{\l_L^{\D_+} - \l_L^{\D_-} \approx O\left({1\over c}\right)}
where $c\sim 1/G_N$. It would be interesting to verify this explicitly, and (if true) to determine the sign of the correction.

\sec{Discussion}\label{s7}

We conclude with some additional directions for future work, and final reflections on the potential power of chaos in classifying conformal field theories. 

\sssec*{Strings from chaos, and the CFT landscape}

We have explored the way in which $\l_L$ depends on the OPE data, and seen that demanding $\l_L\leq {2\pi/\b}$ constrains the spectrum of higher spin currents. What about other higher spin operators? String theory and AdS/CFT suggest that, at least for a wide class of CFTs, primary operators may be arranged in Regge trajectories. Given the sensitivity of $\l_L$ to the spin spectrum, it seems possible that demanding Reggeization of OTO correlators would lead to a CFT derivation of this picture.\footnote{We thank Tom Hartman, Dan Roberts and Douglas Stanford for conversations on this topic.}%

Quite generally, it would be extremely useful to discover the path of least action between the set of spectral data and $\l_L$. This would create a manifest link between the Euclidean bootstrap built on crossing symmetry, and the Lorentzian bootstrap built on causality and chaos.\footnote{Some recent papers in this direction in the context of rational 2d CFT are \cite{Gu:2016hoy, Caputa:2016tgt}.}

Recently, it has been argued that a kind of ``average'' measure of chaotic behavior of general 2d CFTs can be directly related to the second R\'enyi entropy, $S_2$, for two disjoint intervals separated in space and time \cite{Hosur:2015ylk}. Given that $S_2$ for two intervals is known to be proportional to the torus partition function, this implies that the spectrum alone, and not the OPE coefficients, can determine $\l_L$ and perhaps other broad features of chaos.\footnote{In contrast, entanglement entropy in the vacuum is not directly related to $\l_L$: while the Ryu-Takayanagi result for intervals in vacuum follows from sparseness and a mild assumption about large $c$ growth of OPE coefficients \cite{Hartman:2013mia}, $\l_L={2\pi/ \b}$ does not.} It would be fruitful to better understand the relation between $S_2$ and chaos in purely CFT terms. On the other hand, while an average notion of chaos would be useful, so would understanding the distribution of leading Lyapunov exponents over the space of possible OTO correlators in a given CFT. In generic CFTs, unlike in holographic CFTs, the value of $\l_L$ does depend on the choice of $V$ and $W$. 

If $\l_L$ is hiding in $S_2$, where else is it hiding? How does $\l_L$ relate more generally to entanglement measures? In 2d CFT, do correlators on surfaces of higher topology contain complementary information about chaos? 

We also would like to understand the evolution of chaotic data as we move through the space of CFTs via RG flows or marginal deformations. Along a conformal manifold, $\l_L$ will be a smooth function of the moduli. In the D1-D5 CFT with $T^4$ or $K3$ target, for instance, there is a marginal deformation which triggers passage from the orbifold point to the supergravity point in the moduli space. Along these lines of fixed points, $\l_L$ is a function of the marginal coupling $\l$, interpolating between Sym$^N$ and supergravity regimes: $0\leq \l_L(\l)\leq {2\pi/\b}$. Obtaining $\l_L(\l)$ in closed form is a distant goal, but a perturbative calculation near $\l=0$ seems within reach. (See \cite{Gaberdiel:2015uca} for a recent perturbative spectral calculation.) It is inspirational to consider what role OTO correlators could play in revealing the emergence of a classical spacetime description from CFT. 

\sssec*{Higher spin AdS$_3$/CFT$_2$}
While $SL(N)$-type higher spin gravities may capture some crude aspect of how stringy geometry works, their acausality and dual CFT non-unitarity render them unfit for studying dynamical processes, the black hole information paradox, singularity resolution, and so on. Still, it is perhaps worth quantifying in some detail what goes wrong when trying to directly construct an effective gravitational action coupling $SL(N)$ gauge fields to matter. For comments based on experience with the Vasiliev formalism, see \cite{Kessel:2015kna}. Seeking explicit violations of causality via two-sided Wilson lines in higher spin shock wave backgrounds would also be worthwhile.

We have also shown that 3D Vasiliev theory has imaginary scattering for all $\l>2$. There are reasons to suspect that the range $1< \l<2$ is inconsistent too, based on representation theory \cite{Monnier:2014tfa} and the absence of known, unitary holographic CFTs in this range of $\l$.\footnote{We thank Matthias Gaberdiel for discussions on this point.} Such an inconsistency must come from the scalar coupling to the higher spins. For this reason and others, it would be useful to compute the four-point function of the 3D Vasiliev scalar. Expanding it in conformal blocks would presumably reveal any non-unitarity.

A natural question raised by our results is how large the space of higher spin 2d CFTs actually is. For example, at large $c$, are there CFTs with $\winf$ symmetry besides the $W_N$ minimal models in the `t Hooft limit?

Another natural question is whether demanding $\l_L\leq 2\pi/\b$ uniquely determines the higher spin algebra of a single infinite tower of currents, with one current at each spin $s\geq 2$, to be $\winf$. This is a baby version, phrased in 2d CFT, of the question of whether string theory is unique. 

\sssec*{Slightly broken higher spin chaos}
We would like to check whether \eqr{sbhs} is correct. If so, then computing the leading nonzero term in $\l_L$ in slightly broken higher spin CFTs would seem to require knowing connected correlators at $\O(1/\tilde N^2)$. This is a tall order: even in the critical $O(N)$ model, this is not known. A concrete calculation would be to adapt the ladder diagram techniques of \cite{Stanford:2015owe}, where $1/\tilde N$ is the small parameter. In principle, $\l_L$ should be extractable directly from the spectrum of anomalous dimensions and OPE coefficients. In this way, determining $\l_L$ in these theories would be connected to the slightly broken higher spin bootstrap of \cite{Alday:2015ota}.

\sssec*{A prediction for shock wave scattering in AdS}

The result \eqr{feta12} makes a prediction for the bulk scattering problem of $W$ and $V$ quanta in the background of the hyperbolic AdS black hole at $\b=2\pi$, or the planar BTZ black hole at arbitrary $\b$. An integral representation for $\la VWVW\ra$ was derived in \cite{Shenker:2014cwa}. It was only explicitly evaluated for heavy operators. More precisely, there are two approximations used in the evaluation of the integral in \cite{Shenker:2014cwa, Roberts:2014ifa}: one, that $\D_w\gg\D_v$, and two, that $\D_v\gg 1$. The former permits an interpretation of $V$ moving in a fixed shock wave background generated by $W$ of sharply peaked momentum; the latter allows a geodesic approximation to $\la VV\ra$ evaluated in the shock wave background. For $\D_v,\D_w\sim \O(1)$, neither of those assumptions holds. It would be worthwhile to try to evaluate the overlap integral for light fields and match the functional form of the CFT prediction, and to see whether it also extends to planar black holes in $d>2$.

\section*{Acknowledgments}

We thank Nathan Benjamin, Xi Dong, Ethan Dyer, Liam Fitzpatrick, Matthias Gaberdiel, Simone Giombi, Guy Gur-Ari, Jared Kaplan, Aitor Lewkowycz, Juan Maldacena, Alex Maloney, Mark Mezei, Mukund Rangamani, Steve Shenker, Charlotte Sleight, Herman Verlinde, Ran Yacoby, an anonymous referee, and especially Tom Hartman, Dan Roberts and Douglas Stanford, for valuable discussions. We also thank the Stanford Institute for Theoretical Physics and the UC Davis Physics Department for hospitality in the final stages of this project. Support comes from the Department of Energy under Grant No. DE-FG02-91ER40671.

\begin{appendix}

\section{Conformal blocks in the Regge limit}\label{appa}
Here we present the leading Regge behavior of $d$-dimensional conformal blocks for symmetric tensor exchange. These can be derived in a number of ways, but most easily by solving the conformal Casimir equation in the Regge limit \cite{Cornalba:2006xm}. The result is
\e{regblockapp}{G_{\D,s}^\Reg(z,\eta) = i z^{1-s}\Gc_{\D,s}(\eta)}
where
\e{}{\Gc_{\D,s}(\eta)\equiv C(\D,s)\eta^{\D-s\over 2}\hyp\left({d-2\over 2},\D-1,\D-{d-2\over 2},\eta\right)  }
We have defined
\e{}{C(\D,s) \equiv {\Gamma(d-2)\Gamma\left(s+{d-2\over 2}\right)\over \Gamma\left({d-2\over 2}\right)\Gamma(s+d-2)}C_0(\D+s)}
where
\e{}{ C_0(x) \equiv 2\pi {\Gamma(x)\Gamma(x-1)\over \Gamma^4\left({x\over 2}\right)}}
Note that $C_0(x)$ equals $-i$ times the off-diagonal entry of the monodromy matrix of $\hyp\left({x\over 2},{x\over 2},x,z\right)$ around $z=1$; see \eqr{monod}. Here we are assuming a certain normalization of the blocks which is made explicit below. 

One way to understand why $G_{\D,s}^\Reg\sim z^{1-s}$ is that conformal blocks $G_{\D,s}$ in CFT$_d$ are equivalent to geodesic Witten diagrams in AdS$_{d+1}$ for spin-$s$ exchange \cite{Hijano:2015zsa}. The Regge limit of an ordinary Witten diagram for spin-$s$ exchange scales as $z^{1-s}$. The geodesic Witten diagram, being essentially a restricted amplitude, has the same high energy behavior.

For $d=2,4$, the Regge blocks are given by \eqr{regblockapp} with 
\es{}{d=2:&\quad \Gc_{\D,s}(\eta) = {1+\delta_{s,0}\over 2}\eta^{\D-s\over 2}C_0(\D+s)\\
d=4:&\quad \Gc_{\D,s}(\eta) = {1\over 1+s}{\eta^{\D-s\over 2}\over 1-\eta}C_0(\D+s)}
These results may be easily checked using the closed-form expressions for the blocks, in conjunction with the hypergeometric monodromy \eqr{monod}:
\es{evenblock}{d=2:\quad G_{\D,s}(z,\zb)&= \half\left(g_{h}(z)g_{\hb}(\zb)+(z\leftrightarrow\zb)\right)\\
d=4:\quad G_{\D,s}(z,\zb)&= {1\over s+1}{1\over z-\zb}\left(z g_h(z)\zb g_{\hb-1}(\zb)+(z\leftrightarrow\zb)\right)}
where $g_h(z)$ is the $SL(2,\mathbb{R})$ global block. 
Similar simplification occurs in all even $d$. The fact that $G_{\D,s}$ has a hypergeometric representation in even $d$ gives a natural explanation for the appearance of the $C_0(\D+s)$ factor; interestingly, this factor appears in all $d$.

There is one exception to the above formulas: in $d=2$ when $s=0$, \eqr{regblockapp} is actually not correct when $\D=2h< 1$. Looking at the full blocks \eqr{evenblock} and the monodromy of the hypergeometric function, one sees that the term in \eqr{regblockapp} is actually subleading. Instead, when $\D<1$, the Regge limit is simply
\e{}{d=2: \quad G_{\D<1,0}^\Reg(z,\eta) \approx (z\zb)^h(1+\O(z^{1-\D})) \approx z^{\D}\eta^{\D\over 2}}

\sssec*{A peculiarity in $3d$ CFT}
Note that for $2<d<4$, $\Gc_{\D,0}(\eta)$ is negative for some portion of the region $0\leq\eta<1$ when ${d-2\over 2}<\D<1$, as allowed by unitarity. In particular, this includes $d=3$. On the other hand, in looking at examples of 3d CFTs with $\D<1$ scalars -- the Ising model, critical Gross-Neveu model, and all $O(N)$ models with exactly one relevant scalar in each of the $O(N)$ singlet and vector representations \cite{Kos:2015mba} -- such exchanges do not appear in the OPE of identical operators. It is not clear whether $\D<1$ scalars can ever appear in an $\O\times \O$ OPE for an arbitrary scalar operator $\O$.\footnote{An exception is the line of parity-breaking Chern-Simons-matter fixed points connecting the free $O(N)$ and critical Gross-Neveu models at large $N$. 3d bosonization \cite{Aharony:2012nh} implies that at $O(1/N)$, the scalar bilinear must have $\D<1$ for at least some range of $\l$, so as to smoothly match onto the Gross-Neveu result. Then a result of \cite{Aharony:2012nh} implies a nonzero OPE coefficient.} It would be nice to understand this better.

\sec{Chaos in $\N=4$ super-Yang-Mills}\label{appb}

In this section, we derive an OTO four-point function in Rindler space in planar $\N=4$ SYM at large $\l$, by analytic continuation of the vacuum four-point function of the \2. Regge limits of $\N=4$ correlators have been studied in some detail before (e.g. \cite{Brower:2006ea, Cornalba:2007fs, Cornalba:2008qf, Costa:2012cb, Costa:2013zra}). Here we wish only to present a result in the language of chaos. Our calculation explicitly exhibits the position-dependence ascertained in \eqr{feta12} on general grounds.

We take $V$ and $W$ to be the 1/2-BPS scalar operator in the \2 of $SU(4)$, with $\D_{\bf 20'}=2$. Its vacuum four-point function was computed using supergravity in \cite{Arutyunov:2000py}. We introduce only the most basic aspects of formalism needed to present the result. We follow the conventions of \cite{Dolan:2004iy}, see also \cite{Alday:2014qfa} for a streamlined review. The \2 transforms in the [0,2,0] representation of $SU(4)\cong SO(6)$, and is typically written as a symmetric traceless rank-two tensor of $SO(6)$. Introducing a null vector $t^i$, where $i=1\ldots 6$ and $t^2=0$, we define
\e{}{\O_{\2}(x,t) \equiv \O_{\2;ij}(x)t^it^j}
The four-point function is written as
\e{}{\langle \O_{\2}(x_1,t_1)\O_{\2}(x_2,t_2)\O_{\2}(x_3,t_3)\O_{\2}(x_4,t_4)\ra = \left({t_1\cdot t_2t_3\cdot t_4\over x_{12}^2x_{34}^2}\right)^2{\cal G}(z,\zb;\a,\ab)}
where $z,\zb$ are the usual coordinates in \eqr{uvdef}, $\a,\bar\a$ are defined in terms of the $SU(4)$ invariants
\e{}{\a\ab \equiv {t_1\cdot t_3\,t_2\cdot t_4\over t_1\cdot t_2\,t_3\cdot t_4}~,\quad  (1-\a)(1-\ab) \equiv {t_1\cdot t_4\,t_2\cdot t_3\over t_1\cdot t_2\,t_3\cdot t_4}}
and the $t$ subscript refers to the $n$'th operator. The function ${\cal G}(z,\zb;\a,\ab)$ is constrained by superconformal symmetry to take the following form:
\es{}{&{\cal G}(z,\zb;\a,\ab) = (\a z-1)(\ab z-1)(\a \zb-1)(\ab\zb-1)\H(z,\zb)\\&+{(\ab z-1)(\a \zb-1)(F(z,\a)+F(\zb,\ab))-(\a z-1)(\ab \zb-1)(F(z,\ab)+F(\zb,\a))\over (\a-\ab)(z-\zb)}-k}
where $k$ is a constant. The second line is fixed solely by the exchange of SUSY-protected operators, hence is independent of the coupling. The first line depends on a free function $\H(z,\zb)$, which receives contributions from both protected and unprotected operator exchanges. The conformal block decomposition of $\H(z,\zb)$ includes exchanges of $SU(4)$ singlets only, which is why $\H(z,\zb)$ does not depend on $\a,\ab$.\footnote{For generic four-point functions of, say, the higher $SU(4)$ representations $[0,p,0]$, $\H$ also depends polynomially on $\a,\ab$, with degree fixed by the size of the representations of $SU(4)$ allowed in the $[0,p,0]\times [0,p,0]$ OPE. }

The correlator $\la \O_{\2}\O_{\2}\O_{\2}\O_{\2}\ra$ at large $N$ and large $\l$ was computed from supergravity in \cite{Arutyunov:2000py}. The \2 is the lowest KK mode of a linear combination of $g_{\mu\nu}$ and $C_{\mu\nu\rho\sigma}$ with legs along $S^5$ \cite{Kim:1985ez, D'Hoker:2002aw}. Focusing on $\H(z,\zb)$, the result is \cite{Dolan:2001tt}
\e{}{\H(z,\zb) = -{4\over N^2} (z\zb)^2 \Db_{2422}(z,\zb)}
where $\Db_{2422}(z,\zb)$ is the reduced $D$-function. It may be given a closed-form expression by using $D$-function identities (e.g. \cite{Arutyunov:2002fh}) to write
\e{2422}{\Db_{2422}(z,\zb) = \p_u\p_v(1+u\p_u+v\p_v)\Db_{1111}(z,\zb)}
and using the explicit expression for $\Db_{1111}(z,\zb)$,
\e{}{\bar D_{1111}(z,\zb) = {1\over z-\zb}\left(2\text{Li}_2(z)-2\text{Li}_2(\zb)+\log z\zb\log{1-z\over 1-\zb}\right)}
Recall that $u=z\zb$ and $v=(1-z)(1-\zb)$ were defined in \eqr{uvdef}

We now take the Regge limit of \eqr{2422}. Using the monodromies
\es{}{\text{Li}_2(z) &\,\rar \,\Li_2(z)+2\pi i\log z\\
\log(1-z) &\,\rar\, \log(1-z)-2\pi i~,}
analytic continuation to the second sheet yields
\es{}{\Db_{2422}(z,\zb) \rar \Db_{2422}(z,\zb)+ \p_u\p_v(1+u\p_u+v\p_v)\left[{1\over z-\zb}2\pi i \log\left({z\over \zb}\right)\right]}
Taking the Regge limit and expressing the result in terms of $x$ and $t$ (and recalling that $\b=2\pi$),
\es{}{\H^\Reg(x,t)&= -{\pi i\over 8N^2\eps_{12}^*\eps_{34}}{e^{t} \over \sinh^{7} x}\left(e^{4x}+28e^{2x}-28e^{-2x}-e^{-4x}-{24}x\left(e^{{2}x}+3+e^{{-2}x}\right)\right)}
This passes the obvious consistency checks: namely, $\l_L=2\pi/\b=1$, and $\text{Re}(\H^\Reg(x,t))<0$ for operators in the Euclidean time configuration \eqr{e26}. Note that $v_B=1/3$, consistent with \eqr{i5}. 

Expanding at large $x$, each term can be explained by an accounting of the $SU(4)$ singlet, spin-2 operators appearing in the $\O_\2 \times \O_\2$ OPE at $\O(1/N^2)$. The list of such operators is relatively short: including their twists \cite{Hoffmann:2000dx, Alday:2014tsa},
\es{}{T_{\mu\nu}:&\quad \t=2\\
[\O_\2\O_\2]_{n,2}:&\quad \t=4+2n-{2(1+n)(2+n)(3+n)\over 3N^2}}
This matches the general structure of $f(\eta)$ in Section \ref{s3}.

\sec{More on chaos in $W_N$ CFTs}\label{appc}
We repeat the $W_3$ calculation of Section \ref{s4} for $W_4$. We also do a computation at arbitrary $N$. In both cases, we find chaos bound-violating behavior consistent with \eqr{lams}. 

\ssec{$N=4$}
The semiclassical $W_4$ vacuum block was derived in \cite{Hegde:2015dqh} for general charges, following the derivation in \cite{Castro:2014mza} for the uncharged case. $\F_{\vac,4}(z)$ is
\e{}{\F_{\vac,4}(z) = ((1-z)^5 m_1m_2m_3)^{- h_v/5}\left({m_1\over m_3}\right)^{q_v^{(3)}/2}\left({m_2\over(m_1m_3)^{2/3}}\right)^{q_v^{(4)}}}
where the $m_i$ are
\es{}{m_1 &= -6\left({(1-z)^{-n_1}\over n_{12}n_{13}n_{14}}+\text{cyclic}\right)\\
m_2 &= 12\Bigg({(1-z)^{-n_1-n_2}\over n_{13}n_{14}n_{23}n_{24}} + {(1-z)^{-n_1-n_3}\over n_{12}n_{14}n_{32}n_{34}}+{(1-z)^{-n_1-n_4}\over n_{12}n_{13}n_{42}n_{43}}+{(1-z)^{-n_2-n_3}\over n_{21}n_{24}n_{31}n_{34}}\\&+{(1-z)^{-n_2-n_4}\over n_{21}n_{23}n_{41}n_{43}}+{(1-z)^{-n_3-n_4}\over n_{31}n_{32}n_{41}n_{42}}\Bigg)\\
 m_3&= -6\left({(1-z)^{n_1}\over n_{12}n_{13}n_{14}}+\text{cyclic}\right)}
with $n_{ij} \equiv n_i-n_j$. The $n_i$ are roots of the quartic equation
\be\label{pn}
n^4-{5\a^2\over 2}n^2 + 24 q_3 n + 36 q_4+{9\over 16}\a^4=0
\ee
where we have defined rescaled charges,
\e{}{q_3 \equiv {6\over c}q_w^{(3)}~, \quad q_4 \equiv {6\over c}q_w^{(4)}}
and $\a=\sqrt{1-4\eps}$ was defined in \eqr{alpha}. 

For simplicity, we take $V$ to be uncharged, so that $q_v^{(3)} = q_v^{(4)}=0$. To streamline the result, we turn on either $q_3$ or $q_4$, but not both at once. Let's also define 
\e{}{Z(\eps)\equiv z-4\pi i \eps}
For $q_3\neq 0$, we find
\e{W4a}{\A^\Reg(z,\eta) =  z^{2h_v}\left(-\frac{1}{Z(\eps)^2 \left(Z(\eps)^4-1728\pi^2q_3^2\right) \left(Z(\eps)^4+5184\pi^2q_3^2\right)}\right)^{h_v/5}}
For $q_4\neq 0$, we find
\e{W4b}{\A^\Reg(z,\eta) =  z^{2h_v}\left(\frac{Z(\eps)^9+345600 \pi ^2 q_4^2 Z(\eps)^3-55296000  \pi ^3i q_4^3}{Z(\eps)^{10} \left(Z(\eps)^3-240 \pi i q_4\right)^2 \left(Z(\eps)^3+480 \pi i  q_4\right)}\right)^{h_v/5}}
Both results have been obtained by resummation of perturbation theory through $\O(q^{16})$.

The features that plagued the $W_3$ result are also present here. In equation \eqr{W4a}, every power of $q_3$ comes with a $1/cz^2$, and the correlator is non-analytic in parts of the half-strip. In \eqr{W4a}, every power of $q_4$ comes with a $1/cz^3$, which implies a spin-4 Lyapunov exponent and associated scrambling time
\e{}{\l_L^{(4)}={6\pi\over \b}~, \quad t_*^{(4)}={\b\over 6\pi}\log c}
For various choices of $\eps_{12}^*\eps_{34}$ within the half-strip, the correlator diverges at times $t\approx x+t_*^{(4)}$. For instance, $(\eps_{12}^*\eps_{34})^3 = i$ leads to a divergence from the $Z(\eps)^3-240 i \pi  q_4$ denominator factor.

\ssec{Arbitrary $N$}

We can also extend these arguments to arbitrary $N$. Consider an uncharged probe ($q_v^{(s>2)}=0$) and allow $W$ to have arbitrary higher spin charges $q_w^{(s)}$. Expanding $\F_{\vac,N}(z)$ perturbatively in the $q_w^{(s)}$ using the results of \cite{Hegde:2015dqh}, one finds\footnote{This comes from eq. 6.30 of \cite{Hegde:2015dqh}. First, pass to the plane. Then use $S_{EE} \approx -\log \F_0 -\F_2/\F_0$, where the subscript denotes the order in $q_w^{(3)}$, and $h_v = c/12$ for the twist field in the conventions of \cite{Hegde:2015dqh}.} the result for general $N$ is
\es{}{&\F_{\vac,N}(z) \approx  \F_{\vac}(z) \Bigg[1+{N^2-4\over 5}\frac{ 6h_v }{\a^6(z^{\a}-1)^4} \Big(6 \alpha ^2 \left(z^{2 \alpha }+1\right) z^{\alpha } \log ^2 z\\& +\alpha  \left(z^{4 \alpha }-14 z^{3 \alpha }+14 z^{\alpha }-1\right) \log z-\left(z^{\alpha }-1\right)^2 \left(5 z^{2 \alpha }-22 z^{\alpha }+5\right)\Big)(q_w^{(3)})^2+\O((q_w^{(4)})^2)\Bigg]_{z\rar 1-z}}
Comparing to \eqr{e429}, the only $N$-dependence is in a coefficient. We conclude that when $W$ carries spin-3 charge in a $W_N$ CFT, the OTO correlator evolves in time with $\l_L^{(3)}=4\pi/\b$, independent of $N$. It follows from the general analysis in \cite{Hegde:2015dqh} that the same notion of $N$-independence is true under a perturbation of any spin.

\section{$\winf$ vacuum block for $V={\bf f}$, $W=$ {\bf asym}$_2$}\label{appd}
Here we compute $\finf(z|\l)$ for $V$ and $W$ in the representations indicated. This is a supplement to Section \ref{s51}. 

The $V$ charges may be read off from \eqr{qqn}. The $W$ charges were derived in \cite{Hegde:2015dqh}; with the normalization\footnote{That is, 
\e{}{N_s = {1\over (1-\l^2)\Gamma(1-\l)\Gamma(1+\l)}{\Gamma^2(s)\Gamma(s-\l)\Gamma(s+\l)\over \Gamma(2s-1)}}} such that $N_2=1/2$, one has
\e{}{q_s({\bf asym}_2) = {1\over \G{2+\l}}{(s^2-s+2(1+\l))\G{s}^2\G{s+\l}\over \G{2s-1}}}
We now want to compute the $W_{\infty}[\l]$ vacuum block at $\O(1/c)$ for these charges:
\es{}{\finf(z|\l) &= \sum_{s=2}^{\infty}{q_v^{(s)}({\bf f}) q_w^{(s)}({\bf asym}_2)\over N_s} z^s \hyp(s,s,2s,z)}
where we think of this as running in the s-channel of $\la VVWW\ra$. As in the $V=W=$ {\bf f} case, we evaluate this by using \eqr{hypint} and swapping the sum and integral. We evaluate this in closed form for certain rational $\l$. With inspiration from the $V=W=$ {\bf f} case, we infer the following structure:
\es{fasym}{\cF_{\vac,\infty}^{(1)}(z|\l) &= 2(1-\l^2) \left(z\,{}_3F_2(3,1,1;2,1-\l;z)+\log(1-z)\right)}
Presumably this can be proven using generalized hypergeometric identities. For reference, we give the results at $\l=0,1/2, 1$:
\es{}{\cF_{{\rm vac},\infty}^{(1)}(z|0) &={1\over (1-z)^2}-1+2\log(1-z)\\
\cF_{{\rm vac},\infty}^{(1)}\left(z\Big|\half\right) &={3 (5-2 z) z^{3/2} \arcsin\sqrt{z}\over 8(1-z)^{5/2}}+{z(4-z)\over8 (1-z)^2}+\frac{3}{2} \log (1-z)\\
\cF_{\vac,\infty}^{(1)}(z|1) &= {2z^2(3-z)\over (1-z)^3}}

To take the Regge limit, we need to know the monodromy of ${}_3F_2$ around $z=1$. Moving around the branch point yields a linear combination of the three linearly-independent solutions of the hypergeometric equation near $z=0$. For parameters $\lbrace a_1,a_2,a_3; b_1,b_2\rbrace$, the solutions are
\es{}{&{}_3F_2(a_1,a_2,a_3; b_1,b_2; z)~,\\z^{1-b_1}\,&{}_3F_2(a_1+1-b_1,a_2+1-b_1,a_3+1-b_1; 2-b_1, b_2+1-b_1; z)~,\\z^{1-b_2}\,&{}_3F_2(a_1+1-b_2,a_2+1-b_2,a_3+1-b_2; 2-b_2, b_1+1-b_2; z)}
In our case, the parameters are $\lbrace 3,1,1;2,1-\l\rbrace$, and the three solutions are
\es{}{&{}_3F_2(3,1,1;2,1; z)~,\\z^{-1}\,&{}_3F_2(2,0,0; 0, -\l; z)=0~,\\
z^{\l}\,&{}_2F_1(3+\l, 1+\l; 2+\l; z)}
Expanding near $z=0$, it follows that \eqr{fasym} has the same scaling as in the $V=W=$ {\bf f} case: the $-2\pi i$ from the log dominates. This yields $\l_L=0$. Note that at $\l=1$, the solution has trivial monodromy again. One can confirm the result at $\l=1/2$ using the monodromy $\arcsin\sqrt{z} \rar -\arcsin\sqrt{z}+\pi$ around $z=1$.

\section{$W^{\rm cl}_\i[\l>2]$ is complex}\label{appe}
Here we derive \eqr{3s1}. The $W_{\i}^{\rm cl}[\l]$ structure constants are known in the so-called primary basis -- i.e. the diagonal basis \eqr{jnorm} of primary currents $J_s$ -- in closed, but very complicated, form \cite{Campoleoni:2011hg, Prochazka:2014gqa}. Rather than analyze the expression for $C_{3s}^{s+1}$ directly, we will deduce its form from general principles.

We begin with the relation between $W_{\infty}[\l]$ and $W_N$: namely, 
\e{}{W_{\infty}[\pm N]/\chi_N \cong W_N}
The quotient algebra sets all currents of spins $s>N$ to zero. More importantly for us, it also means that when we set $\l=\pm N$, the OPE of two currents of spins $s_1,s_2\leq N$ must reduce to the OPE of $W_N$. (Indeed, this is essentially how $\g^2$ was derived in \cite{Gaberdiel:2012ku}.) This implies that any structure constant $C_{s_1 s_2}^{s_3}$ for which ${\rm max}(s_1,s_2,s_3) = s_3$ must vanish at integer values of $\l$ in the range ${\rm max}(s_1,s_2)\leq \l \leq s_3-1$. The algebra is invariant under $\l \rar -\l$, so
\e{}{(C_{s_1 s_2}^{s_3})^2 \propto  (\l^2 - ({\rm max}(s_1,s_2))^2) \ldots  (\l^2-(s_3-1)^2)}
To determine whether/how this becomes negative, we need to know what the denominator looks like. 

We now focus on $C_{3s}^{s+1}$. Let us write this as
\e{}{(C_{3s}^{s+1})^2 = (\l^2-s^2) \times{f_s(\l)}}
We want to determine $f_s(\l)$ using the following facts about the $W_{\i}^{\rm cl}[\l]$ algebra:
\vs
{\bf i)} In the normalization of \cite{Gaberdiel:2012ku}, all $(C_{s_1s_2}^{s_3})^2$ are rational functions of $\l^2$. 
\vs
{\bf ii)} For $\l\in\R$, the only degeneration points of $W_{\i}^{\rm cl}[\l]$ are at $\l\in\Z$.
\vs
{\bf iii)} The denominator of $(C_{3s}^{s+1})^2$ includes a $\l^2-4$ factor. This reflects the fact that at $\l=\pm 2$, all of the generators $J_3,J_s,J_{s+1}$ are in the ideal. This is clear from the analysis of \cite{Gaberdiel:2012ku}.
\vs

The above properties imply that the zero at $\l=\pm s$, and the pole at $\l=\pm 2$, are the {\it only} real zeros and poles, respectively, of $(C_{3s}^{s+1})^2$. Putting these facts together, we can write
\e{cs3}{(C_{3s}^{s+1})^2 = {\l^2-s^2\over \l^2-4}\times {f_s(\l^2)}}
where $f_s(\l^2)$ is a rational function which has no zeroes or poles for $\l\in\mathbb{R}$. This further implies its sign-definiteness for $\l\in\R$. 

To complete the proof, we need to show that $f_s(\l^2)>0$ for all $\l\in\R$. Since $f_s(\l^2)$ is sign-definite for $\l\in\R$, it suffices to evaluate its sign for a single real value of $\l$. At $\l=1$, we have the isomorphism $W_{\i}^{\rm cl}[1]\cong W_{\i}^{\rm PRS}$, and the latter has real structure constants \cite{Gaberdiel:2012ku}. Since ${\l^2-s^2\over \l^2-4}$ is positive at $\l=1$, this implies that
\e{}{f_s(\l^2)>0}
for all $s$ and $\l\in\R$. Actually, $f_3(\l^2)$ and $f_4(\l^2)$ are constant, which strongly suggests that $f_s(\l^2)$ is constant for all $s$. 

In any case, having established positivity of $f_s(\l^2)$, it directly follows from \eqr{cs3} that
\e{}{(C_{3s}^{s+1})^2 <0~\text{when}~2<\l<s}
\vs

A final comment: $W_{\i}^{\rm cl}[\l]$ inherits a triality symmetry from the quantum $W_{\infty}[\l]$ algebra \cite{Gaberdiel:2012ku}, which under which algebras with three different values of $\l$ are isomorphic. One might wonder whether this plays a hidden role in invalidating our conclusions: namely, whether for $\l>2$, either of the triality images of $\l$ is less than 2. But they aren't. If we denote $\mathbb{T}(\l)$ as the triality orbit of the quantum $W_{\i}[\l]$ algebra, then it follows from \cite{Gaberdiel:2012ku}, equation 2.20, that
\e{}{\lim_{c\rar\i}\mathbb{T}(\l) = \lbrace \l, -\l,-c\rbrace}
This is related to property {\bf ii)}.

\end{appendix}

\bibliographystyle{ssg}
\bibliography{refs}

\begingroup\raggedright\begin{thebibliography}{100}

\bibitem{Shenker:2013pqa}
S.~H. Shenker and D.~Stanford, ``{Black holes and the butterfly effect},'' {\em
  JHEP} {\bf 03} (2014) 067, \href{http://xxx.lanl.gov/abs/1306.0622}{{\tt
  1306.0622}}.

\bibitem{Maldacena:2015waa}
J.~Maldacena, S.~H. Shenker, and D.~Stanford, ``{A bound on chaos},''
  \href{http://xxx.lanl.gov/abs/1503.01409}{{\tt 1503.01409}}.

\bibitem{Shenker:2013yza}
S.~H. Shenker and D.~Stanford, ``{Multiple Shocks},'' {\em JHEP} {\bf 12}
  (2014) 046, \href{http://xxx.lanl.gov/abs/1312.3296}{{\tt 1312.3296}}.

\bibitem{Leichenauer:2014nxa}
S.~Leichenauer, ``{Disrupting Entanglement of Black Holes},'' {\em Phys. Rev.}
  {\bf D90} (2014), no.~4 046009, \href{http://xxx.lanl.gov/abs/1405.7365}{{\tt
  1405.7365}}.

\bibitem{kitaev}
A.~Kitaev, “A simple model of quantum holography,” KITP string seminar and
Entanglement 2015 program (Feb. 12, April 7, and May 27, 2015).

\bibitem{Roberts:2014isa}
D.~A. Roberts, D.~Stanford, and L.~Susskind, ``{Localized shocks},'' {\em JHEP}
  {\bf 03} (2015) 051, \href{http://xxx.lanl.gov/abs/1409.8180}{{\tt
  1409.8180}}.

\bibitem{Roberts:2014ifa}
D.~A. Roberts and D.~Stanford, ``{Two-dimensional conformal field theory and
  the butterfly effect},'' {\em Phys. Rev. Lett.} {\bf 115} (2015), no.~13
  131603, \href{http://xxx.lanl.gov/abs/1412.5123}{{\tt 1412.5123}}.

\bibitem{Jackson:2014nla}
S.~Jackson, L.~McGough, and H.~Verlinde, ``{Conformal Bootstrap, Universality
  and Gravitational Scattering},'' {\em Nucl. Phys.} {\bf B901} (2015)
  382--429, \href{http://xxx.lanl.gov/abs/1412.5205}{{\tt 1412.5205}}.

\bibitem{Shenker:2014cwa}
S.~H. Shenker and D.~Stanford, ``{Stringy effects in scrambling},'' {\em JHEP}
  {\bf 05} (2015) 132, \href{http://xxx.lanl.gov/abs/1412.6087}{{\tt
  1412.6087}}.

\bibitem{Polchinski:2015cea}
J.~Polchinski, ``{Chaos in the black hole S-matrix},''
  \href{http://xxx.lanl.gov/abs/1505.08108}{{\tt 1505.08108}}.

\bibitem{Caputa:2015waa}
P.~Caputa, J.~Simón, A.~Štikonas, T.~Takayanagi, and K.~Watanabe,
  ``{Scrambling time from local perturbations of the eternal BTZ black hole},''
  {\em JHEP} {\bf 08} (2015) 011,
  \href{http://xxx.lanl.gov/abs/1503.08161}{{\tt 1503.08161}}.

\bibitem{Hosur:2015ylk}
P.~Hosur, X.-L. Qi, D.~A. Roberts, and B.~Yoshida, ``{Chaos in quantum
  channels},'' {\em JHEP} {\bf 02} (2016) 004,
  \href{http://xxx.lanl.gov/abs/1511.04021}{{\tt 1511.04021}}.

\bibitem{Stanford:2015owe}
D.~Stanford, ``{Many-body chaos at weak coupling},''
  \href{http://xxx.lanl.gov/abs/1512.07687}{{\tt 1512.07687}}.

\bibitem{Gur-Ari:2015rcq}
G.~Gur-Ari, M.~Hanada, and S.~H. Shenker, ``{Chaos in Classical D0-Brane
  Mechanics},'' {\em JHEP} {\bf 02} (2016) 091,
  \href{http://xxx.lanl.gov/abs/1512.00019}{{\tt 1512.00019}}.

\bibitem{Berkowitz:2016znt}
E.~Berkowitz, M.~Hanada, and J.~Maltz, ``{Chaos in Matrix Models and Black Hole
  Evaporation},'' \href{http://xxx.lanl.gov/abs/1602.01473}{{\tt 1602.01473}}.

\bibitem{Polchinski:2016xgd}
J.~Polchinski and V.~Rosenhaus, ``{The Spectrum in the Sachdev-Ye-Kitaev
  Model},'' \href{http://xxx.lanl.gov/abs/1601.06768}{{\tt 1601.06768}}.

\bibitem{Fitzpatrick:2016thx}
A.~L. Fitzpatrick and J.~Kaplan, ``{A Quantum Correction To Chaos},''
  \href{http://xxx.lanl.gov/abs/1601.06164}{{\tt 1601.06164}}.

\bibitem{Michel:2016kwn}
B.~Michel, J.~Polchinski, V.~Rosenhaus, and S.~J. Suh, ``{Four-point function
  in the IOP matrix model},'' \href{http://xxx.lanl.gov/abs/1602.06422}{{\tt
  1602.06422}}.

\bibitem{larkin}
A.~Larkin and Y.~N. Ovchinnikov, ``{Quasiclassical method in the theory of
  superconductivity},'' {\em JETP} {\bf 28} (1969) 1200.

\bibitem{Sekino:2008he}
Y.~Sekino and L.~Susskind, ``{Fast Scramblers},'' {\em JHEP} {\bf 10} (2008)
  065, \href{http://xxx.lanl.gov/abs/0808.2096}{{\tt 0808.2096}}.

\bibitem{Camanho:2014apa}
X.~O. Camanho, J.~D. Edelstein, J.~Maldacena, and A.~Zhiboedov, ``{Causality
  Constraints on Corrections to the Graviton Three-Point Coupling},'' {\em
  JHEP} {\bf 02} (2016) 020, \href{http://xxx.lanl.gov/abs/1407.5597}{{\tt
  1407.5597}}.

\bibitem{Bekaert:2010hw}
X.~Bekaert, N.~Boulanger, and P.~Sundell, ``{How higher-spin gravity surpasses
  the spin two barrier: no-go theorems versus yes-go examples},'' {\em Rev.
  Mod. Phys.} {\bf 84} (2012) 987--1009,
  \href{http://xxx.lanl.gov/abs/1007.0435}{{\tt 1007.0435}}.

\bibitem{Maldacena:2011jn}
J.~Maldacena and A.~Zhiboedov, ``{Constraining Conformal Field Theories with A
  Higher Spin Symmetry},'' {\em J. Phys.} {\bf A46} (2013) 214011,
  \href{http://xxx.lanl.gov/abs/1112.1016}{{\tt 1112.1016}}.

\bibitem{Vasiliev:1990en}
M.~A. Vasiliev, ``{Consistent equation for interacting gauge fields of all
  spins in (3+1)-dimensions},'' {\em Phys. Lett.} {\bf B243} (1990) 378--382.

\bibitem{vasstar}
M.~Vasiliev, ``{Higher spin gauge theories: Star product and AdS space},''
  \href{http://xxx.lanl.gov/abs/hep-th/9910096}{{\tt hep-th/9910096}}.

\bibitem{Prokushkin:1998bq}
S.~F. Prokushkin and M.~A. Vasiliev, ``{Higher spin gauge interactions for
  massive matter fields in 3-D AdS space-time},'' {\em Nucl. Phys.} {\bf B545}
  (1999) 385, \href{http://xxx.lanl.gov/abs/hep-th/9806236}{{\tt
  hep-th/9806236}}.

\bibitem{Giombi:2012ms}
S.~Giombi and X.~Yin, ``{The Higher Spin/Vector Model Duality},'' {\em J.
  Phys.} {\bf A46} (2013) 214003, \href{http://xxx.lanl.gov/abs/1208.4036}{{\tt
  1208.4036}}.

\bibitem{Didenko:2014dwa}
V.~E. Didenko and E.~D. Skvortsov, ``{Elements of Vasiliev theory},''
  \href{http://xxx.lanl.gov/abs/1401.2975}{{\tt 1401.2975}}.

\bibitem{Giombi:2011kc}
S.~Giombi, S.~Minwalla, S.~Prakash, S.~P. Trivedi, S.~R. Wadia, and X.~Yin,
  ``{Chern-Simons Theory with Vector Fermion Matter},'' {\em Eur. Phys. J.}
  {\bf C72} (2012) 2112, \href{http://xxx.lanl.gov/abs/1110.4386}{{\tt
  1110.4386}}.

\bibitem{Aharony:2011jz}
O.~Aharony, G.~Gur-Ari, and R.~Yacoby, ``{d=3 Bosonic Vector Models Coupled to
  Chern-Simons Gauge Theories},'' {\em JHEP} {\bf 03} (2012) 037,
  \href{http://xxx.lanl.gov/abs/1110.4382}{{\tt 1110.4382}}.

\bibitem{Sundborg:2000wp}
B.~Sundborg, ``{Stringy gravity, interacting tensionless strings and massless
  higher spins},'' {\em Nucl. Phys. Proc. Suppl.} {\bf 102} (2001) 113--119,
  \href{http://xxx.lanl.gov/abs/hep-th/0103247}{{\tt hep-th/0103247}}.
  [,113(2000)].

\bibitem{Gaberdiel:2014cha}
M.~R. Gaberdiel and R.~Gopakumar, ``{Higher Spins \& Strings},'' {\em JHEP}
  {\bf 11} (2014) 044, \href{http://xxx.lanl.gov/abs/1406.6103}{{\tt
  1406.6103}}.

\bibitem{Gaberdiel:2015mra}
M.~R. Gaberdiel and R.~Gopakumar, ``{Stringy Symmetries and the Higher Spin
  Square},'' {\em J. Phys.} {\bf A48} (2015), no.~18 185402,
  \href{http://xxx.lanl.gov/abs/1501.07236}{{\tt 1501.07236}}.

\bibitem{Gaberdiel:2015wpo}
M.~R. Gaberdiel and R.~Gopakumar, ``{String Theory as a Higher Spin Theory},''
  \href{http://xxx.lanl.gov/abs/1512.07237}{{\tt 1512.07237}}.

\bibitem{Achucarro:1987vz}
A.~Achucarro and P.~K. Townsend, ``{A Chern-Simons Action for Three-Dimensional
  anti-De Sitter Supergravity Theories},'' {\em Phys. Lett.} {\bf B180} (1986)
  89.

\bibitem{Witten:1988hc}
E.~Witten, ``{(2+1)-Dimensional Gravity as an Exactly Soluble System},'' {\em
  Nucl. Phys.} {\bf B311} (1988) 46.

\bibitem{Campoleoni:2010zq}
A.~Campoleoni, S.~Fredenhagen, S.~Pfenninger, and S.~Theisen, ``{Asymptotic
  symmetries of three-dimensional gravity coupled to higher-spin fields},''
  {\em JHEP} {\bf 11} (2010) 007, \href{http://xxx.lanl.gov/abs/1008.4744}{{\tt
  1008.4744}}.

\bibitem{Pope:1989sr}
C.~N. Pope, L.~J. Romans, and X.~Shen, ``{$W$(infinity) and the Racah-wigner
  Algebra},'' {\em Nucl. Phys.} {\bf B339} (1990) 191--221.

\bibitem{Henneaux:2010xg}
M.~Henneaux and S.-J. Rey, ``{Nonlinear $W_{infinity}$ as Asymptotic Symmetry
  of Three-Dimensional Higher Spin Anti-de Sitter Gravity},'' {\em JHEP} {\bf
  12} (2010) 007, \href{http://xxx.lanl.gov/abs/1008.4579}{{\tt 1008.4579}}.

\bibitem{Gaberdiel:2011wb}
M.~R. Gaberdiel and T.~Hartman, ``{Symmetries of Holographic Minimal Models},''
  {\em JHEP} {\bf 05} (2011) 031, \href{http://xxx.lanl.gov/abs/1101.2910}{{\tt
  1101.2910}}.

\bibitem{Gutperle:2011kf}
M.~Gutperle and P.~Kraus, ``{Higher Spin Black Holes},'' {\em JHEP} {\bf 05}
  (2011) 022, \href{http://xxx.lanl.gov/abs/1103.4304}{{\tt 1103.4304}}.

\bibitem{Ammon:2012wc}
M.~Ammon, M.~Gutperle, P.~Kraus, and E.~Perlmutter, ``{Black holes in three
  dimensional higher spin gravity: A review},'' {\em J. Phys.} {\bf A46} (2013)
  214001, \href{http://xxx.lanl.gov/abs/1208.5182}{{\tt 1208.5182}}.

\bibitem{Bunster:2014mua}
C.~Bunster, M.~Henneaux, A.~Perez, D.~Tempo, and R.~Troncoso, ``{Generalized
  Black Holes in Three-dimensional Spacetime},'' {\em JHEP} {\bf 05} (2014)
  031, \href{http://xxx.lanl.gov/abs/1404.3305}{{\tt 1404.3305}}.

\bibitem{deBoer:2014fra}
J.~de~Boer and J.~I. Jottar, ``{Boundary Conditions and Partition Functions in
  Higher Spin AdS$_3$/CFT$_2$},'' \href{http://xxx.lanl.gov/abs/1407.3844}{{\tt
  1407.3844}}.

\bibitem{Ammon:2013hba}
M.~Ammon, A.~Castro, and N.~Iqbal, ``{Wilson Lines and Entanglement Entropy in
  Higher Spin Gravity},'' {\em JHEP} {\bf 10} (2013) 110,
  \href{http://xxx.lanl.gov/abs/1306.4338}{{\tt 1306.4338}}.

\bibitem{deBoer:2013vca}
J.~de~Boer and J.~I. Jottar, ``{Entanglement Entropy and Higher Spin Holography
  in AdS$_3$},'' {\em JHEP} {\bf 04} (2014) 089,
  \href{http://xxx.lanl.gov/abs/1306.4347}{{\tt 1306.4347}}.

\bibitem{Chen:2013dxa}
B.~Chen, J.~Long, and J.-j. Zhang, ``{Holographic Rényi entropy for CFT with W
  symmetry},'' {\em JHEP} {\bf 04} (2014) 041,
  \href{http://xxx.lanl.gov/abs/1312.5510}{{\tt 1312.5510}}.

\bibitem{Perlmutter:2013paa}
E.~Perlmutter, ``{Comments on Renyi entropy in AdS$_3$/CFT$_2$},'' {\em JHEP}
  {\bf 05} (2014) 052, \href{http://xxx.lanl.gov/abs/1312.5740}{{\tt
  1312.5740}}.

\bibitem{Datta:2014ska}
S.~Datta, J.~R. David, M.~Ferlaino, and S.~P. Kumar, ``{Higher spin
  entanglement entropy from CFT},'' {\em JHEP} {\bf 06} (2014) 096,
  \href{http://xxx.lanl.gov/abs/1402.0007}{{\tt 1402.0007}}.

\bibitem{Long:2014oxa}
J.~Long, ``{Higher Spin Entanglement Entropy},'' {\em JHEP} {\bf 12} (2014)
  055, \href{http://xxx.lanl.gov/abs/1408.1298}{{\tt 1408.1298}}.

\bibitem{deBoer:2014sna}
J.~de~Boer, A.~Castro, E.~Hijano, J.~I. Jottar, and P.~Kraus, ``{Higher spin
  entanglement and $ {\mathcal{W}}_{\mathrm{N}} $ conformal blocks},'' {\em
  JHEP} {\bf 07} (2015) 168, \href{http://xxx.lanl.gov/abs/1412.7520}{{\tt
  1412.7520}}.

\bibitem{Hegde:2015dqh}
A.~Hegde, P.~Kraus, and E.~Perlmutter, ``{General Results for Higher Spin
  Wilson Lines and Entanglement in Vasiliev Theory},'' {\em JHEP} {\bf 01}
  (2016) 176, \href{http://xxx.lanl.gov/abs/1511.05555}{{\tt 1511.05555}}.

\bibitem{Afshar:2013vka}
H.~Afshar, A.~Bagchi, R.~Fareghbal, D.~Grumiller, and J.~Rosseel, ``{Spin-3
  Gravity in Three-Dimensional Flat Space},'' {\em Phys. Rev. Lett.} {\bf 111}
  (2013), no.~12 121603, \href{http://xxx.lanl.gov/abs/1307.4768}{{\tt
  1307.4768}}.

\bibitem{Haehl:2014yla}
F.~M. Haehl and M.~Rangamani, ``{Permutation orbifolds and holography},'' {\em
  JHEP} {\bf 03} (2015) 163, \href{http://xxx.lanl.gov/abs/1412.2759}{{\tt
  1412.2759}}.

\bibitem{Belin:2014fna}
A.~Belin, C.~A. Keller, and A.~Maloney, ``{String Universality for Permutation
  Orbifolds},'' {\em Phys. Rev.} {\bf D91} (2015), no.~10 106005,
  \href{http://xxx.lanl.gov/abs/1412.7159}{{\tt 1412.7159}}.

\bibitem{Hartman:2015lfa}
T.~Hartman, S.~Jain, and S.~Kundu, ``{Causality Constraints in Conformal Field
  Theory},'' \href{http://xxx.lanl.gov/abs/1509.00014}{{\tt 1509.00014}}.

\bibitem{Hartman:2016dxc}
T.~Hartman, S.~Jain, and S.~Kundu, ``{A New Spin on Causality Constraints},''
  \href{http://xxx.lanl.gov/abs/1601.07904}{{\tt 1601.07904}}.

\bibitem{Costa:2012cb}
M.~S. Costa, V.~Goncalves, and J.~Penedones, ``{Conformal Regge theory},'' {\em
  JHEP} {\bf 12} (2012) 091, \href{http://xxx.lanl.gov/abs/1209.4355}{{\tt
  1209.4355}}.

\bibitem{Heemskerk:2009pn}
I.~Heemskerk, J.~Penedones, J.~Polchinski, and J.~Sully, ``{Holography from
  Conformal Field Theory},'' {\em JHEP} {\bf 10} (2009) 079,
  \href{http://xxx.lanl.gov/abs/0907.0151}{{\tt 0907.0151}}.

\bibitem{ElShowk:2011ag}
S.~El-Showk and K.~Papadodimas, ``{Emergent Spacetime and Holographic CFTs},''
  {\em JHEP} {\bf 10} (2012) 106, \href{http://xxx.lanl.gov/abs/1101.4163}{{\tt
  1101.4163}}.

\bibitem{Maldacena:2015iua}
J.~Maldacena, D.~Simmons-Duffin, and A.~Zhiboedov, ``{Looking for a bulk
  point},'' \href{http://xxx.lanl.gov/abs/1509.03612}{{\tt 1509.03612}}.

\bibitem{Hartman:2013mia}
T.~Hartman, ``{Entanglement Entropy at Large Central Charge},''
  \href{http://xxx.lanl.gov/abs/1303.6955}{{\tt 1303.6955}}.

\bibitem{Hartman:2014oaa}
T.~Hartman, C.~A. Keller, and B.~Stoica, ``{Universal Spectrum of 2d Conformal
  Field Theory in the Large c Limit},'' {\em JHEP} {\bf 09} (2014) 118,
  \href{http://xxx.lanl.gov/abs/1405.5137}{{\tt 1405.5137}}.

\bibitem{Asplund:2014coa}
C.~T. Asplund, A.~Bernamonti, F.~Galli, and T.~Hartman, ``{Holographic
  Entanglement Entropy from 2d CFT: Heavy States and Local Quenches},'' {\em
  JHEP} {\bf 02} (2015) 171, \href{http://xxx.lanl.gov/abs/1410.1392}{{\tt
  1410.1392}}.

\bibitem{Castro:2014mza}
A.~Castro and E.~Llabrés, ``{Unravelling Holographic Entanglement Entropy in
  Higher Spin Theories},'' {\em JHEP} {\bf 03} (2015) 124,
  \href{http://xxx.lanl.gov/abs/1410.2870}{{\tt 1410.2870}}.

\bibitem{Papallo:2015rna}
G.~Papallo and H.~S. Reall, ``{Graviton time delay and a speed limit for small
  black holes in Einstein-Gauss-Bonnet theory},'' {\em JHEP} {\bf 11} (2015)
  109, \href{http://xxx.lanl.gov/abs/1508.05303}{{\tt 1508.05303}}.

\bibitem{Gaberdiel:2011zw}
M.~R. Gaberdiel, R.~Gopakumar, T.~Hartman, and S.~Raju, ``{Partition Functions
  of Holographic Minimal Models},'' {\em JHEP} {\bf 08} (2011) 077,
  \href{http://xxx.lanl.gov/abs/1106.1897}{{\tt 1106.1897}}.

\bibitem{Gaberdiel:2012ku}
M.~R. Gaberdiel and R.~Gopakumar, ``{Triality in Minimal Model Holography},''
  {\em JHEP} {\bf 07} (2012) 127, \href{http://xxx.lanl.gov/abs/1205.2472}{{\tt
  1205.2472}}.

\bibitem{Gaberdiel:2013jca}
M.~R. Gaberdiel, K.~Jin, and E.~Perlmutter, ``{Probing higher spin black holes
  from CFT},'' {\em JHEP} {\bf 10} (2013) 045,
  \href{http://xxx.lanl.gov/abs/1307.2221}{{\tt 1307.2221}}.

\bibitem{Campoleoni:2013lma}
A.~Campoleoni, T.~Prochazka, and J.~Raeymaekers, ``{A note on conical solutions
  in 3D Vasiliev theory},'' {\em JHEP} {\bf 05} (2013) 052,
  \href{http://xxx.lanl.gov/abs/1303.0880}{{\tt 1303.0880}}.

\bibitem{Grumiller:2008qz}
D.~Grumiller and N.~Johansson, ``{Instability in cosmological topologically
  massive gravity at the chiral point},'' {\em JHEP} {\bf 07} (2008) 134,
  \href{http://xxx.lanl.gov/abs/0805.2610}{{\tt 0805.2610}}.

\bibitem{Anninos:2011ui}
D.~Anninos, T.~Hartman, and A.~Strominger, ``{Higher Spin Realization of the
  dS/CFT Correspondence},'' \href{http://xxx.lanl.gov/abs/1108.5735}{{\tt
  1108.5735}}.

\bibitem{Perlmutter:2012ds}
E.~Perlmutter, T.~Prochazka, and J.~Raeymaekers, ``{The semiclassical limit of
  $W_N$ CFTs and Vasiliev theory},'' {\em JHEP} {\bf 05} (2013) 007,
  \href{http://xxx.lanl.gov/abs/1210.8452}{{\tt 1210.8452}}.

\bibitem{Vafa:2014iua}
C.~Vafa, ``{Non-Unitary Holography},''
  \href{http://xxx.lanl.gov/abs/1409.1603}{{\tt 1409.1603}}.

\bibitem{Amati:1987wq}
D.~Amati, M.~Ciafaloni, and G.~Veneziano, ``{Superstring Collisions at
  Planckian Energies},'' {\em Phys. Lett.} {\bf B197} (1987) 81.

\bibitem{Amati:1987uf}
D.~Amati, M.~Ciafaloni, and G.~Veneziano, ``{Classical and Quantum Gravity
  Effects from Planckian Energy Superstring Collisions},'' {\em Int. J. Mod.
  Phys.} {\bf A3} (1988) 1615--1661.

\bibitem{Amati:1988tn}
D.~Amati, M.~Ciafaloni, and G.~Veneziano, ``{Can Space-Time Be Probed Below the
  String Size?},'' {\em Phys. Lett.} {\bf B216} (1989) 41.

\bibitem{Maldacena:2012sf}
J.~Maldacena and A.~Zhiboedov, ``{Constraining conformal field theories with a
  slightly broken higher spin symmetry},'' {\em Class. Quant. Grav.} {\bf 30}
  (2013) 104003, \href{http://xxx.lanl.gov/abs/1204.3882}{{\tt 1204.3882}}.

\bibitem{zamo2}
A.~Zamolodchikov, ``Conformal symmetry in two-dimensional space: Recursion
  representation of conformal block,'' {\em Theoretical and Mathematical
  Physics} {\bf 73} (1987), no.~1 1088--1093.

\bibitem{Cornalba:2006xm}
L.~Cornalba, M.~S. Costa, J.~Penedones, and R.~Schiappa, ``{Eikonal
  Approximation in AdS/CFT: Conformal Partial Waves and Finite N Four-Point
  Functions},'' {\em Nucl. Phys.} {\bf B767} (2007) 327--351,
  \href{http://xxx.lanl.gov/abs/hep-th/0611123}{{\tt hep-th/0611123}}.

\bibitem{Pappadopulo:2012jk}
D.~Pappadopulo, S.~Rychkov, J.~Espin, and R.~Rattazzi, ``{OPE Convergence in
  Conformal Field Theory},'' {\em Phys. Rev.} {\bf D86} (2012) 105043,
  \href{http://xxx.lanl.gov/abs/1208.6449}{{\tt 1208.6449}}.

\bibitem{Hijano:2015zsa}
E.~Hijano, P.~Kraus, E.~Perlmutter, and R.~Snively, ``{Witten Diagrams
  Revisited: The AdS Geometry of Conformal Blocks},'' {\em JHEP} {\bf 01}
  (2016) 146, \href{http://xxx.lanl.gov/abs/1508.00501}{{\tt 1508.00501}}.

\bibitem{Liu:1998th}
H.~Liu, ``{Scattering in anti-de Sitter space and operator product
  expansion},'' {\em Phys. Rev.} {\bf D60} (1999) 106005,
  \href{http://xxx.lanl.gov/abs/hep-th/9811152}{{\tt hep-th/9811152}}.

\bibitem{Fitzpatrick:2011dm}
A.~L. Fitzpatrick and J.~Kaplan, ``{Unitarity and the Holographic S-Matrix},''
  {\em JHEP} {\bf 10} (2012) 032, \href{http://xxx.lanl.gov/abs/1112.4845}{{\tt
  1112.4845}}.

\bibitem{Arutyunov:2002fh}
G.~Arutyunov, F.~A. Dolan, H.~Osborn, and E.~Sokatchev, ``{Correlation
  functions and massive Kaluza-Klein modes in the AdS / CFT correspondence},''
  {\em Nucl. Phys.} {\bf B665} (2003) 273--324,
  \href{http://xxx.lanl.gov/abs/hep-th/0212116}{{\tt hep-th/0212116}}.

\bibitem{Komargodski:2016gci}
Z.~Komargodski, M.~Kulaxizi, A.~Parnachev, and A.~Zhiboedov, ``{Conformal Field
  Theories and Deep Inelastic Scattering},''
  \href{http://xxx.lanl.gov/abs/1601.05453}{{\tt 1601.05453}}.

\bibitem{douglasdan}
D.~Roberts and D.~Stanford, {unpublished}.

\bibitem{Maldacena:2016hyu}
J.~Maldacena and D.~Stanford, ``{Comments on the Sachdev-Ye-Kitaev model},''
  \href{http://xxx.lanl.gov/abs/1604.07818}{{\tt 1604.07818}}.

\bibitem{Abbott:2016blz}
{\bf Virgo, LIGO Scientific} Collaboration, B.~P. Abbott {\em et.~al.},
  ``{Observation of Gravitational Waves from a Binary Black Hole Merger},''
  {\em Phys. Rev. Lett.} {\bf 116} (2016), no.~6 061102,
  \href{http://xxx.lanl.gov/abs/1602.03837}{{\tt 1602.03837}}.

\bibitem{Maloney:2007ud}
A.~Maloney and E.~Witten, ``{Quantum Gravity Partition Functions in Three
  Dimensions},'' {\em JHEP} {\bf 02} (2010) 029,
  \href{http://xxx.lanl.gov/abs/0712.0155}{{\tt 0712.0155}}.

\bibitem{Fitzpatrick:2014vua}
A.~L. Fitzpatrick, J.~Kaplan, and M.~T. Walters, ``{Universality of
  Long-Distance AdS Physics from the CFT Bootstrap},'' {\em JHEP} {\bf 08}
  (2014) 145, \href{http://xxx.lanl.gov/abs/1403.6829}{{\tt 1403.6829}}.

\bibitem{Fitzpatrick:2015zha}
A.~L. Fitzpatrick, J.~Kaplan, and M.~T. Walters, ``{Virasoro Conformal Blocks
  and Thermality from Classical Background Fields},'' {\em JHEP} {\bf 11}
  (2015) 200, \href{http://xxx.lanl.gov/abs/1501.05315}{{\tt 1501.05315}}.

\bibitem{Campoleoni:2011hg}
A.~Campoleoni, S.~Fredenhagen, and S.~Pfenninger, ``{Asymptotic W-symmetries in
  three-dimensional higher-spin gauge theories},'' {\em JHEP} {\bf 09} (2011)
  113, \href{http://xxx.lanl.gov/abs/1107.0290}{{\tt 1107.0290}}.

\bibitem{Ammon:2011nk}
M.~Ammon, M.~Gutperle, P.~Kraus, and E.~Perlmutter, ``{Spacetime Geometry in
  Higher Spin Gravity},'' {\em JHEP} {\bf 10} (2011) 053,
  \href{http://xxx.lanl.gov/abs/1106.4788}{{\tt 1106.4788}}.

\bibitem{Castro:2016ehj}
A.~Castro, N.~Iqbal, and E.~Llabrés, ``{Eternal Higher Spin Black Holes: a
  Thermofield Interpretation},'' \href{http://xxx.lanl.gov/abs/1602.09057}{{\tt
  1602.09057}}.

\bibitem{Edelstein:2016nml}
J.~D. Edelstein, C.~Gomez, E.~Kilicarslan, M.~Leoni, and B.~Tekin, ``{Causality
  in 3D Massive Gravity Theories},''
  \href{http://xxx.lanl.gov/abs/1602.03376}{{\tt 1602.03376}}.

\bibitem{Costa:2014kfa}
M.~S. Costa, V.~Gonçalves, and J.~Penedones, ``{Spinning AdS Propagators},''
  {\em JHEP} {\bf 09} (2014) 064, \href{http://xxx.lanl.gov/abs/1404.5625}{{\tt
  1404.5625}}.

\bibitem{Honda:2015mel}
M.~Honda, N.~Iizuka, A.~Tanaka, and S.~Terashima, ``{Exact Path Integral for 3D
  Higher Spin Gravity},'' \href{http://xxx.lanl.gov/abs/1511.07546}{{\tt
  1511.07546}}.

\bibitem{Ammon:2011ua}
M.~Ammon, P.~Kraus, and E.~Perlmutter, ``{Scalar fields and three-point
  functions in D=3 higher spin gravity},'' {\em JHEP} {\bf 07} (2012) 113,
  \href{http://xxx.lanl.gov/abs/1111.3926}{{\tt 1111.3926}}.

\bibitem{opac-b1078126}
A.~P. Prudnikov, I.~A. Brychkov, O.~I. Marichev, and G.~G. Gould, {\em
  Integrals and series. Volume 3. , More special functions}.
\newblock Gordon and Breach science publ, Amsterdam, Paris, New York, 1990.

\bibitem{Kraus:2011ds}
P.~Kraus and E.~Perlmutter, ``{Partition functions of higher spin black holes
  and their CFT duals},'' {\em JHEP} {\bf 11} (2011) 061,
  \href{http://xxx.lanl.gov/abs/1108.2567}{{\tt 1108.2567}}.

\bibitem{Pope:1989ew}
C.~N. Pope, L.~J. Romans, and X.~Shen, ``{The Complete Structure of
  W(Infinity)},'' {\em Phys. Lett.} {\bf B236} (1990) 173.

\bibitem{Dolan:2003hv}
F.~A. Dolan and H.~Osborn, ``{Conformal partial waves and the operator product
  expansion},'' {\em Nucl. Phys.} {\bf B678} (2004) 491--507,
  \href{http://xxx.lanl.gov/abs/hep-th/0309180}{{\tt hep-th/0309180}}.

\bibitem{douglas}
D.~Stanford, {private communication}.

\bibitem{Gaberdiel:2010pz}
M.~R. Gaberdiel and R.~Gopakumar, ``{An AdS$_3$ Dual for Minimal Model CFTs},''
  {\em Phys. Rev.} {\bf D83} (2011) 066007,
  \href{http://xxx.lanl.gov/abs/1011.2986}{{\tt 1011.2986}}.

\bibitem{Gaberdiel:2012uj}
M.~R. Gaberdiel and R.~Gopakumar, ``{Minimal Model Holography},'' {\em J.
  Phys.} {\bf A46} (2013) 214002, \href{http://xxx.lanl.gov/abs/1207.6697}{{\tt
  1207.6697}}.

\bibitem{Giombi:2013fka}
S.~Giombi and I.~R. Klebanov, ``{One Loop Tests of Higher Spin AdS/CFT},'' {\em
  JHEP} {\bf 12} (2013) 068, \href{http://xxx.lanl.gov/abs/1308.2337}{{\tt
  1308.2337}}.

\bibitem{Papadodimas:2011pf}
K.~Papadodimas and S.~Raju, ``{Correlation Functions in Holographic Minimal
  Models},'' {\em Nucl. Phys.} {\bf B856} (2012) 607--646,
  \href{http://xxx.lanl.gov/abs/1108.3077}{{\tt 1108.3077}}.

\bibitem{Chang:2011vka}
C.-M. Chang and X.~Yin, ``{Correlators in $W_N$ Minimal Model Revisited},''
  {\em JHEP} {\bf 10} (2012) 050, \href{http://xxx.lanl.gov/abs/1112.5459}{{\tt
  1112.5459}}.

\bibitem{Hijano:2013fja}
E.~Hijano, P.~Kraus, and E.~Perlmutter, ``{Matching four-point functions in
  higher spin AdS$_3$/CFT$_2$},'' {\em JHEP} {\bf 05} (2013) 163,
  \href{http://xxx.lanl.gov/abs/1302.6113}{{\tt 1302.6113}}.

\bibitem{Beem:2014kka}
C.~Beem, L.~Rastelli, and B.~C. van Rees, ``{$ \mathcal{W} $ symmetry in six
  dimensions},'' {\em JHEP} {\bf 05} (2015) 017,
  \href{http://xxx.lanl.gov/abs/1404.1079}{{\tt 1404.1079}}.

\bibitem{Bouwknegt:1988sv}
P.~Bouwknegt, ``{EXTENDED CONFORMAL ALGEBRAS},'' {\em Phys. Lett.} {\bf B207}
  (1988) 295.

\bibitem{Headrick:2015gba}
M.~Headrick, A.~Maloney, E.~Perlmutter, and I.~G. Zadeh, ``{Rényi entropies,
  the analytic bootstrap, and 3D quantum gravity at higher genus},'' {\em JHEP}
  {\bf 07} (2015) 059, \href{http://xxx.lanl.gov/abs/1503.07111}{{\tt
  1503.07111}}.

\bibitem{Witten:2007kt}
E.~Witten, ``{Three-Dimensional Gravity Revisited},''
  \href{http://xxx.lanl.gov/abs/0706.3359}{{\tt 0706.3359}}.

\bibitem{Asplund:2015eha}
C.~T. Asplund, A.~Bernamonti, F.~Galli, and T.~Hartman, ``{Entanglement
  Scrambling in 2d Conformal Field Theory},'' {\em JHEP} {\bf 09} (2015) 110,
  \href{http://xxx.lanl.gov/abs/1506.03772}{{\tt 1506.03772}}.

\bibitem{Lunin:2000yv}
O.~Lunin and S.~D. Mathur, ``{Correlation functions for M**N / S(N)
  orbifolds},'' {\em Commun. Math. Phys.} {\bf 219} (2001) 399--442,
  \href{http://xxx.lanl.gov/abs/hep-th/0006196}{{\tt hep-th/0006196}}.

\bibitem{Pakman:2009zz}
A.~Pakman, L.~Rastelli, and S.~S. Razamat, ``{Diagrams for Symmetric Product
  Orbifolds},'' {\em JHEP} {\bf 10} (2009) 034,
  \href{http://xxx.lanl.gov/abs/0905.3448}{{\tt 0905.3448}}.

\bibitem{Burrington:2012yn}
B.~A. Burrington, A.~W. Peet, and I.~G. Zadeh, ``{Twist-nontwist correlators in
  $M^N/S_N$ orbifold CFTs},'' {\em Phys. Rev.} {\bf D87} (2013), no.~10 106008,
  \href{http://xxx.lanl.gov/abs/1211.6689}{{\tt 1211.6689}}.

\bibitem{Leonhardt:2003du}
T.~Leonhardt and W.~Ruhl, ``{The Minimal conformal O(N) vector sigma model at d
  = 3},'' {\em J. Phys.} {\bf A37} (2004) 1403--1413,
  \href{http://xxx.lanl.gov/abs/hep-th/0308111}{{\tt hep-th/0308111}}.

\bibitem{Gu:2016hoy}
Y.~Gu and X.-L. Qi, ``{Fractional Statistics and the Butterfly Effect},''
  \href{http://xxx.lanl.gov/abs/1602.06543}{{\tt 1602.06543}}.

\bibitem{Caputa:2016tgt}
P.~Caputa, T.~Numasawa, and A.~Veliz-Osorio, ``{Scrambling without chaos in
  RCFT},'' \href{http://xxx.lanl.gov/abs/1602.06542}{{\tt 1602.06542}}.

\bibitem{Gaberdiel:2015uca}
M.~R. Gaberdiel, C.~Peng, and I.~G. Zadeh, ``{Higgsing the stringy higher spin
  symmetry},'' {\em JHEP} {\bf 10} (2015) 101,
  \href{http://xxx.lanl.gov/abs/1506.02045}{{\tt 1506.02045}}.

\bibitem{Kessel:2015kna}
P.~Kessel, G.~Lucena~Gómez, E.~Skvortsov, and M.~Taronna, ``{Higher Spins and
  Matter Interacting in Dimension Three},'' {\em JHEP} {\bf 11} (2015) 104,
  \href{http://xxx.lanl.gov/abs/1505.05887}{{\tt 1505.05887}}.

\bibitem{Monnier:2014tfa}
S.~Monnier, ``{Finite higher spin transformations from exponentiation},'' {\em
  Commun. Math. Phys.} {\bf 336} (2015), no.~1 1--26,
  \href{http://xxx.lanl.gov/abs/1402.4486}{{\tt 1402.4486}}.

\bibitem{Alday:2015ota}
L.~F. Alday and A.~Zhiboedov, ``{Conformal Bootstrap With Slightly Broken
  Higher Spin Symmetry},'' \href{http://xxx.lanl.gov/abs/1506.04659}{{\tt
  1506.04659}}.

\bibitem{Kos:2015mba}
F.~Kos, D.~Poland, D.~Simmons-Duffin, and A.~Vichi, ``{Bootstrapping the O(N)
  Archipelago},'' {\em JHEP} {\bf 11} (2015) 106,
  \href{http://xxx.lanl.gov/abs/1504.07997}{{\tt 1504.07997}}.

\bibitem{Aharony:2012nh}
O.~Aharony, G.~Gur-Ari, and R.~Yacoby, ``{Correlation Functions of Large N
  Chern-Simons-Matter Theories and Bosonization in Three Dimensions},'' {\em
  JHEP} {\bf 12} (2012) 028, \href{http://xxx.lanl.gov/abs/1207.4593}{{\tt
  1207.4593}}.

\bibitem{Brower:2006ea}
R.~C. Brower, J.~Polchinski, M.~J. Strassler, and C.-I. Tan, ``{The Pomeron and
  gauge/string duality},'' {\em JHEP} {\bf 12} (2007) 005,
  \href{http://xxx.lanl.gov/abs/hep-th/0603115}{{\tt hep-th/0603115}}.

\bibitem{Cornalba:2007fs}
L.~Cornalba, ``{Eikonal methods in AdS/CFT: Regge theory and multi-reggeon
  exchange},'' \href{http://xxx.lanl.gov/abs/0710.5480}{{\tt 0710.5480}}.

\bibitem{Cornalba:2008qf}
L.~Cornalba, M.~S. Costa, and J.~Penedones, ``{Eikonal Methods in AdS/CFT: BFKL
  Pomeron at Weak Coupling},'' {\em JHEP} {\bf 06} (2008) 048,
  \href{http://xxx.lanl.gov/abs/0801.3002}{{\tt 0801.3002}}.

\bibitem{Costa:2013zra}
M.~S. Costa, J.~Drummond, V.~Goncalves, and J.~Penedones, ``{The role of
  leading twist operators in the Regge and Lorentzian OPE limits},'' {\em JHEP}
  {\bf 04} (2014) 094, \href{http://xxx.lanl.gov/abs/1311.4886}{{\tt
  1311.4886}}.

\bibitem{Arutyunov:2000py}
G.~Arutyunov and S.~Frolov, ``{Four point functions of lowest weight CPOs in
  N=4 SYM(4) in supergravity approximation},'' {\em Phys. Rev.} {\bf D62}
  (2000) 064016, \href{http://xxx.lanl.gov/abs/hep-th/0002170}{{\tt
  hep-th/0002170}}.

\bibitem{Dolan:2004iy}
F.~A. Dolan and H.~Osborn, ``{Conformal partial wave expansions for N=4 chiral
  four point functions},'' {\em Annals Phys.} {\bf 321} (2006) 581--626,
  \href{http://xxx.lanl.gov/abs/hep-th/0412335}{{\tt hep-th/0412335}}.

\bibitem{Alday:2014qfa}
L.~F. Alday and A.~Bissi, ``{Generalized bootstrap equations for $
  \mathcal{N}=4 $ SCFT},'' {\em JHEP} {\bf 02} (2015) 101,
  \href{http://xxx.lanl.gov/abs/1404.5864}{{\tt 1404.5864}}.

\bibitem{Kim:1985ez}
H.~J. Kim, L.~J. Romans, and P.~van Nieuwenhuizen, ``{The Mass Spectrum of
  Chiral N=2 D=10 Supergravity on S**5},'' {\em Phys. Rev.} {\bf D32} (1985)
  389.

\bibitem{D'Hoker:2002aw}
E.~D'Hoker and D.~Z. Freedman, ``{Supersymmetric gauge theories and the AdS /
  CFT correspondence},'' in {\em {Strings, Branes and Extra Dimensions: TASI
  2001: Proceedings}}, pp.~3--158, 2002.
\newblock \href{http://xxx.lanl.gov/abs/hep-th/0201253}{{\tt hep-th/0201253}}.

\bibitem{Dolan:2001tt}
F.~A. Dolan and H.~Osborn, ``{Superconformal symmetry, correlation functions
  and the operator product expansion},'' {\em Nucl. Phys.} {\bf B629} (2002)
  3--73, \href{http://xxx.lanl.gov/abs/hep-th/0112251}{{\tt hep-th/0112251}}.

\bibitem{Hoffmann:2000dx}
L.~Hoffmann, L.~Mesref, and W.~Ruhl, ``{Conformal partial wave analysis of AdS
  amplitudes for dilaton axion four point functions},'' {\em Nucl. Phys.} {\bf
  B608} (2001) 177--202, \href{http://xxx.lanl.gov/abs/hep-th/0012153}{{\tt
  hep-th/0012153}}.

\bibitem{Alday:2014tsa}
L.~F. Alday, A.~Bissi, and T.~Lukowski, ``{Lessons from crossing symmetry at
  large N},'' {\em JHEP} {\bf 06} (2015) 074,
  \href{http://xxx.lanl.gov/abs/1410.4717}{{\tt 1410.4717}}.

\bibitem{Prochazka:2014gqa}
T.~Procházka, ``{Exploring $ {\mathcal{W}}_{\infty } $ in the quadratic
  basis},'' {\em JHEP} {\bf 09} (2015) 116,
  \href{http://xxx.lanl.gov/abs/1411.7697}{{\tt 1411.7697}}.

\end{thebibliography}\endgroup

\end{document}